\documentclass[11pt]{article}
\usepackage{amsmath}
\usepackage{amssymb}
\usepackage{amsthm,amsxtra}
\usepackage{epsf}
\usepackage{eepic}
\newlength{\mytopmargin}
\newlength{\myleftmargin}
\newtheorem{thm}{Theorem}
\newtheorem{cor}{Corollary}
\newtheorem{lemma}{Lemma}
\newtheorem{prop}{Proposition}
\renewcommand{\theequation}{\thesection.\arabic{equation}}

\newcommand{\zz}{\mathbb Z}

\begin{document}
\vspace{4cm}
\noindent
{\bf Interpretations of some parameter dependent generalizations
of classical matrix ensembles}

\vspace{5mm}
\noindent
Peter J.~Forrester${}^*$ and Eric M.~Rains${}^\dagger$

\noindent
${}^*$Department of Mathematics and Statistics,
University of Melbourne, \\
Victoria 3010, Australia ;
${}^\dagger$AT\&T Research, Florham Park, NJ 07932, USA  \\
(Present address: 
Center for Communications 
Research, Princeton, NJ 08540, USA)

\small
\begin{quote}
Two types of parameter dependent generalizations of classical matrix
ensembles are defined by their probability density functions (PDFs).
As the parameter is varied, one interpolates between the
eigenvalue PDF for the superposition of two classical ensembles with
orthogonal symmetry and the eigenvalue PDF for a single classical
ensemble with unitary symmetry, while the other interpolates between
a classical ensemble with orthogonal symmetry and a classical
ensemble with symplectic symmetry. We give interpretations of these PDFs
in terms of probabilities associated to the continuous 
Robinson-Schensted-Knuth correspondence between matrices, with entries chosen
from certain exponential distributions, and non-intersecting lattice paths,
and in the course of this probability measures on partitions and
pairs of partitions are identified. The latter are generalized by using
Macdonald polynomial theory, and a particular continuum limit --- the
Jacobi limit --- of the resulting measures is shown to give PDFs related
to those appearing in the work of Anderson on the Selberg integral. By
interpreting Anderson's work as giving the PDF for the zeros of a certain
rational function, it is then possible to  identify random matrices
whose eigenvalue PDFs realize the original parameter dependent PDFs.
This line of theory allows sampling of the original
parameter dependent PDFs, their Anderson-type generalizations and
associated marginal distributions, from the zeros of certain polynomials
defined in terms of random three term recurrences. 
\end{quote}

\section{Introduction}
This paper is a companion to our work \cite{FR02}. In \cite{FR02} we
studied the correlation functions for the 
probability density functions (PDFs)
\begin{equation}\label{1.1}
{1 \over C} \prod_{j=1}^{2n} e^{-x_j/2} \prod_{j=1}^n e^{A(x_{2j-1} - x_{2j})/2}\prod_{1 \le j < k \le 2n}(x_j - x_k)
\end{equation}
\begin{equation}\label{1.5}
{1 \over C} \prod_{j=1}^{2n} e^{-x_j/2}
\prod_{j=1}^n e^{A(x_{2j-1} - x_{2j})/2}
\prod_{1 \le j < k \le n} (x_{2j-1} - x_{2k-1})(x_{2k} - x_{2j}),
\end{equation}
where $C$ is the normalization (throughout $C$ will be used to
denote {\it some} normalization;
we remark too that it is required $A < 1$ for (\ref{1.1})
to be normalizable) and
\begin{equation}\label{1.2}
x_1 > x_2 > \cdots > x_{2n} \ge 0.
\end{equation}
The first of these was isolated
\cite{BR01a} in the context of a study of generalizations
of Ulam's problem --- the computation of the distribution of the length
of the longest increasing subsequence in a random permutation. 
The second, although not given explicitly in the same paper that (\ref{1.1})
was noted, has its origin in a particular model introduced in
\cite{BR01a}. 

In \cite{FR02} we also studied the correlation functions for the
particular  parameter dependent PDFs
\begin{equation}\label{1.8a}
{1 \over C} \prod_{j=1}^{2n}  x_j^{(a-1)/2}
\prod_{l=1}^n \Big ( {x_{2l} \over  x_{2l-1}} \Big )^{-A/2}
\prod_{1 \le j < k \le 2n} (x_j - x_k),
\end{equation}
\begin{equation}\label{1.8b}
{1 \over C} \prod_{j=1}^{2n}  x_j^{(a-1)/2}
\prod_{l=1}^n \Big ( {x_{2l} \over x_{2l-1}} \Big )^{-A/2}
\prod_{1 \le j < k \le n} (x_{2j-1} - x_{2k-1})(x_{2j} - x_{2k}),
\end{equation}
where the condition $A < a+1$ is required for (\ref{1.8a}) and
(\ref{1.8b}) to be normalizable and
$$
1 > x_1 > x_2 > \cdots > x_{2n} > 0.
$$
(In \cite{FR02} these PDFs were defined on $(-1,1)$ rather than $(0,1)$
as done here; to define the former simply change variables
$x_j \mapsto (1 - x_{2n+1-j})/2$ in the above.) The PDFs (\ref{1.8a})
and (\ref{1.8b}) were identified in  \cite{FR02} as the only parameter
dependent extensions of classical matrix ensembles with an even
number of eigenvalues, in addition to (\ref{1.1}) and (\ref{1.5}), with
the special property that after integrating over every second
eigenvalue the eigenvalue PDF of a matrix ensemble with symplectic and
unitary symmetry respectively results. As noted in \cite{FR02}, by
scaling the variables and parameters
\begin{equation}\label{1.6a}
x_j \mapsto x_j/L, \quad a \mapsto L, \quad A  \mapsto L A
\end{equation}
and taking the limit $L \to \infty$, (\ref{1.8a}) and (\ref{1.8b})
reduce to (\ref{1.1}) and (\ref{1.5}) respectively.

Let us remark at this point how the above PDFs relate to the
classical matrix ensembles. Following the notation of \cite{FR01},
we specify a matrix ensemble with orthogonal $(\beta = 1$), unitary
$(\beta = 2)$ or symplectic $(\beta = 4)$ symmetry by the eigenvalue
PDFs
\begin{equation}\label{1.2a}
{1 \over C} \prod_{l=1}^N g(x_l) \prod_{1 \le j < k \le N}
|x_k - x_j |^\beta.
\end{equation}
A classical matrix ensemble then refers to an eigenvalue PDF of the
form (\ref{1.2a}) with $g(x)$ a classical weight function ---
Gaussian $(e^{-x^2})$, Laguerre $(x^a e^{-x})$, Jacobi $(x^a(1-x)^b)$ or
Cauchy $((1+x^2)^{-\alpha})$. Thus we recognize (\ref{1.1}) in the
case $A=0$ as the Laguerre orthogonal ensemble with $a=0$ and
$2n$ eigenvalues which we denote as LOE${}_{2n}|_{a=0}$, while
we recognize (\ref{1.8a}) in the case $A=0$ as the
Jacobi orthogonal ensemble with $a \mapsto (a-1)/2$,
$b=0$ and $2n$ eigenvalues which we denote as JOE${}_{2n}|_{a \mapsto
(a-1)/2 \atop b=0}$. 
In the limit $A \to - \infty$ of (\ref{1.1}), (\ref{1.8a})
each pair of coordinates collapses, and after
appropriate rescaling and
renaming of the collapsed pairs the limiting PDFs are 
\begin{eqnarray}\label{1.3}
&&{1 \over C} \prod_{j=1}^{n} e^{-x_j} \prod_{1 \le j < k \le n}
(x_j - x_k)^4, \\
&& {1 \over C} \prod_{j=1}^{n} x^{a+1}_j
\prod_{1 \le j < k \le n} (x_j - x_k)^4. \label{1.3a}
\end{eqnarray}
The first of these is the Laguerre symplectic ensemble with parameter $a=0$,
denoted as LSE${}_n|_{a=0}$, while the second is the Jacobi symplectic ensemble
with $a \mapsto a+1$, $b=0$, denoted as JSE${}_n |_{a \mapsto a+1 \atop
b=0}$. Consequently we have that (\ref{1.1}) and (\ref{1.8a}) interpolate
between particular orthogonal ensembles and symplectic ensembles. 
Regarding the PDFs (\ref{1.5}) and (\ref{1.8b}), we require the fact
\cite{FR01} that superimposing two orthogonal ensembles as specified by
(\ref{1.2a}) with $\beta = 1$ at random gives the eigenvalue PDF
\begin{equation}\label{1.3b}
{1 \over C} \prod_{j=1}^{2n} g(x_j) \prod_{1 \le j < k \le n}
(x_{2j-1} - x_{2k-1})(x_{2k} - x_{2j}).
\end{equation}
Thus with $A=0$, (\ref{1.5}) and (\ref{1.8b}) are recognized as the
superimposed ensembles LOE${}_n|_{a=0} \cup {\rm LOE}_n |_{a=0}$ 
and JOE${}_n|_{a\mapsto (a-1)/2 \atop b=0} 
\cup {\rm JOE}_n|_{a\mapsto (a-1)/2 \atop b=0}$ respectively. In the
$A \to - \infty$ limit (\ref{1.5}) and (\ref{1.8b})
effectively reduce to \cite{FR02}
\begin{eqnarray}
&& {1 \over C} \prod_{j=1}^n e^{-x_j}
\prod_{1 \le j < k \le n} (x_k - x_j)^2 \label{1.6} \\
&&{1 \over C} \prod_{j=1}^n x_j^a \prod_{1 \le j < k \le n} 
(x_k - x_j)^2 \label{1.6aa}
\end{eqnarray}
respectively, which specify the Laguerre unitary ensemble with $a=0$,
denoted LUE${}_n|_{a =0}$, and the Jacobi unitary ensemble with $b=0$,
denoted JUE${}_n|_{b=0}$. Consequently (\ref{1.5}) and (\ref{1.8b})
interpolate between particular superimposed orthogonal ensembles and unitary
ensembles.

Our objective in this paper is to give interpretations of each of
(\ref{1.1}), (\ref{1.5}), (\ref{1.8a}) and (\ref{1.8b}) as probability
densities relating to longest increasing subsequence problems in
settings analogous to that already known for (\ref{1.1}), and also to
specify parameter dependent random matrices which have these distributions
as their eigenvalue PDF. We will see that these two pursuits are
intimately related via a conditional PDF to be referred to as the
Anderson density. To establish this link requires first generalizing the
most immediate interpretation of the PDFs in the setting of longest
increasing subsequence problems/ last passage percolation, discussed in
Section 2, to a measure on pairs of partitions $\lambda$, $\kappa$
with $\lambda/\kappa$ a horizontal strip suggested by Macdonald polynomial
theory. This is done in Section 3. Taking a particular continuum limit,
already known from Section 2 as the Jacobi limit, gives rise to the 
Anderson density. The crucial point in making the link with random
matrix theory is an interpretation of the workings of Anderson's
paper \cite{An91} as giving the density of zeros of a certain random 
rational function. This same random rational function occurs as part of the
characteristic equation for the random projection of a fixed matrix
with in general degenerate eigenvalues. Such a random projection is
used in Section 4.1 to give the construction of random matrices
with eigenvalue PDFs (\ref{1.8a}) and (\ref{1.8b}). It is shown in
Section 4.2 that the intricacies of the relationship between the
Anderson density and the random rational function allow a 
generalization of the joint densities (\ref{1.8a}) and (\ref{1.8b})
--- as well as an associated marginal density --- to be sampled from
the zeros of a polynomial generated by a random three term recurrence.

The Anderson density and the corresponding random rational function
have a well defined Laguerre limit giving rise to a generalization
of the joint densities (\ref{1.1}) and (\ref{1.5}). As noted in
Section 5.1 the Laguerre limit of the random rational function occurs
as part of the eigenvalue equation for a random rank 1 projection of
a fixed matrix with in general degenerate eigenvalues, allowing for the
construction of random matrices with eigenvalue PDFs 
(\ref{1.1}) and (\ref{1.5}). Furthermore it is shown in Section 5.2 that the
Laguerre limit of the random three term recurrences of
Section 4.2 generates polynomials from which a generalization of
(\ref{1.1}) and (\ref{1.5}), and an  associated marginal density
corresponding to the Laguerre limit of the Selberg integral,
can be sampled.

In Section 6 we carry out a further limiting analysis of the results of
Section 4, this time characterized by the Jacobi type weights associated
to the Anderson density degenerating to Gaussian weights. A special
case of the resulting parameter dependent Gaussian ensemble is known
from Section 2.3 as the joint probability density for a particular
last passage percolation model.

In the course of our study we encounter the need to further develop/
reformulate some existing theory, in particular the 
Robinson-Schensted-Knuth correspondence. We also encounter some
consequences of our findings by way of new insights into some existing
results. Such points are presented in the Appendices.

\section{Last passage percolation and tableaux coordinates}
\setcounter{equation}{0}
The generalizations of Ulam's problem of interest to us can in
turn be regarded as generalizations of a last passage percolation
model introduced by Johansson \cite{Jo00}. To define the latter
consider the right quadrant square lattice $\{(i,j): \,
i,j \in \zz^+\}$. Associate with each lattice site $(i,j)$ a
random non-negative integer variable $x_{i,j}$, chosen from the
geometric distribution with parameter $a_i b_j$ so that
\begin{equation}\label{6.1}
{\rm Pr}(x_{i,j} = k) = (1 - a_i b_j) (a_i b_j)^k.
\end{equation}
For given non-negative parameters $a_1,a_2,\dots,b_1,b_2,\dots$ the
quantity of interest is the distribution of the so-called last
passage time
\begin{equation}\label{6.2}
L(n_1,n_2) := \max \sum_{(1,1) {\rm u/rh} (n_1,n_2)} x_{i,j}
\end{equation}
where the notation $(1,1) {\rm u/rh} (n_1,n_2)$ denotes that the sum
is over all lattice points in a path starting at $(1,1)$ and finishing
at $(n_1,n_2)$, with segments which are either up or right horizontal
(such a path is said to be weakly increasing). The celebrated
Robinson-Schensted-Knuth (RSK) correspondence (see e.g.~\cite{Fu97,vL96})
gives a bijection between
$n_1 \times n_2$ non-negative integer matrices and pairs of
semi-standard tableaux of the same shape $\mu = (\mu_1,\mu_2,\dots,\mu_n)$
say with the crucial feature that $\mu_1 = L(n_1,n_2)$. It is these variables
which after rescaling give rise to the parameter dependent PDFs listed
in the Introduction.

We require some (mostly)
known facts about the RSK correspondence in the situation
that there is a measure on the space of non-negative integer matrices as
implied by (\ref{6.1}). First, the probability an $n_1 \times n_2$
non-negative integer matrix with this measure corresponds to a pair of
semi-standard tableaux with shape $\mu$, one of
content $n_1$, the other of content $n_2$ is given by \cite{Kn70}
\begin{equation}\label{6.3}
\prod_{i=1}^{n_1} \prod_{j=1}^{n_2} (1 - a_i b_j)
s_\mu(a_1,\dots,a_{n_1}) s_\mu(b_1,\dots,b_{n_2}),
\end{equation}
where $s_\mu$ denotes the Schur polynomial. 
Second, for $n_2 \ge n_1$, the joint probability that an
$n_1 \times (n_2 + 1)$ non-negative integer matrix with measure
implied by (\ref{6.1}) corresponds to a pair of
semi-standard tableaux with shape $\mu$, content $n_1$ and
$n_2 + 1$, and that the $n_1 \times n_2$ bottom left
sub-block corresponds to a pair of semi-standard tableaux with
shape $\kappa$, content $n_1$ and $n_2$ is
\begin{equation}\label{2.19}
\prod_{i=1}^{n_1} \prod_{j=1}^{n_2+1} (1 - a_i b_j)
s_\mu(a_1,\dots,a_{n_1}) s_\kappa(b_1,\dots,b_{n_2})
b_{n_2+1}^{\sum_{j=1}^{n_1} \mu_j - \kappa_j}
\end{equation}
where
\begin{equation}\label{2.11}
\mu_1 \ge \kappa_1 \ge \mu_2 \ge \kappa_2 \ge \cdots
\ge \mu_{n_1} \ge \kappa_{n_1} \ge 0.
\end{equation}
Note that by the assumption $n_1 \le n_2$, we require
$\ell(\mu)$ --- the number of non-zero parts of $\mu$ --- to
be less than or equal to $n_1$ for (\ref{2.19}) to be non-zero.
If instead $n_1 > n_2$, then (\ref{2.19}) holds with
\begin{equation}\label{2.11'}
b_{n_2+1}^{\sum_{j=1}^{n_1} \mu_j - \kappa_j} \mapsto
b_{n_2+1}^{\sum_{j=1}^{n_2} (\mu_j - \kappa_j) + \mu_{n_2+1}}
\end{equation}
and (\ref{2.11}) must be modified to read
\begin{equation}\label{2.11a}
\mu_1 \ge \kappa_1 \ge \mu_2 \ge \kappa_2 \ge \cdots
\ge \mu_{n_2} \ge \kappa_{n_2} \ge \mu_{n_2+1} \ge 0.
\end{equation}
Third, in the case of
matrices symmetric about $i=j$ (here $i$ denotes the row counted from the
bottom), when the RSK correspondence maps the matrix to a single
semi-standard tableau, with the parameters of the geometric
distribution chosen so that
\begin{equation}\label{6.3a}
{\rm Pr}(x_{i,j} = k) = (1 - a_i a_j) (a_i a_j)^k, \: \: i < j \qquad
{\rm Pr}(x_{i,i}= k) = (1 - a_i) a_i^k,
\end{equation}
the derivation of (\ref{6.3}) given in \cite{Kn70} implies the
probability that the tableau has shape $\mu$ is given by
\begin{equation}\label{6.4}
\prod_{i=1}^n (1 - a_i)
\prod_{1 \le i < j \le n} (1 - a_i a_j)
s_\mu(a_1,\dots,a_n).
\end{equation}
Fourth, in this latter situation, the correspondence is such that
\begin{equation}\label{6.5}
\sum_{j=1}^n x_{j,j} = \sum_{j=1}^n (-1)^{j-1} \mu_j.
\end{equation}
In \cite{Kn70}, (\ref{6.5}) is given in the form
\begin{equation}\label{6.6}
\sum_{j=1}^n x_{j,j} = 
\sum_{j=1}^{\ell(\mu)} {1 \over 2} (1 - (-1)^{\mu_j'})
\end{equation}
where $\mu_j'$ denotes the length of the $j$th column of the
diagram of $\mu$, which is the sum of the number of odd columns;
simple reasoning shows that (\ref{6.6}) is equivalent to (\ref{6.5}).
In Appendix A we will give a self contained derivation of
(\ref{6.5}) which is in keeping with the interpretation of
the RSK correspondence as a cascade of growth models given
in \cite{Jo02}. The same ideas will be used to derive
(\ref{2.19}).

It follows from (\ref{6.4}) and (\ref{6.5}) that if (\ref{6.3a})
is modified so that the second equation reads
\begin{equation}\label{6.7}
{\rm Pr}(x_{i,i} = k) = (1 - \alpha a_i) (\alpha a_i)^k
\end{equation}
then the probability that the tableau has shape $\mu$ is given by
\begin{equation}\label{6.8}
\prod_{i=1}^n (1 - \alpha a_i)
\prod_{1 \le i < j \le n} (1 - a_i a_j)
\alpha^{\sum_{j=1}^n (-1)^{j-1} \mu_j}
s_\mu(a_1,\dots,a_n).
\end{equation}
We will show that (\ref{1.1}) and (\ref{1.8a}) result from this
probability. The PDFs
(\ref{1.5}) and (\ref{1.8b}) will be derived from (\ref{2.19}).

Before undertaking the derivations, we note that from the origin of 
(\ref{2.19}) as a joint probability with the corresponding marginal
density given or implied by (\ref{6.3}), the former must satisfy
special identities with respect to summation over $\kappa$ and
summation over $\mu$. Thus let $R$ denote the region
(\ref{2.11}) in the case $n_2 \ge n_1$, and
the region (\ref{2.11a}) in the case $n_1 > n_2$. Then since
summing (\ref{2.19}) over $\kappa \in R$ must give (\ref{6.3})
with $n_2 \mapsto n_2+1$, it follows
\begin{equation}\label{3.39}
\sum_{\kappa: \kappa \in R} s_\kappa(b_1,\dots,b_{n_2})
b_{n_2 + 1}^{|\mu|-|\kappa|} =
s_\mu(b_1,\dots,b_{n_2 + 1})
\end{equation}
where $|\mu| = \sum_{i=1}^{\ell(\mu)} \mu_i$ and similarly the
meaning of $|\kappa|$. In fact (\ref{3.39}) is a well known
recurrence satisfied by the Schur polynomials
\cite[special case of (5.10) pg.~72]{Ma95}.
Similarly, summing (\ref{2.19}) over $\mu$ must give (\ref{6.3})
with $\mu \mapsto \kappa$ and so
\begin{equation}\label{3.39a}
\prod_{i=1}^{n_1}(1 - a_i b_{n_2+1})
\sum_{\mu: \mu \in R}
s_\mu(a_1,\dots,a_{n_1}) b_{n_2 + 1}^{|\mu|-|\kappa|} =
s_\kappa(a_1,\dots,a_{n_1})
\end{equation}
which is also a known Schur polynomial identity
\cite[special case of (1) pg.~93]{Ma95}.
 
\subsection{Jacobi limit}
The parameter dependent Jacobi ensembles (\ref{1.8a}), (\ref{1.8b})
are obtained from (\ref{6.8}), (\ref{2.19}) respectively by
specializing the parameters $\{a_i\}$, $\{b_j\}$. In
(\ref{6.8}) we choose
\begin{equation}\label{8.0a}
(a_1,\dots,a_n) = (z,zt, zt^2,\dots, z t^{n-1})
\end{equation}
while in (\ref{2.19}) we choose
\begin{equation}\label{8.0b}
(a_1,\dots,a_{n_1}) = (z,zt, zt^2,\dots, z t^{n_1-1}), \qquad
(b_1,\dots,b_{n_2}) = (z,zt, zt^2,\dots, z t^{n_2-1}).
\end{equation}
The probabilities then assume an explicit form in terms of the parts
of $\mu$ and $\kappa$ due to the evaluation formula
\cite{Ma95}
\begin{eqnarray}\label{8.1}
s_\lambda(1,t,\dots,t^{n-1}) & = & t^{\sum_{i=1}^n (i-1) \lambda_i}
\prod_{1 \le i < j \le n}
{1 - t^{\lambda_i - \lambda_j - i + j} \over 1 - t^{j-i} }
\nonumber \\
& = &
{ t^{-\sum_{j=1}^n (j-1) (n^*-j)} 
\over \prod_{l=1}^{n-1} (t;t)_l}
\prod_{1 \le i < j \le n} (t^{h_j} - t^{h_i})
\end{eqnarray}
where in the second line, which follows from the first by simple
manipulation,  $(t;t)_l := (1-t)(1-t^2)
\cdots (1 - t^l)$, $h_j := \lambda_j + n^* - j$ and $n^*$
is arbitrary. 

Substituting (\ref{8.1}) with $n^* = n$ in (\ref{6.8}) we
deduce the following result.

\begin{prop}\label{p1}
On each site $(i,j)$ of the $n \times n$ square lattice, specify a
non-negative integer $x_{i,j}$ according to the probability
distribution
\begin{eqnarray*}
{\rm Pr}(x_{i,j} = k) & = & (1 - z^2 t^{i+j-2})
(z^2 t^{i+j-2})^k \quad i < j \\
{\rm Pr}(x_{i,i} = k) & = & (1 - \alpha z t^{i-1})
(\alpha z t^{i-1})^k
\end{eqnarray*}
and impose the symmetry constraint that $x_{i,j} = x_{j,i}$ for
$i > j$. Then the probability that a configuration $[x_{i,j}]$
gives a tableau of shape $\mu$ under the RSK correspondence is
equal to
\begin{equation}\label{8.8}
c_n(z,\alpha,t) z^{\sum_{j=1}^n h_j}
\alpha^{\sum_{j=1}^n (-1)^{j-1} h_j}
\prod_{1 \le i < j \le n} (t^{h_j} - t^{h_i}),
\end{equation}
\begin{eqnarray}
c_n(z,\alpha,t) & := & z^{-\sum_{j=1}^n (n-j)} \alpha^{-[n/2]}
{t^{-\sum_{j=1}^n (j-1)(n-j)} \over \prod_{l=1}^{n-1}(t;t)_l}  \nonumber \\
&  & \times \prod_{i=1}^n(1 - \alpha z t^{i-1})
\prod_{1 \le i < j \le n}
(1 - z^2 t^{i+j-2}),
\end{eqnarray}
where $h_j := \mu_j + n - j$ and thus
$h_1 > h_2 > \cdots > h_n \ge 0$.
 Furthermore in the scaled (Jacobi)
limit
\begin{equation}\label{8.8a}
t = e^{-1/L}, \: \: z= e^{-a/L}, \: \:
\alpha = e^{-a_1/L}, \: \: h_j/L = x_j, \: \: L \to \infty,
\end{equation}
when each lattice site $(i,j)$ specifies a non-negative continuous
exponential random variable with site dependent variance
\begin{eqnarray}
{\rm Pr}(x_{i,j} \in [y,y+dy]) & = & (i+j-2+2a) 
e^{-y(i+j-2+2a)}dy, \qquad
i < j \nonumber \\
{\rm Pr}(x_{i,i} \in [y,y+dy]) & = & (i-1+a+a_1) e^{-y(i-1+a+a_1)}dy,
\end{eqnarray}
the probability (\ref{8.8}) multiplied by $L^n$ tends to the
PDF
\begin{eqnarray}\label{8.9}
&&\tilde{c}_n(a,a_1) e^{-a \sum_{j=1}^{n} x_j} e^{-a_1 \sum_{j=1}^n
(-1)^{j-1} x_j} \prod_{1 \le i < j \le n}
(e^{-x_j} - e^{-x_i}),  \\
&& \qquad \tilde{c}_n(a,a_1)  := {\Gamma(a + a_1 + n) \over \Gamma(a + a_1)}
{1 \over \prod_{l=1}^{n-1} l!}
\prod_{i=1}^{n-1} {\Gamma (2a + i + n - 1) \over
\Gamma(2a + 2i - 1)}
\end{eqnarray}
where $x_1 > x_2 > \cdots > x_n > 0$.
\end{prop}

\noindent
After the change of variables and replacement of parameters
\begin{equation}\label{2.29}
e^{-x_j} \mapsto  x_{n+1-j}, \: \:
a \mapsto (a+1)/2, \: \: a_1 \mapsto - A/2, \: \:
n \mapsto 2n
\end{equation}
we see that (\ref{8.9}) coincides with (\ref{1.8a}).

Let us now consider the specialization (\ref{8.0b}) in (\ref{2.19}).
We must first give the form of (\ref{8.1}) in the case that
$n \mapsto n_2$, $\lambda \mapsto \kappa$, $\ell(\kappa) = n_1$
with $n_1 \le n_2$ so that $\kappa_{n_1+1} = \cdots =
\kappa_{n_2} = 0$. Then with $n^* = n_1$ 
and $r_j := \kappa_j + n_1 - j$, manipulation of
(\ref{8.1}) shows
\begin{eqnarray}\label{2.28}
&&
s_\kappa(1,t,\dots,t^{n_2-1})  =  
t^{-\sum_{j=1}^{n_2-n_1} j(j-1)}
t^{-n_1 \sum_{j=1}^{n_2 - n_1} j}
{ t^{-\sum_{j=1}^{n_2} (j-1) (n_1-j)} 
\over \prod_{l=1}^{n_2-1} (t;t)_l}
\nonumber \\
&& \qquad \times \prod_{i=1}^{n_2-n_1-1} (t;t)_i
\prod_{i=1}^{n_1} {(t;t)_{r_i + n_2 - n_1} \over
(t;t)_{r_i} }
\prod_{1 \le i < j \le n_1} (t^{r_j} - t^{r_i})
\end{eqnarray}
where the first product in the second line must be replaced by unity
if $n_2=n_1, n_1+1$.
Substituting this result, and (\ref{8.1}) with $n \mapsto n_1$, $n^* = n_1$, 
$\lambda \mapsto \mu$ in (\ref{2.19}) we deduce an interpretation
of the PDF (\ref{1.5}) in the context of a last passage
percolation model.

\begin{prop}\label{p2}
Let $n_2 \ge n_1$. On each site of the $n_1 \times (n_2 +1)$ square
lattice specify a non-negative integer $x_{i,j}$ according to
the probability distribution
\begin{eqnarray}\label{2.28'}
{\rm Pr}(x_{i,j} = k) & = & (1 - z^2 t^{i+j-2}) (z^2 t^{i+j-2})^k,
\quad j \ne n_2 +1 \nonumber \\
{\rm Pr}(x_{i,{n_2+1}} = k) & = & (1 - \alpha z t^{i-1})
(\alpha z t^{i-1})^k.
\end{eqnarray}
The joint probability that a configuration $[x_{i,j}]$ gives,
under the RSK correspondence, a pair of tableaux of shape $\mu$,
one of content $n_1$ and the other of content $n_2 + 1$,
and that the subconfiguration $[x_{i,j}]_{i=1,\dots,n_1 \atop
j=1,\dots, n_2}$ gives a  pair of tableaux of shape $\kappa$,
one of content $n_1$ and the other of content $n_2$, is non-zero
if and only if
\begin{equation}\label{2.28a}
h_1 \ge r_1 > h_2 \ge r_2 > \cdots > h_{n_1} \ge r_{n_1} \ge 0,
\end{equation}
where $h_j := \mu_j +n_1 - j$  and $r_j := \kappa_j + n_1 - j$.
Furthermore the joint
probability then has the explicit form
\begin{equation}\label{2.33}
k_{n_1,n_2}(z,\alpha,t) z^{\sum_{j=1}^{n_1} (h_j+r_j)}
\alpha^{\sum_{j=1}^{n_1}( h_j - r_j)}
\prod_{i=1}^{n_1} {(t;t)_{r_i+n_2 - n_1} \over
(t;t)_{r_i} }
\prod_{1 \le i < j \le n_1} (t^{h_j} - t^{h_i})
(t^{r_j} - t^{r_i}),
\end{equation}
\begin{eqnarray}\label{2.34}
k_{n_1,n_2}(z,\alpha,t) & := & z^{-2 \sum_{j=1}^{n_1} (n_1 - j)}
{t^{-\sum_{j=1}^{n_1}(j-1)(n_1-j)}  
\over \prod_{l=1}^{n_1-1}(t;t)_l}
\nonumber \\&& \times
t^{-\sum_{j=1}^{n_2-n_1} j(j-1)}
t^{-n_1 \sum_{j=1}^{n_2 - n_1} j}
{ t^{-\sum_{j=1}^{n_2} (j-1) (n_1-j)} 
\over \prod_{l=1}^{n_2-1} (t;t)_l} \nonumber \\
&& \times \prod_{l=1}^{n_2-n_1-1} (t;t)_l
\prod_{i=1}^{n_1} \prod_{j=1}^{n_2}
(1 - z^2 t^{i+j-2}) \prod_{i=1}^{n_1} (1 - \alpha z t^{i-1}).
\end{eqnarray}
In the Jacobi limit (\ref{8.8a}) (with the additional scaled quantity 
$r_j/L=:y_j$), (\ref{2.33}) multiplied
by $L^{2n_1}$ tends to the PDF
\begin{eqnarray}\label{2.33a}
&&\tilde{k}_{n_1,n_2}(a,a_1) \prod_{i=1}^{n_1}(1-e^{-y_i})^{n_2-n_1}
\prod_{1 \le i < j \le n_1}(e^{-y_j} - e^{-y_i})(e^{-x_j} - e^{-x_i})
\nonumber \\
&& \qquad \times e^{-a \sum_{j=1}^{n_1}(x_j+y_j)}
e^{-a_1 \sum_{j=1}^{n_1}(x_j-y_j)},
\end{eqnarray}
\begin{equation}\label{2.34a}
\tilde{k}_{n_1,n_2}(a,a_1) :=
{\prod_{l=1}^{n_2-n_1-1} l! \over (\prod_{l=1}^{n_1-1} l!)
(\prod_{l=1}^{n_2-1} l!) }
{\Gamma(a+a_1+n_1) \over \Gamma(a+a_1)}
\prod_{i=1}^{n_1} {\Gamma(2a+i-1+n_2) \over
\Gamma(2a+i-1)},
\end{equation}
where it is required that
\begin{equation}\label{2.34b}
x_1 > y_1 > x_2 > y_2 > \cdots > x_{n_1} > y_{n_1} > 0.
\end{equation} 
\end{prop}

\noindent
Analogous to (\ref{2.29}), after the change of variables and
replacement of parameters
\begin{equation}\label{2.39}
e^{-x_j} \mapsto  x_{2n+1-2j}, \: \:
e^{-y_j}  \mapsto x_{2n+2-2j}, \: \:
a \mapsto (a+1)/2, \: \: a_1 \mapsto - A/2
\end{equation}
we see that with $n_2 = n_1 = n$ (\ref{2.33a}) coincides with (\ref{1.8b}).

To specialize (\ref{2.19}) modified by the replacement (\ref{2.11'})
according to (\ref{8.0b}), we note that for this to be non-zero
(\ref{2.11a}) gives we require $\ell(\mu) \le n_2 + 1$.
Making the replacements $\kappa \mapsto \mu$, $n_2 \mapsto n_1$,
$n_1 \mapsto n_2 + 1$, $r_j \mapsto h_j :=
\kappa_j + n_2 + 1 - j$ in (\ref{2.28}) shows
\begin{eqnarray}\label{2.28b}
&&
s_\mu(1,t,\dots,t^{n_1-1})  =
t^{-\sum_{j=1}^{n_1-n_2-1} j(j-1)}
t^{-(n_2 + 1) \sum_{j=1}^{n_2 - (n_1 + 1)} (j - 1)}
{ t^{-\sum_{j=1}^{n_1} (j-1) (n_2+1-j)} 
\over \prod_{l=1}^{n_1-1} (t;t)_l}
\nonumber \\
&& \qquad \times
\prod_{l=1}^{n_1-n_2-2}(t;t)_l
\prod_{i=1}^{n_2+1} {(t;t)_{h_i + n_1 - (n_2+1) } \over
(t;t)_{h_i} }
\prod_{1 \le i < j \le n_2+1} (t^{h_j} - t^{h_i}).
\end{eqnarray} 
Substituting this result, and (\ref{8.1}) with $n \mapsto n_2$, $n^* = n_2$,
$\lambda \mapsto \kappa$, $h_j \mapsto r_j := \kappa_j + n_2 - j$ we
deduce the analogue of Proposition \ref{p2} in the case $n_1 > n_2$.

\begin{prop}\label{p3}
Let $n_1 > n_2$. On each site of the $n_1 \times (n_2 +1)$ square
lattice specify a non-negative integer $x_{i,j}$ according to
the probability distribution (\ref{2.28'}).
The joint probability that a configuration $[x_{i,j}]$ gives,
under the RSK correspondence, a pair of tableaux of shape $\mu$,
one of content $n_1$ and the other of content $n_2 + 1$,
and that the subconfiguration $[x_{i,j}]_{i=1,\dots,n_1 \atop
j=1,\dots, n_2}$ gives a  pair of tableaux of shape $\kappa$,
one of content $n_1$ and the other of content $n_2$, is non-zero
if and only if
\begin{equation}\label{2.29a}
h_1 \ge r_1 > h_2 \ge r_2 > \cdots > h_{n_2} \ge r_{n_2} > h_{n_2+1} \ge 0,
\end{equation}
where $h_j := \mu_j +n_2 +1 - j$  and $r_j := \kappa_j + n_2+1 - j$.
Furthermore the joint probability has the explicit form
\begin{eqnarray}\label{2.37}
&&
K_{n_1,n_2}(\alpha,z,t) z^{\sum_{j=1}^{n_2} (h_j+r_j)}
\alpha^{\sum_{j=1}^{n_2}( h_j - r_j)} (z \alpha)^{h_{n_2 + 1}}
\prod_{i=1}^{n_2+1} {(t;t)_{h_i+n_1 - (n_2+1)} \over
(t;t)_{h_i} }
\prod_{1 \le i < j \le n_2+1} (t^{h_j} - t^{h_i}) \nonumber \\&&
\hspace{5cm}
\times
\prod_{1 \le i < j \le n_2} (t^{r_j} - t^{r_i}),
\end{eqnarray}
\begin{eqnarray}
K_{n_1,n_2}(\alpha,z,t) & := & z^{-2 \sum_{j=1}^{n_2} (n_2 - j)}
{t^{-\sum_{j=1}^{n_2}(j-1)(n_2-j)} 
\over \prod_{l=1}^{n_2-1}(t;t)_l}
\nonumber \\&& \times
t^{-\sum_{j=1}^{n_1-n_2-1} j(j-1)}
t^{-(n_2+1) \sum_{j=1}^{n_1 - n_2 -1} j}
{ t^{-\sum_{j=1}^{n_1} (j-1) (n_2+1-j)} 
\over \prod_{l=1}^{n_1-1} (t;t)_l} \nonumber \\
&& \times 
\prod_{l=1}^{n_1-n_2-2}(t;t)_l
\prod_{i=1}^{n_1} \prod_{j=1}^{n_2}
(1 - z^2 t^{i+j-2}) \prod_{i=1}^{n_1} (1 - \alpha z t^{i-1}).
\end{eqnarray}
In the Jacobi limit, (\ref{2.37}) multiplied by $L^{2n_2+1}$ tends to the
PDF
\begin{eqnarray}\label{2.43a}
&&\tilde{K}_{n_1,n_2}(a,a_1) \prod_{i=1}^{n_2+1}(1-e^{-x_i})^{n_1-(n_2+1)}
\prod_{1 \le i < j \le n_2+1}(e^{-x_j} - e^{-x_i})
\prod_{1 \le i < j \le n_2} (e^{-y_j} - e^{-y_i})
\nonumber \\
&& \qquad \times e^{-a \sum_{j=1}^{n_2}(x_j+y_j)}
e^{-a_1 \sum_{j=1}^{n_2}(x_j-y_j)} e^{-(a+a_1) x_{n_2+1}},
\end{eqnarray}
\begin{equation}\label{2.44a}
\tilde{K}_{n_1,n_2}(a,a_1) :=
{\prod_{l=1}^{n_1-n_2-2} l! \over (\prod_{l=1}^{n_1-1} l!)
(\prod_{l=1}^{n_2-1} l!) }
{\Gamma(a+a_1+n_1) \over \Gamma(a+a_1)}
\prod_{i=1}^{n_1} {\Gamma(2a+i+n_2-1) \over
\Gamma(2a+i-1)},
\end{equation}
where it is required that
\begin{equation}\label{2.44b}
x_1 > y_1 > x_2 > y_2 > \cdots > x_{n_2} > y_{n_2} > x_{n_2+1} > 0.
\end{equation}
\end{prop}

\noindent
With $n_1=n_2+1$, making the change of variables and replacements
(\ref{2.39}) in (\ref{2.43a}) gives the natural generalization
of (\ref{1.8a}) to the case of an odd number of coordinates. 

In the case $n_1=n_2=:n$, there is yet another combinatorial
interpretation of the joint probability (\ref{2.33}), which
relates to a particular model (model (v)) introduced in \cite{BR01c}.
Thus consider the $2n \times 2n$ square lattice of sites $(i,j)$,
$1 \le i,j \le 2n$. For the triangular shaped region specified by
$1 \le i \le 2n-1$, $i \ge j$, $i \le 2n+1-j$ associate with each
lattice site a non-negative integer $x_{i,j}$ chosen according to
the probability distributions
\begin{eqnarray*}
{\rm Pr}(x_{i,j}=k) & = & {\rm Pr}(x_{i,2n+1-j}=k) \: = \:
(1 - z^2 t^{i+j-2})(z^2 t^{i+j-2})^k, \qquad i,j \le n \: \: (i \ne j) \\
{\rm Pr}(x_{i,i}=k) & = & (1 - \alpha z t^{i-1}) (\alpha z t^{i-1})^k, \\
{\rm Pr}(x_{i, 2n+1-i = k}) & = & 0.
\end{eqnarray*}
With the $x_{i,j}$ in this region thus chosen, specify $x_{i,j}$ at the
remaining lattice sites in the square by the symmetry requirements
\begin{equation}\label{nv1}
x_{j,i} = x_{i,j}, \qquad x_{2n+1-i,2n+1-j} = x_{i,j}.
\end{equation}
The first of the symmetries in (\ref{nv1}) implies that under the RSK
correspondence the integer matrix maps to a single semi-standard tableau
$\mu$ (of content $2n$), while the second symmetry implies that each row is
of even length. At a more sophisticated level, the resulting tableau is
constrained to be self-dual (invariant under Sch\"utzenberger involution).
Although we don't present the details, using ideas from \cite{BR01a},
from this one can show that with $h_j = \mu_{2j-1}/2 + n -j$ and
$r_j = \mu_{2j}/2 + n -j$ the probability that $[x_{i,j}]$ maps to
the semi-standard tableau $\mu$ is given by (\ref{2.33}) with
$n_1 = n_2 = n$.

We remark at this point that the special case $a=1$, $A=0$ of the
Jacobi parameter dependent PDF (\ref{1.8b}) occurs as various
probabilities in the work of Ciucu \cite{Ci97},
and Krattenthaler \cite{Kr97} on
perfect matchings (tilings) on the Aztec lattice with removed
sites.

\subsection{Laguerre limit}
It has already been remarked that after writing $x_j \mapsto
{1 \over 2}(x_j + 1)$, $(j=1,\dots,2n)$, then scaling the
variables and parameters according to (\ref{1.6a}) and taking the
limit $L \to \infty$, the parameter dependent Jacobi ensembles
(\ref{1.8a}) and (\ref{1.8b}) reduce to the parameter
dependent Laguerre ensembles (\ref{1.1}) and (\ref{1.5})
respectively. As first noticed by Johansson \cite{Jo00},
Laguerre ensembles can be obtained directly from the Schur
measure (\ref{6.3}) by first setting all the variables equal
(and thus choosing $t=1$ in (\ref{8.0b})), then scaling the
remaining parameters and variables as in (\ref{8.8a}). We
thus obtain the following interpretation of the parameter
dependent Laguerre ensembles.

\begin{prop}
First, on each site $(i,j)$ of the $n\times n$ square lattice, specify
a continuous exponential random variable
\begin{eqnarray*}
{\rm Pr}(x_{i,j} \in [y,y+dy]) & = & 2a e^{-2a y} \, dy, \qquad i < j \\
{\rm Pr}(x_{i,i} \in [y,y+dy]) & = & (a+a_1) e^{-(a+{a}_1)y} \, dy
\end{eqnarray*}
and impose the symmetry constraint that $x_{i,j} = x_{j,i}$ for
$i > j$. Then the probability density that a configuration
$[x_{i,j}]$ gives, under the continuous RSK correspondence of
Appendix A, a non-intersecting path configuration with maximum
displacement $x_l$ at level-$l$, is given by (see also \cite{Ba02})
\begin{equation}
{(a+a_1)^n (2a)^{n(n-1)/2} \over \prod_{l=1}^{n-1} l!}
e^{-a\sum_{j=1}^n x_j} e^{-a_1 \sum_{j=1}^n (-1)^{j-1} x_j}
\prod_{1 \le i < j \le n} (x_i - x_j)
\end{equation}
where $x_1 > x_2 > \cdots > x_n > 0$. For $n \mapsto 2n$ this is
equivalent to (\ref{1.1}).

Second,  on each site $(i,j)$ of the $n_1\times(n_2+1)$ square lattice
specify a continuous exponential random variable
\begin{eqnarray}\label{aa1}
{\rm Pr}(x_{ij} \in [y,y+dy]) & = & 2a e^{-2a y} \, dy, \qquad j \ne n_2 +1
\nonumber \\
{\rm Pr}(x_{i\, n_2+1} \in [y,y+dy]) & = & (a+a_1) e^{-(a+{a}_1)y} \, dy.
\end{eqnarray}
Then for $n_2 \ge n_1$ the joint probability density that a configuration
$[x_{i,j}]$ gives, under the continuous RSK correspondence of
Appendix A, a non-intersecting path configuration with maximum
displacement $x_l$ at level-$l$, and that the subconfiguration
$[x_{i,j}]_{i=1,\dots,n_1 \atop j=1,\dots,n_2}$ gives a
non-intersecting path configuration with maximum
displacement $y_l$ at level-$l$ is non-zero if and only if the
interlacing condition (\ref{2.34b}) holds, when it has the explicit form
\begin{eqnarray}
&&
{(2a)^{n_1n_2}(a+a_1)^{n_1} \prod_{l=1}^{n_2-n_1-1} l! \over
\prod_{l=1}^{n_1-1}l! \prod_{l=1}^{n_2-1}l! }
e^{-a\sum_{j=1}^{n_1} (x_j+y_j)} 
e^{-a_1 \sum_{j=1}^{n_1} (x_j-y_j)}
\prod_{i=1}^{n_1} y_i^{n_2-n_1} \nonumber \\
&& \qquad \times
\prod_{1 \le i < j \le n_1} (x_i - x_j) (y_i - y_j)
\end{eqnarray}
In the case $n_1=n_2$ this is equivalent to the PDF (\ref{1.5}). For
$n_2 < n_1$, the same joint probability density is non-zero if and only
if the interlacing condition (\ref{2.44b}) holds, when it has the 
explicit form
\begin{eqnarray}\label{aa3}
&&
{(2a)^{n_1n_2}(a+a_1)^{n_1} \prod_{l=1}^{n_1-n_2-2} l! \over
\prod_{l=1}^{n_1-1}l! \prod_{l=1}^{n_2-1}l! }
e^{-a\sum_{j=1}^{n_2} (x_j+y_j)} 
e^{-a_1 \sum_{j=1}^{n_2}  (x_j-y_j)}
\prod_{i=1}^{n_2+1} y_i^{n_2-n_1} \nonumber \\
&& \qquad \times
\prod_{1 \le i < j \le n_2+1} (x_i - x_j) 
\prod_{1 \le i < j \le n_2}(y_i - y_j)
\end{eqnarray}
\end{prop}

\subsection{Gaussian limit}
It was pointed out by Baryshnikov \cite{Ba01}, upon interpreting a result
of Glynn and Whitt \cite{GW91}, that for $x_{i,j}$ i.i.d.~random
variables with finite variance, the quantity $L(n_1,n_2)$ specified
by (\ref{6.2}) has a universal scaled form in the limit
$n_1 \to \infty$, independent of the details of the distribution.
This universal form is the PDF for the distribution of the largest
eigenvalue in the GUE of $n_2\times n_2$ random complex
Hermitian matrices, which have the joint eigenvalue
probability density
\begin{equation}\label{2.45'}
{1 \over C} \prod_{l=1}^{n_2} e^{-x_l^2} 
\prod_{1 \le j < k \le n_2} (x_k - x_j)^2.
\end{equation}
It follows that with $a=a_1$ in (\ref{aa1}) so as to obtain
i.i.d.~random variables, we can expect to obtain a Gaussian type
ensemble by taking the scaled $n_1 \to \infty$ limit in the joint
probability (\ref{aa3}). To see that this occurs requires nothing
more than the classical transition between the Laguerre and
Gaussian weights,
$$
\lim_{c \to \infty} e^c e^{-cx} x^c
\Big |_{x \mapsto 1 + x\sqrt{2/c}} = e^{-x^2},
$$
allowing us to derive the following result.

\begin{prop}
In (\ref{aa3}) taking the Gaussian limit by setting
$$
a=a_1, \: \: 2a=n_1-(n_2+1) =: c, \: \:
x_i \mapsto 1 + x_i \sqrt{2/c}, \: \:
y_i  \mapsto 1 + y_i \sqrt{2/c}, \: \: n_1 \to \infty,
$$
gives the PDF
\begin{equation}\label{bary}
{2^{n_2(n_2+1)/2} \over \pi^{(n_2+1)/2}}
\prod_{i=1}^{n_2+1} e^{-x_i^2}
\prod_{1 \le i < j \le n_2+1} (x_i - x_j)
\prod_{1 \le i < j \le n_2}(y_i-y_j)
\end{equation}
where it is required that
$$
\infty > x_1 > y_1  > x_2 > y_2 > \cdots > x_{n_2} >
y_{n_2} > x_{n_2+1} > - \infty. 
$$
\end{prop} 
\subsection{Limit to a biorthogonal Jacobi ensemble}
In the joint probability (\ref{2.19}) let us generalize the
specialization (\ref{8.0b}) so that it involves two
distinct sets of $t$ variables and two distinct $z$ variables,
and thus choose
\begin{equation}\label{sps1}
(a_1,\dots,a_{n_1})=(z_1,z_1t_1,z_1t_1^2,\dots,z_1t_1^{n_1-1}), \quad
(b_1,\dots,b_{n_2})=(z_2,z_2t_2,z_2t_2^2,\dots,z_2t_2^{n_2-1}).
\end{equation}
Using the Schur function evaluation formulas (\ref{8.1}) and (\ref{2.28})
we can readily write down the generalization of (\ref{2.33}) and
(\ref{2.37}). Furthermore, the Jacobi limit of these generalizations can
be computed. Let us make note of the explicit form in the case of
(\ref{2.33}).

\begin{prop}
Consider the generalization of (\ref{2.33}) obtained by specializing
(\ref{2.19}) by (\ref{sps1}). Let
$$
z_1=e^{-a/L}, z_2 =e^{-\bar{a}/L}, t_1 = e^{-1/L}, t_2 = e^{-c/L},
\alpha = e^{-a_1/L}, h_j/L = x_j, r_j/L=y_j.
$$
Then as $L \to \infty$, this probability multiplied by $L^{2n_1}$ tends to
the PDF
\begin{eqnarray}\label{2.53a}
&&\tilde{k}_{n_1,n_2}^*(a,\bar{a},a_1,c) 
\prod_{i=1}^{n_1}(1-e^{-c y_i})^{n_2-n_1}
\prod_{1 \le i < j \le n_1}(e^{-c y_j} - e^{-c y_i})(e^{-x_j} - e^{-x_i})
\nonumber \\
&& \qquad \times e^{-a \sum_{j=1}^{n_1}x_j}
 e^{-\bar{a} \sum_{j=1}^{n_1}y_j}
e^{-a_1 \sum_{j=1}^{n_1}(x_j-y_j)},
\end{eqnarray}
\begin{equation}\label{2.54a}
\tilde{k}_{n_1,n_2}^*(a,\bar{a},a_1,c) :=
{\prod_{l=1}^{n_2-n_1-1} c^l l! \over (\prod_{l=1}^{n_1-1} l!)
(\prod_{l=1}^{n_2-1} c^l l!) }
{\Gamma(a+a_1+n_1) \over \Gamma(a+a_1)}
\prod_{j=1}^{n_2} {\Gamma(a+\bar{a}+c(j-1)+n_2) \over
\Gamma(a+\bar{a}+j-1)}.
\end{equation}
\end{prop}

The Jacobi limit of the Schur function identity (\ref{3.39a}) tells
us that if we integrate (\ref{2.53a}) over $x_1,\dots,x_{n_1}$
we obtain the $a_1 \to \infty$ scaled limit (scaled by $a_1^{-n_1}$)
of the same PDF, and thus
\begin{eqnarray}\label{2.49a}
&& \Big (\lim_{a_1 \to \infty} a_1^{-n_1}
k_{n_1,n_2}^*(a,\bar{a},a_1,c) \Big )
e^{-(a + \bar{a})\sum_{j=1}^{n_1} y_j}
\prod_{i=1}^{n_1}(1 - e^{-c y_i} )^{n_2 - n_1} \nonumber \\&&
\qquad \qquad \times
\prod_{1 \le i < j \le n_1}(e^{-cy_j} - e^{-cy_i})
(e^{-y_j} - e^{-y_i}).
\end{eqnarray}
In the case $n_1=n_2=n$ this same PDF was derived in 
(\ref{crsk1}) of Appendix A
from the continuous version of the RSK
correspondence, 
giving the PDF for the event that the $n_1 \times
n_2$ lattice
of non-negative random variables $[x_{i,j}]$ distributed according to
$$
{\rm Pr}(x_{ij} = y)  =  (i-1+c(j-1)+a+\bar{a}) 
e^{-y(i-1+c(j-1)+a+\bar{a})}, 
$$
gives a polynuclear growth model with 
height of the level-$l$ path $y_l$. 
After the change of variables and replacement of parameters
$e^{-y_j} \mapsto y_{n+1-j}$, $(a+\bar{a}) \mapsto (\alpha + 1)$
we obtain the PDF
\begin{equation}\label{bob}
 \Big (\lim_{a_1 \to \infty} a_1^{-n_1}
k_{n_1,n_2}^*(a,\bar{a},a_1,c) \Big )
\prod_{i=1}^{n_1}y_i^\alpha (1 - y_i^c )^{n_2 - n_1} 
\prod_{1 \le i < j \le n_1}(y_j^c - y_i^c)
(y_j - y_i)
\end{equation}
where $1>y_1>\dots>y_n>0$. For general $\alpha>-1$, $c>0$ 
and with $n_1=n_2$ the $k$-point
distribution corresponding to (\ref{bob}) 
has been computed by Borodin \cite{Bo99}. As the method required to
accomplish this task made use of ideas from the theory of
biorthogonal systems, it was referred to as the
biorthogonal Jacobi ensemble. The PDF (\ref{2.53a}) represents a
more general biorthogonal Jacobi ensemble, but the correlations
for both sets of variables are yet to be computed.

We remark that from (\ref{bob}) we can take a Laguerre and a
Gaussian limit. The correlations for both cases were also
computed in \cite{Bo99}.

\subsection{Distribution functions}
Let us denote by $E(j;I;{\rm PDF})$ the probability that the interval
$I$ of the specified PDF (for this we will use the corresponding equation
number) contains exactly $j$ eigenvalues (for the sake of definiteness
in terminology we will regard the PDFs as measures for eigenvalues).
Similarly, let us denote by $E^{(\cdot)}(j;I;{\rm PDF})$ the same
quantity except that only $(\cdot)=$(e)ven labelled or
$(\cdot)=$(o)dd labelled eigenvalues are being observed. Let us suppose
now that $I=(s,\infty)$ where $s$ is inside the support of the PDF.
Then as discussed in \cite{FR02}, knowledge of $\{E(j;I;{\rm PDF})\}$
is equivalent to knowledge of $\{E^{(\cdot)}(j;I;{\rm PDF})\}$.
Furthermore $p(k-1;s;{\rm ME})$ --- the distribution function of the
$k$th eigenvalue from the right --- is determined by
$\{E(j;I;{\rm PDF})\}$. The quantities $E(j;I;{\rm PDF})$, 
$E^{(\cdot)}(j;I;{\rm PDF})$, $p(k-1;s;{\rm ME})$ for the PDFs
(\ref{1.1}), (\ref{1.5}), (\ref{1.8a}), (\ref{1.8b}) are discussed in
\cite{FR02}, as are the scaled limits of these quantities. Here we
want to use knowledge of the so called hard edge scalings from
\cite{FR02} to identify scales associated to the large eigenvalues
of the particular continuous RSK measures (\ref{8.9}) and (\ref{2.33a}).

In (\ref{1.8a}), (\ref{1.8b}) we first change variables $x_j \mapsto
(1 - x_{2n+1-j})/2$ for consistency with \cite{FR02}. We know from
\cite{FR02} that then, with $2n=N$ in (\ref{1.8a}) and $n=N$ in
(\ref{1.8b}), the large eigenvalues have a well defined scaled limit
obtained by setting
$$
x_j = 1 - {X_j \over 2 N^2}, \qquad A = 4N^2 \bar{\alpha}
$$
(we use $\bar{\alpha}$ rather than $\alpha$ as used in \cite{FR02} to
avoid confusion with $\alpha$ as used in (\ref{6.7})) and taking
$N \to \infty$. In particular
\begin{eqnarray}
\lim_{N \to \infty} E^{(\rm e)}\Big ( p; (1 - {s^2 \over 2N^2},1);
(\ref{1.8a}) \Big |_{A=4N^2  \bar{\alpha}} \Big )& = &
E(p;(0,s);{\rm SE}^{{\rm hard}, a+1}) \label{ihm.1}
\\
\lim_{N \to \infty} E^{(\rm e)}\Big ( p; (1 - {s^2 \over 2N^2},1);
(\ref{1.8b}) \Big |_{A=4N^2  \bar{\alpha}} \Big )& = &
E(p;(0,s);{\rm UE}^{{\rm hard}, a}) \label{ihm.2}
\\
\lim_{N \to \infty} E^{(\rm o)}\Big ( p; (1 - {s^2 \over 2N^2},1);
(\ref{1.8a}) \Big |_{A=4N^2  \bar{\alpha}} \Big ) & = &
E^{(\rm o)}(p;(0,s);{\rm OE}^{ \bar{\alpha}, a}) \label{ihm.3}
\\
\lim_{N \to \infty} E^{(\rm e)}\Big ( p; (1 - {s^2 \over 2N^2},1);
(\ref{1.8b}) \Big |_{A=4N^2  \bar{\alpha}} \Big ) & = &
E(p;(0,s);({\rm OE}\cup{\rm OE})^{\bar{\alpha}, a}) \label{ihm.4}
\end{eqnarray}
Notice that (\ref{ihm.1}), (\ref{ihm.2}) are independent of the parameter
$\bar{\alpha}$. We have already commented that the PDFs (\ref{8.9})
and (\ref{2.33a}) (the latter with $n_1=n_2=n$) are related to
(\ref{1.8a}) and (\ref{1.8b}) by a change of variables. Hence it
follows that
\begin{eqnarray}\label{ihm.5}
\lim_{N \to \infty} E^{(\cdot)}\Big ( p;
(2 \log N - \log {s \over 4}, \infty); (\ref{8.9}) 
\Big |_{n=N, a \mapsto (a+1)/2 \atop a_1 = - 2N^2 \bar{\alpha}} \Big )
& = & E^{(\cdot)}(p;(0,s);({\rm OE})^{\bar{\alpha}, a}) \nonumber \\
\lim_{N \to \infty} E^{(\cdot)}\Big ( p;
(2 \log N - \log {s \over 4}, \infty); (\ref{2.33a}) 
\Big |_{n_1=n_2=N, a \mapsto (a+1)/2 \atop a_1 = - 2N^2 \bar{\alpha}} \Big )
& = & E^{(\cdot)}(p;(0,s);({\rm OE}\cup{\rm OE})^{\bar{\alpha}, a}).
\end{eqnarray}

Also of interest is the scaled form of the gap probability for the
continuous RSK measure (\ref{2.49a}). Now results in \cite{Bo99}
imply
\begin{equation}
\lim_{n_1 \to \infty} E\Big (p;(0, {s \over 4 N^{1+1/c}};
(\ref{bob})  \Big |_{n_2=n_1} \Big ) =
{(-1)^p \over p!} {\partial^p \over \partial \xi^p}
\det (1 - \xi K^{(\alpha,c)}) \Big |_{\xi = 1}
\end{equation}
where $K^{(\alpha,c)}$ is the integral operator supported on
$(0,s)$ with the kernel
\begin{eqnarray*}
K^{(\alpha,c)}(x,y) & = & {c \over 4} \int_0^1
J_{(\alpha+1)/c,1/c}(xt/4) J_{\alpha+1,c}((yt/4)^c) t^{\alpha} \, dt, \\
J_{a,b}(x) & := & \sum_{m=0}^\infty {(-x)^m \over m! \Gamma(a+bm)}.
\end{eqnarray*}
It follows from the relationship between (\ref{bob}) and (\ref{2.49a})
that
\begin{equation}
\lim_{n_1 \to \infty} E\Big (p;((1+1/c)\log N - \log {s \over 4}, \infty);
(\ref{2.49a}) \Big |_{n_2=n_1 \atop
a+\bar{a} = \alpha +1} \Big ) =
{(-1)^p \over p!} {\partial^p \over \partial \xi^p}
\det (1 - \xi K^{(\alpha,c)}) \Big |_{\xi = 1}.
\end{equation}
Notice that in the case $c=1$ the scaled interval is the same as that in
(\ref{ihm.5}).

\section{Interpolating ensembles from Macdonald polynomial theory}
\setcounter{equation}{0}
The recurrence (\ref{3.39})
satisfied by the Schur polynomials is a special case of 
a more general  recurrence satisfied by the Macdonald polynomials,
as is the marginal probability (\ref{6.3}) and the evaluation
formula (\ref{8.1}). This then allows a generalization of the
joint probability (\ref{2.19}) to the Macdonald setting. 
We will see that the probability
(\ref{8.8}), which in the last passage percolation
problem results from imposing the symmetry constraint
$x_{i,j} = x_{j,i}$ on the waiting times, is also a special case of the
generalized joint probability.

The recurrence (\ref{3.39}) can be used to define the Schur polynomials.
Likewise we can define the (monic) Macdonald  polynomials
$P_\kappa(b_1,\dots,b_{n_2};q,t)$ by the recurrence
\cite[pg.~348]{Ma95}
\begin{equation}\label{rr1}
\sum_{\kappa: \kappa \in R}
\psi_{\mu/\kappa}(q,t) P_\kappa(b_1,\dots,b_{n_2};q,t)
b_{n_2+1}^{|\mu|-|\kappa|}
= P_\mu(b_1,\dots,b_{n_2+1};q,t)
\end{equation}
where  with $f(x) := (tx;q)_\infty/(qx;q)_\infty$ 
\begin{equation}\label{rr2}
\psi_{\mu/\kappa}(q,t) :=
\prod_{1 \le i \le j \le \ell(\kappa)}
{f(q^{\kappa_i - \kappa_j} t^{j-i})
f(q^{\mu_i - \mu_{j+1}} t^{j-i}) \over
f(q^{\mu_i - \kappa_j} t^{j-i})
f(q^{\kappa_i - \mu_{j+1}} t^{j-i})}.
\end{equation}
The Schur polynomials are the
special case $q=t$ of the Macdonald  polynomials (note that with
$q=t$, $f(x) = 1$ and so $\psi_{\mu/\kappa}(q,q) = 1$).

In the Macdonald theory, the generalization of the marginal probability
(\ref{6.3}) is
$$
\prod_{i=1}^{n_1} \prod_{j=1}^{n_2}
{(a_i b_j ; q)_\infty \over (ta_i b_j; q)_\infty}
Q_\mu(a_1,\dots, a_{n_1}; q,t)
P_\mu(b_1,\dots,b_{n_2}; q,t)
$$
where
$$
Q_\mu(a_1,\dots, a_{n_1}; q,t) =
\langle P_\mu, P_\mu \rangle^{-1} P_\mu(a_1,\dots,a_{n_1};q,t)
$$
and $\langle \cdot, \cdot \rangle$ is a particular power sum inner
product \cite[pg.~309]{Ma95}. Thus the natural generalization of
(\ref{2.19}) is
\begin{equation}\label{mac2}
\prod_{i=1}^{n_1} \prod_{j=1}^{n_2+1}
{(a_i b_j;q)_\infty \over (ta_i b_j; q)_\infty}
Q_\mu(a_1,\dots, a_{n_1}; q,t) 
P_\kappa(b_1,\dots,b_{n_2}; q,t)
\psi_{\mu/\kappa}(q,t) b_{n_2+1}^{|\mu| - |\kappa|}
\end{equation}
where the parts of $\mu$ and $\kappa$ are restricted by
(\ref{2.11}) in the case $n_2 \ge n_1$, and
(\ref{2.11a}) in the case $n_1 > n_2$.
Because of the identity (\ref{rr1}), and the further Macdonald
polynomial identity \cite[special case of 6(a) pg.~352]{Ma95}
\begin{equation}\label{rr7}
\prod_{i=1}^{n_1}{(a_i b_{n_2+1};q)_\infty \over
(ta_ib_{n_2+1};q)_\infty}
\sum_{\mu: \mu \in R}
Q_\mu(a_1,\dots,a_{n_1}) \psi_{\mu/\kappa}(q,t) b_{n_2+1}^{|\mu|-|\kappa|}
= Q_\kappa(a_1,\dots,a_{n_1})
\end{equation}
which generalizes (\ref{3.39a}), this furthermore has the 
interpretation as a joint probability density function on tableaux of
shape $\mu$ and shape $\kappa$ when $\mu/\kappa$ is a horizontal strip.

The Macdonald polynomials exhibit a generalization of the
evaluation formula (\ref{8.1}). Let $p_r : = \sum_{j=1}^n
x_j^r$ denote the  power sum of degree $r$. Define a
homomorphism
\begin{equation}\label{rr3}
\varepsilon_{u,t}(p_r) = {1 - u^r \over 1 - t^r}
\end{equation}
Then for any symmetric function $f$ analytic in $x_1,
\dots,x_n$ it is easy to see that
\begin{equation}\label{3.5a}
\varepsilon_{t^n,t}(f) = f(1,t,\dots,t^{n-1}),
\end{equation}
so the evaluation of $\varepsilon_{u,t}(P_\mu)$ includes the
generalization of (\ref{8.1}) for the Macdonald polynomials.
From \cite[pg.~343]{Ma95} we have with $t=q^k$ and
$\ell(\lambda) \le n$
\begin{eqnarray}\label{3.7}
\varepsilon_{u,t}(P_\lambda) & = &
t^{\sum_{i=1}^{n}(i-1)\lambda_i}
\prod_{1 \le i < j \le n}
(q^{\lambda_i - \lambda_j} t^{j-i};q)_k
\prod_{i=1}^n {(ut^{1-i};q)_{\lambda_i} \over
(t;q)_{\lambda_i + k(n-i)}} \label{mac6} \\
\varepsilon_{u,t}(Q_\lambda) & = &
t^{\sum_{i=1}^{n}(i-1)\lambda_i}
\prod_{1 \le i < j \le n}
(q^{\lambda_i - \lambda_j+1} t^{j-i-1};q)_k
\prod_{i=1}^n {(ut^{1-i};q)_{\lambda_i} \over
(q;q)_{\lambda_i + k(n-i)}} \label{mac7}
\end{eqnarray}
Note that for $n$ fixed these quantities are strictly positive for
$u$ small enough, but may become zero or negative if $n$ is
unrestricted. In fact a special choice of $u$ in each case shows that
the action of $\varepsilon_{u,t}$ may annihilate $P_\lambda$ or
$Q_\lambda$ for all $\lambda$ of length greater than a fixed size.
Thus we see that
\begin{equation}
\varepsilon_{t^n,t}(P_\lambda)=0, \quad
\varepsilon_{qt^{n-1},t}(Q_\lambda)=0, \qquad
{\rm for} \quad \ell(\lambda) > n,
\end{equation}
while these same quantities are strictly positive for
$\ell(\lambda) \le n$ (in the first case corresponding to the
evaluation (\ref{3.5a})).
 
In (\ref{mac2}) we write $b_{n_2+1} =: \alpha$, replace $a_i$
by $z_1 a_i$ ($i=1,\dots,n_1$),
$b_j$ by $z_2 b_j$ $(j=1,\dots,n_2)$ and take the limit $n_1, n_2 \to
\infty$. We would  now like to apply 
the homomorphism $\varepsilon_{u,t}$ to the functions of
$\{a_i\}$ and the homomorphism
$\varepsilon_{w,t}$ to the functions of
$\{b_j\}$. For the Macdonald polynomials, after
factoring out the $z$-dependence using homogeneity, this is
done using the formulas 
(\ref{mac6}) and (\ref{mac7}). For the infinite products we recall the
formula \cite[pg.~310]{Ma95}
$$
\prod_{i,j=1}^{\infty} 
{(z_1 z_2 a_i b_j ; q)_\infty \over
(t z_1 z_2 a_i b_j; q)_\infty} =
\prod_{n=1}^\infty \exp \Big ( - {(z_1 z_2)^{n} \over n}
{1 - t^n \over 1 - q^n} p_n(a) p_n(b) \Big ),
$$
from which the action (\ref{rr3}) immediately implies 
\begin{eqnarray*}
\varepsilon_{u,t}^{\{a_i\}}
\varepsilon_{w,t}^{\{b_i\}}
\Big (
\prod_{i,j=1}^{\infty} 
{(z_1 z_2 a_i b_j ; q)_\infty \over
(t z_1 z_2 a_i b_j; q)_\infty} \Big ) & = &
\prod_{n=1}^\infty \exp \Big ( - {(z_1z_2)^{n} \over n}
{(1-u^n)(1-w^n) \over (1-q^n)(1-t^n)} \Big ) \\
 & = & \prod_{p=0}^\infty
{(z_1z_2 t^p;q)_\infty (uwz_1z_2 t^p;q)_\infty \over
(uz_1z_2 t^p;q)_\infty (wz_1z_2  t^p;q)_\infty} \\
\varepsilon_{u,t}^{\{a_i\}}
\prod_{i=1}^{\infty} {(z_1 a_i b_{n_2+1} ; q)_\infty \over
(t z_1 a_i b_{n_2+1}; q)_\infty}  & = &
{(z_1b_{n_2+1};q)_\infty \over (uz_1b_{n_2+1};q)_\infty}.
\end{eqnarray*}
Consequently the image of the joint probability
(\ref{mac2}) is given by 
\begin{equation}\label{mac10}
 {(z_1\alpha ;q)_\infty \over (uz_1\alpha;q)_\infty}
 \prod_{p=0}^\infty
{(z_1z_2 t^p;q)_\infty (uwz_1z_2 t^p;q)_\infty \over
(uz_1z_2 t^p;q)_\infty (wz_1z_2  t^p;q)_\infty} 
\varepsilon_{u,t}(Q_\mu) 
\varepsilon_{w,t}(P_\kappa)  
\psi_{\mu/\kappa}(q,t) 
z_1^{|\mu|} z_2^{|\kappa|} \alpha^{|\mu| - |\kappa|}
\end{equation}

For general $u$ and $w$ (\ref{mac10}) is not itself a meaningful joint
probability on partitions $\mu, \kappa$ because when the number of
parts becomes large enough it will become negative. However, according to
(\ref{mac6}) and (\ref{mac7})
for the special choice $u=t^{n}$ or $w=t^{n-1}$ this
does not happen but rather (\ref{mac10})
vanishes when $\ell(\mu) > n$. Thus
we are naturally led to two distinct joint probabilities on partitions
$\mu, \kappa$ with $\mu/\kappa$ a horizontal strip,
\begin{eqnarray}\label{pre}
{\rm Pr_e}(\mu,\kappa) & := &
{(z_1\alpha;q)_\infty \over (t^{n}z_1 \alpha;q)_\infty}
 \prod_{p=0}^\infty
{(z_1z_2 t^p;q)_\infty (wz_1z_2 t^{p+n};q)_\infty \over
(z_1z_2 t^{p+n};q)_\infty (wz_1z_2	 t^{p} ;q)_\infty}
\nonumber \\
&& \times \varepsilon_{t^{n},t}(Q_\mu)
\varepsilon_{w,t}(P_\kappa) \psi_{\mu/\kappa}(q,t)
z_1^{|\mu|} z_2^{|\kappa|} \alpha^{|\mu| - |\kappa|}
\end{eqnarray}
for which
\begin{equation}\label{pro1}
\mu_1 \ge \kappa_1 \ge \mu_2  \ge \kappa_2 \ge \cdots \ge \mu_n \ge
\kappa_n \ge 0,
\end{equation}
and 
\begin{eqnarray}\label{pro}
{\rm Pr_o}(\mu,\kappa) & := &
{(z_1\alpha;q)_\infty \over (uz_1 \alpha;q)_\infty}
 \prod_{p=0}^\infty
{(z_1z_2 t^p;q)_\infty (uz_1z_2 t^{p+n-1};q)_\infty \over
(uz_1z_2 t^p;q)_\infty (z_1z_2	 t^{p+n-1} ;q)_\infty}
\nonumber \\
&& \times \varepsilon_{u,t}(Q_\mu)
\varepsilon_{t^{n-1},t}(P_\kappa) \psi_{\mu/\kappa}(q,t)
z_1^{|\mu|} z_2^{|\kappa|} \alpha^{|\mu| - |\kappa|}
\end{eqnarray}
for which
\begin{equation}\label{pro2}
\mu_1 \ge \kappa_1 \ge \mu_2  \ge \kappa_2 \ge \cdots \ge \mu_{n-1} \ge
\kappa_{n-1} \ge \mu_n \ge 0.
\end{equation}

A short calculation shows that setting 
$ w = q^{n_2-n_1+1} t^{n-1}, z_1=z_2=z, t=q$ in
${\rm Pr_e}(\mu,\kappa)$ gives (\ref{2.33}), while setting
$u=q^{n_1 - (n_2+1)}  t^{n-1}, z_1=z_2=z, t=q$ in
${\rm Pr_o}(\mu,\kappa)$ gives (\ref{2.37}). 
Furthermore (\ref{pre}), (\ref{pro})
in the case
$t=q^2$ reclaim (\ref{8.8}). Thus straight forward simplification 
gives the following result.

\begin{prop}
Let $w = q^{-1} t^{n}$,
$z_1=z$, $z_2=qz$,
and $t=q^2$, and write $h_{2j-1}:= \mu_j +2n - (2j-1)$,
$h_{2j} := \kappa_j + 2n - 2j$. Then (\ref{pre}) reduces to
(\ref{8.8}) with $n \mapsto 2n$, $t \mapsto q$ in the latter.
Similarly, let $u=qt^{n-1}$,
$z_1=z$, $z_2=qz$ and $t=q^2$, and write
$h_{2j-1}:= \mu_j + (2n-1) - (2j-1), h_{2j} := \kappa_j + (2n-1) - 2j$.
Then (\ref{pro}) reduces to 
(\ref{8.8}) with $n \mapsto 2n-1$, $t \mapsto q$ in the latter.
\end{prop}

The probabilities
${\rm Pr_e}(\mu,\kappa)$ and ${\rm Pr_o}(\mu,\kappa)$ exhibit a
special property with respect to summation over $\mu$. Thus it
follows from (\ref{rr7}) that
\begin{eqnarray}\label{sps}
\sum_\mu  {\rm Pr_e}(\mu,\kappa)
& = &
 \prod_{p=0}^\infty
{(z_1z_2 t^p;q)_\infty (wz_1z_2 t^{p+n};q)_\infty \over
(z_1z_2 t^{p+n};q)_\infty (wz_1z_2	 t^{p} ;q)_\infty}
\varepsilon_{t^{n},t}(Q_\kappa)
\varepsilon_{w,t}(P_\kappa) 
(z_1z_2)^{|\kappa|} \nonumber \\
\sum_\mu  {\rm Pr_o}(\mu,\kappa)
& = &
 \prod_{p=0}^\infty
{(z_1z_2 t^p;q)_\infty (uz_1z_2 t^{p+n-1};q)_\infty \over
(uz_1z_2 t^p;q)_\infty (z_1z_2	 t^{p+n-1} ;q)_\infty}
\varepsilon_{u,t}(Q_\kappa)
\varepsilon_{t^{n-1},t}(P_\kappa) 
(z_1z_2)^{|\kappa|}
\end{eqnarray}

A special case of ${\rm Pr_o}(\mu,\kappa)$ also exhibits a special property
with respect to summation over $\kappa$. To see this we first note
from (\ref{3.5a}) that
$$
\varepsilon_{t^{n-1},t}(P_\kappa) = P_\kappa(1,t,\dots,t^{n-2})
$$
and so for the $\kappa$ dependent terms in (\ref{pro}) with
$\alpha=1$, $z_2 = t$ we have
\begin{eqnarray}\label{3.14'}
&& \sum_\kappa \psi_{\mu/\kappa}(q,t) t^{|\kappa|}
\varepsilon_{t^{n-1},t}(P_\kappa) =
\sum_\kappa \psi_{\mu/\kappa}(q,t)  P_\kappa(t,t^2,\dots,t^{n-1})
\nonumber \\ &&
\qquad =  P_\mu(1,t,t^2,\dots,t^{n-1}) =
\varepsilon_{t^n,t}( P_\mu)
\end{eqnarray}
where the second equality follows from (\ref{rr1}) and the fact that
$P_\kappa$ is a symmetric function. Thus
\begin{equation}\label{sps4}
\sum_\kappa {\rm Pr_o}(\mu,\kappa) \Big |_{\alpha = 1 \atop
z_2 = t} = {(z_1;q)_\infty \over (uz_1;q)_\infty}
\prod_{p=0}^\infty
{(z_1 t^{p+1}; q)_\infty (uz_1t^{p+n};q)_\infty \over
(uz_1t^{p+1};q)_\infty (z_1t^{p+n};q)_\infty}
\varepsilon_{u,t}(Q_\mu) \varepsilon_{t^{n},t}(P_\mu)
z_1^{|\mu|}.
\end{equation}
To use (\ref{3.14'}) in the case of Pr${}_{\rm e}(\mu,\kappa)$ we
must set $w=t^{n-1}$. This in turn implies $\kappa_n=0$, so we see
that no new identity results, but rather we reclaim the special case
$u=t^n$ of (\ref{sps4}).
As made explicit in Appendix B, the identity (\ref{sps4}) can be recognized
as being equivalent to a special case of a $q$-integral due to
Evans \cite{Ev92}, and also as the $\nu = \emptyset$ case of
Okounkov's $q$-integral representation of the Macdonald polynomial
$P_\nu$ \cite{Ok98}. The structure afforded by (\ref{sps4}) suggests
a simplified derivation of the latter which is given in Appendix B.
We remark too that (\ref{sps4}) can be considered as a particular
$q,t$ generalization of  a class of measures on partitions known as 
$z$-measures \cite{BO99a}.

Because the probabilities  (\ref{2.33}), (\ref{2.37}) and
(\ref{8.8}) can all be derived from (\ref{mac10}), they all exhibit
the special property (\ref{sps}). Of course in the cases of
(\ref{2.33}), (\ref{2.37}) this identity is immediate from their
interpretation as joint probabilities for tableaux of shape
$\mu$ and tableaux of shape $\kappa$ with $\mu/\kappa$ a horizontal
strip. But the probability (\ref{8.8}) has no such interpretation, and
the identity implied by (\ref{sps}),
\begin{eqnarray}
&& {\rm even} \Big ( c_n(\alpha,z,t) 
z^{\sum_{j=1}^n h_j}
\alpha^{\sum_{j=1}^n(-1)^{j-1} h_j}
\prod_{1 \le i < j \le n} (t^{h_j} - t^{h_i}) \Big ) \nonumber \\
&& \qquad
= \Big ( \alpha^{[n/2]} c_n(\alpha,z,t) \Big )
\Big |_{\alpha = 0}
z^{\sum_{j=1}^{[n/2]} h_{2j}}
\prod_{1 \le i < j \le n} (t^{h_j} - t^{h_i})
\Big |_{h_{2j-1} = h_{2j}+1 \: (j=1,\dots,[n/2]) \atop
h_n = 0 \: (n \, {\rm odd})},
\end{eqnarray}
(here the notation even$(\:)$ denotes the distribution of the even
labelled coordinates $h_2,h_4,\dots,h_{2[n/2]}$), 
telling us that there is no dependence on $\alpha$ after summing out the
odd labelled coordinates,
cannot easily be
anticipated.

\subsection{${\rm Pr_{o}}(\mu,\kappa)$ and 
${\rm Pr_{e}}(\mu,\kappa)$ in the Jacobi limit}
Consider (\ref{pro}) with $z_1=z$, $z_2=tq^{-1} \bar{z}$, $u=q^{\beta +1}
t^{n-1}$, $t=q^k$. The Jacobi limit is obtained by setting
\begin{equation}\label{3.16'}
z=e^{-a/L}, \bar{z} = e^{-\bar{a}/L}, \alpha = e^{-a_1/L},
q = e^{-1/L}, \mu_j/L = x_j, \kappa_j/L = y_j,
\end{equation}
multiplying (\ref{pro}) by $L^{2n-1}$ and taking the limit
$L \to \infty$. This gives the PDF
\begin{eqnarray*}
&& C_n(a,\bar{a},a_1,\beta,k)
e^{-a\sum_{i=1}^n x_i} e^{-\bar{a}\sum_{i=1}^{n-1} y_i}
e^{-a_1( \sum_{i=1}^n x_i - \sum_{j=1}^{n-1} y_j)}
\prod_{i=1}^n(1 - e^{-x_i})^\beta \nonumber \\
&& \quad \times
\prod_{1 \le i < j \le n} |e^{-x_j} - e^{-x_i}|
\prod_{1 \le i < j \le n-1}| e^{-y_j} - e^{-y_i} |
\prod_{i=1}^n \prod_{j=1}^{n-1} | e^{-x_j} - e^{-y_i} |^{k-1},  \nonumber \\
 && C_n(a,\bar{a},a_1,\beta,k) :=
{\Gamma (a+a_1+\beta+1+k(n-1)) \over \Gamma(a+a_1) \Gamma(\beta+1)}
\prod_{i=1}^{n-1} {\Gamma(a+\bar{a}+\beta+k(n-1+i)) \over
\Gamma(ki) \Gamma(\beta+ki+1)\Gamma(a+\bar{a}-1+ik)}.
\end{eqnarray*}
Replacing $a,\bar{a}$ by $a+1, \bar{a}+1$ and changing variables
$e^{-x_i} \mapsto x_{n+1-i}$, $e^{-y_i} \mapsto y_{n-i}$, this
reads
\begin{eqnarray}\label{ra}
&& C_n(a+1,\bar{a}+1,a_1,\beta,k)
\prod_{i=1}^n x_i^{a+a_1} (1 - x_i)^{\beta}
\prod_{j=1}^{n-1} y_j^{\bar{a} - a_1} \nonumber \\
&& \qquad \times
\prod_{1 \le i < j \le n} |x_j - x_i|
\prod_{1 \le i < j \le n-1}|y_j - y_i|
\prod_{i=1}^n \prod_{j=1}^{n-1} |x_j-y_i|^{k-1}
\end{eqnarray}
and we require the analogue of the interlacing condition (\ref{pro2}),
\begin{equation}\label{rar}
1 > x_1 > y_1 > x_2 > y_2 > \cdots > y_{n-1} > x_n > 0.
\end{equation}

A natural generalization of (\ref{ra}) is to include a factor
$\prod_{j=1}^{n-1}(1-y_j)^\beta$. To compute the corresponding
normalization, we note that with $R$ denoting the region
(\ref{rar}), the Jacobi limit of the second identity in 
(\ref{sps}) tells us that
\begin{eqnarray}\label{4.9}
&& \int_R dx_1 \cdots dx_n \, \prod_{i=1}^n x_i^\alpha (1 - x_i)^\beta
\prod_{1 \le i < j \le n} |x_i - x_j|
\prod_{i=1}^n \prod_{j=1}^{n-1} | x_j - y_i|^{k-1} \nonumber \\&&
 \qquad = {\Gamma(1+\alpha) \Gamma(1+\beta) (\Gamma(k))^{n-1} \over
\Gamma(2+\alpha+\beta+(n-1)k)}
\prod_{i=1}^{n-1} y_i^{\alpha+k}(1-y_i)^{\beta+k}
\prod_{1 \le i < j \le n-1} |y_i - y_j|^{2k-1}
\end{eqnarray}
This is an integration
formula due to Anderson \cite{An91}. Furthermore, we have the well known
Selberg integral evaluation
\begin{eqnarray}\label{4.9a}
&& \int_0^1 dt_1 \, t_1^{\lambda_1}(1-t_1)^{\lambda_2} \cdots
 \int_0^1 dt_N \, t_N^{\lambda_1}(1-t_N)^{\lambda_2} \,
\prod_{1 \le j < k \le N} |t_k - t_j|^{2\lambda} \nonumber \\
&& \qquad =
\prod_{j=0}^{N-1} {\Gamma(\lambda_1 + 1 + j \lambda)
\Gamma(\lambda_2 + 1 + j \lambda) \Gamma(1+(j+1)\lambda) \over
\Gamma(\lambda_1 + \lambda_2 + 2 + (N+j-1)\lambda) \Gamma(1+\lambda)}
=: S_N(\lambda_1,\lambda_2,\lambda).
\end{eqnarray}
It follows from (\ref{4.9}) and (\ref{4.9a}) that
\begin{eqnarray}\label{J.2}
&& J^{(n,n-1)}_{\rm o}(x,y) := {\Gamma(2+\alpha+\beta+(n-1)k) \over
\Gamma(1+\alpha) \Gamma(1+\beta) (\Gamma(k))^{n-1}}
{1 \over  S_{n-1}(\alpha+\alpha_1+k,\beta+\beta_1+k,k)}
\prod_{i=1}^n x_i^\alpha (1-x_i)^\beta 
\nonumber \\
&& \times \prod_{1 \le i < j \le n}
|x_j - x_i|
\prod_{i=1}^{n-1}y_i^{\alpha_1}(1-y_i)^{\beta_1}
\prod_{1 \le i < j \le n -1} |y_j - y_i|
\prod_{i=1}^n \prod_{j=1}^{n-1} |x_j - y_i|^{k-1}
\end{eqnarray}
is a correctly normalized joint PDF. With $\alpha = a+a_1$,
$\alpha_1 = \bar{a}-a_1$, $\beta_1=0$ it coincides with (\ref{ra}).

Key properties of (\ref{J.2}) are
\begin{eqnarray}\label{4.10a}
&& \int_R dx_1 \cdots dx_n \,  J^{(n,n-1)}_{\rm o}(x,y) 
 =
{1 \over S_{n-1}(\alpha+\alpha_1+k,\beta+\beta_1+k,k)} \nonumber \\&& \qquad
\times \prod_{i=1}^{n-1} y_i^{\alpha+\alpha_1+k}(1-y_i)^{\beta+\beta_1+k}
\prod_{1 \le i < j \le n-1} |y_i-y_j|^{2k} =:
J^{(\_,n-1)}_{\rm o}(y),
\end{eqnarray}
which follows immediately from (\ref{4.9}), and
\begin{eqnarray}\label{4.10b}
&& \int_R dy_1 \cdots dy_{n-1} \,  J^{(n,n-1)}_{\rm o}(x,y)
\Big |_{\alpha_1 = \beta_1=0}\nonumber \\
&& \qquad = {1 \over S_n(\alpha,\beta,k)}
 \prod_{i=1}^n x_i^\alpha (1-x_i)^\beta
\prod_{1 \le i < j \le n} |x_i - x_j|^{2k} =: J^{(n,\_)}_{\rm o}(x).
\end{eqnarray}
which can be deduced from the Jacobi limit of (\ref{sps4}).
Like (\ref{4.9}) (and thus (\ref{4.10a})), (\ref{4.10b}) is an
integration formula due to Anderson \cite{An91}.

Consider now (\ref{pre}), and put $z_1=z$, $z_2=tq^{-1}z$, $w=t^{\beta_2+n}$,
$t=q^k$. The Jacobi limit is obtained by scaling the parameters according to
(\ref{3.16'}), multiplying (\ref{pre}) by $L^{2n}$ and taking the limit
$L \to \infty$. One thus obtains the PDF
\begin{eqnarray*}
&& K_n(a,\bar{a},a_1,\beta,k)
e^{-a\sum_{i=1}^n x_i} e^{-\bar{a}\sum_{i=1}^{n} y_i}
e^{-a_1 \sum_{i=1}^n (x_i - y_i)}
\prod_{i=1}^n(1 - e^{-y_i})^{k \beta_2} \nonumber \\
&& \quad \times
\prod_{1 \le i < j \le n} |e^{-x_j} - e^{-x_i}|
 |e^{-y_j} - e^{-y_i} |
\prod_{i,j=1}^n | e^{-x_j} - e^{-y_i} |^{k-1},	\nonumber \\
 && K_n(a,\bar{a},a_1,\beta,k) :=
{\Gamma (a+a_1+kn)) \over \Gamma(a+a_1)}
{1 \over \prod_{i=1}^n \Gamma(k(\beta_2+i))} 
\prod_{j=1}^{n+\beta_2} {\Gamma(a+\bar{a}-1+k(j+1+n)) \over
\Gamma(a+\bar{a}-1+kj)}.
\end{eqnarray*}
By changing variables $e^{-x_i} \mapsto x_{n+1-i}$, $e^{-y_i} \mapsto y_{n-i}$
and replacing $a,\bar{a}$ by $a+1, \bar{a}+1$ we obtain from this the PDF 
\begin{equation}\label{3.17e}
K_n(a+1,\bar{a}+1,a_1,\beta,k)
\prod_{i=1}^n x_i^{a+a_1}  y_i^{\bar{a} - a_1} (1-y_i)^{k \beta_2}
\prod_{1 \le i < j \le n} |x_j - x_i|
|y_j - y_i|
\prod_{i,j=1}^n  |x_j-y_i|^{k-1}
\end{equation}
and we require the analogue of the interlacing condition (\ref{pro2}),
\begin{equation}\label{3.18e}
1 > y_1 > x_1 > y_2 > x_2 > \cdots > y_{n} > x_n > 0.
\end{equation}

Unlike the situation with (\ref{ra}), it is not a natural generalization
of (\ref{3.17e}) to include an extra factor (here
$\prod_{i=1}^n(1-x_i)^\beta$). Only with this extra factor absent can
we compute the integral of the $x$'s according to the Jacobi limit
of the first identity in (\ref{sps}). Before stating this result, let
us first rename some of the parameters and manipulate the normalization
so that (\ref{3.17e}) reads
\begin{eqnarray}\label{J.2e}
&& J^{(n,n)}_{\rm e}(x,y) := {\Gamma(1+\alpha+nk) \over
\Gamma(1+\alpha)  (\Gamma(k))^{n}}
{1 \over  S_{n}(\alpha+\alpha_1+k,\beta_1,k)}
\prod_{i=1}^n x_i^\alpha y_i^{\alpha_1}(1-y_i)^{\beta_1} 
\nonumber \\
&& \times \prod_{1 \le i < j \le n}
|x_j - x_i| |y_j - y_i|
\prod_{i,j=1}^n  |x_j - y_i|^{k-1}
\end{eqnarray}
In terms of this quantity the Jacobi limit
of the first identity in (\ref{sps}) reads
\begin{eqnarray}\label{4.10ae}
&& \int_{\tilde{R}} dx_1 \cdots dx_n \,  J^{(n,n)}_{\rm e}(x,y)
 =
{1 \over S_{n}(\alpha+\alpha_1+k,\beta_1,k)} \nonumber \\&& \qquad
\times \prod_{i=1}^{n} y_i^{\alpha+\alpha_1+k}(1-y_i)^{\beta_1}
\prod_{1 \le i < j \le n} |y_i-y_j|^{2k} =:
J^{(\_,n)}_{\rm e}(y),
\end{eqnarray}
where $\tilde{R}$ denotes the region (\ref{3.18e}). It follows from
this that we also have
\begin{eqnarray}\label{4.10ae'}
&& \int_{\tilde{R}} dy_1 \cdots dy_n \,	 J^{(n,n)}_{\rm e}(x,y)
\Big |_{\alpha_1 = 0}
 =
{1 \over S_{n}(\alpha,\beta_1+k,k)} \nonumber \\&& \qquad
\times \prod_{i=1}^{n} x_i^{\alpha}(1-x_i)^{\beta_1+k}
\prod_{1 \le i < j \le n} |x_i-x_j|^{2k} =:
J^{(n,\_)}_{\rm e}(x),
\end{eqnarray}

\section{Random matrix interpretation of the Anderson density}
\setcounter{equation}{0}
With $J_{\rm o}^{(n,n-1)}(x,y)$, $J_{\rm o}^{(n,\_)}(x)$ and
$J_{\rm o}^{(\_,n-1)}(y)$ defined by (\ref{J.2}), (\ref{4.10b}) and
(\ref{4.10a}) respectively, we can construct the conditional PDF's
\begin{eqnarray}\label{An.1}
{J_{\rm o}^{(n,n-1)}(x,y) |_{\alpha_1=\beta_1=0} \over J_{\rm o}^{(n,\_)}(x)} & 
= &
{\Gamma(nk) \over (\Gamma(k))^n}
{\prod_{1\le i < j \le n-1}(y_i-y_j) \over
\prod_{1\le i < j \le n} (x_i-x_j)^{2k-1}}
\prod_{i=1}^{n-1} \prod_{j=1}^n |y_i-x_j|^{k-1}, \label{4.1} \\
{J^{(n,n-1)}_{\rm o}(x,y) \over J^{(\_,n-1)}_{\rm o}(y)} & = &
{\Gamma(2+\alpha+\beta+(n-1)k) \over
\Gamma(1+\alpha) \Gamma(1+\beta) (\Gamma(k))^{n-1}}
 \prod_{i=1}^n x_i^\alpha(1-x_i)^\beta
\prod_{i=1}^{n-1} y_i^{-(\alpha+k)}(1-y_i)^{-(\beta+k)}
\nonumber \\
&& \quad \times 
{\prod_{1 \le i < j \le n}(x_i - x_j) \over
\prod_{1 \le i < j \le n} (y_i - y_j)^{2k-1}}
\prod_{i=1}^{n-1} \prod_{j=1}^n |y_i - x_j|^{k-1} \label{4.11}
\end{eqnarray}
where the $x$'s and $y$'s are interlaced according to (\ref{rar}).
The conditional PDF (\ref{An.1}) can be recognized
as the special case $s_1=s_2=\cdots=s_n=k$ of
the conditional density function
\begin{equation}\label{An.2}
{\Gamma(s_1+\cdots+s_n) \over \Gamma(s_1) \cdots \Gamma(s_n)}
{\prod_{1 \le i < j \le n-1} (y_i - y_j) \over
\prod_{1 \le i < j \le n}(x_i-x_j)^{s_i+s_j-1}}
\prod_{i=1}^{n-1} \prod_{j=1}^n|y_i-x_j|^{s_j-1}
\end{equation}
appearing in the work of Anderson \cite{An91} on the Selberg integral.
(Note that the $q$-generalization of this same density appears in
Evans' $q$-integral (\ref{B.3}); also we refer to \cite{RX02} for a
further integration method to verify that the normalization is
correct.) Here we will show that
(\ref{An.2}), with the $s_i$ non-negative integers or half integers,
and thus according to (\ref{An.1}) the conditional PDF associated to
the case $\bar{a}=a_1=0$ of the interpolating ensemble
(\ref{ra}), can be derived from a random matrix problem.

The random matrix problem relates to the corank 1 random
projection of a fixed matrix. Specifically, we seek the
eigenvalue PDF of
\begin{equation}\label{M.1}
M :=  \Pi  A  \Pi, \qquad  \Pi := {\bf 1} -
\vec{x} \vec{x}^{\,\dagger}
\end{equation}
where $A$ is a real symmetric, or complex Hermitian, fixed matrix,
$\vec{x}$ is a real, or complex, normalized Gaussian column
vector of the same number of rows as $A$
and ${\bf 1}$ denotes the identity matrix. The eigenvalue PDF depends
only on the eigenvalues of $A$, which we take to be
$a_1 > a_2 > \cdots > a_n$ with multiplicities 
$m_1,m_2,\dots,m_n$.

All but $n$ eigenvalues of (\ref{M.1}) must coincide with the
eigenvalues $a_i$ of $A$ and must occur in $M$ with multiplicity  $m_i-1$.
For the latter result we make use of the following formula for the
characteristic polynomial of $M$.
 
\begin{lemma}\label{lem8}
We have
\begin{equation}\label{3.18}
\det (M - \lambda {\bf 1}) = - \lambda \det (  A - \lambda
{\bf 1}) {\rm Tr} \Big ( (A - \lambda {\bf 1})^{-1} \vec{x}
\vec{x}^{ \,\dagger} \Big ).
\end{equation}
\end{lemma}

\noindent
Proof. \quad Simple manipulation using (\ref{M.1}) shows
$$
 \det (  M - \lambda {\bf 1}) =
\det (  A - \lambda {\bf 1})
\det({\bf 1} + (  A - \lambda {\bf 1})^{-1}
(- A \vec{x} \vec{x}^{\, \dagger} - \vec{x}
\vec{x}^{\, \dagger}  A +
\vec{x} \vec{x}^{\, \dagger}  A \vec{x} \vec{x}^{\, \dagger})).
$$
The matrix in the second determinant is of the form ${\bf 1} + Y$
where $Y$ has rank 1, and in such a circumstance we have 
$\det({\bf 1}+Y) = 1 + {\rm Tr} \, Y$. Using this fact, then further
manipulation using ${\rm Tr}( \vec{x} \vec{x}^{\, \dagger}) = 1$
(which in turn follows from the assumption that
$\vec{x}$ is normalized) gives (\ref{3.18}). \hfill $\square$

Now (\ref{3.18}) shows there is an eigenvalue $\lambda = 0$,
and that the remaining eigenvalues satisfy
\begin{equation}\label{4.5a}
\prod_{l=1}^n(a_l - \lambda)^{m_l} \sum_{i=1}^n
{\sum_{j=1}^{m_i} u_i^{(j)} \over a_i - \lambda} = 0, \qquad
w_i := \sum_{j=1}^{m_j} u_i^{(j)}
\end{equation}
where the $u_i^{(j)}$ denote the diagonal elements of
$\vec{x} \vec{x}^\dagger$. This shows immediately that $M$ has
eigenvalues $a_i$ with multiplicities $m_i-1$, and the
remaining $n-1$ eigenvalues given by the zeros of the random
rational function
\begin{equation}\label{Rl}
R(\lambda) := \sum_{i=1}^n {w_i \over a_i - \lambda}.
\end{equation}
The fact that the $w_i$ are positive (being equal to sums of squares)
implies that the roots of $R(\lambda)$ are all real (as  must be
since $M$ is Hermitian) and further have the interlacing property
\begin{equation}\label{3.21}
a_1 > \lambda_1 > a_2 > \lambda_2 > \cdots > \lambda_{n-1} > a_{n}
\end{equation}
(c.f.~(\ref{rar})).

We would like to compute the distribution of the roots of $\lambda$
for given $a_1,\dots,a_n$ and the $\{w_i\}$ random. This depends
crucially on the precise distribution of the $\{w_i\}$. Now
$w_i$ has the form
\begin{equation}\label{3.21a}
w_i = X_i/(X_1 + \cdots + X_n)
\end{equation}
where $X_i$ consists of $\beta m_i$ ($\beta = 1$ for
$\vec{x}$ real and $\beta = 2$ for $\vec{x}$ complex) 
independent real Gaussians with
mean zero and standard deviation $\sigma$, and thus has the gamma
distribution $\Gamma(s_i,2\sigma)$, $s_i := \beta m_i/2$.
It follows that the PDF for
$(w_1,w_2,\dots,w_{n-1};w_n)$ is equal to the  Dirichlet
distribution 
\begin{equation}\label{3.22}
{\Gamma(s_1 + \cdots + s_n) \over \Gamma(s_1) \cdots
\Gamma(s_n)} \prod_{i=1}^n w_i^{s_i-1}, \qquad
w_n := 1 - \sum_{j=1}^{n-1} w_j, \quad w_j >0.
\end{equation}

The working in Anderson's paper \cite{An91} shows us that the distribution
of the roots $\lambda_i$ of $R(\lambda)$ 
when the $\{w_i\}$ are distributed according
to (\ref{3.22}) is given by (\ref{An.2}) with
$y_i = \lambda_i$, $x_i = a_i$. As a consequence we can specify the 
sought eigenvalue distribution.

\begin{cor}\label{cor1}
The eigenvalues $\lambda_1,\dots, \lambda_{n-1}$ of $M$ in (\ref{M.1})
differing from the eigenvalues of $A$ and from 0 have the PDF
(\ref{An.2}) with $y_i = \lambda_i$ ($i=1,\dots,n-1$), $x_i = a_i$
($i=1,\dots,n$) and $s_i = \beta m_i/2$
($\beta = 1$ for
$\vec{x}$ real and $\beta = 2$ for $\vec{x}$ complex).
\end{cor}

In the case that all eigenvalues of $A$ are distinct (or doubly
degenerate in the case of $\vec{x}$ complex), the eigenvalues of
$M$ can be interpreted as so called radial Gelfand-Tzetlin
coordinates introduced by Guhr and Kohler \cite{GK02}. Further,
as shown in Appendix C, this observation and Corollary \ref{cor1} can
be used to rederive a recursion formula obtained in  \cite{GK02} for
certain matrix Bessel functions.

\subsection{Construction of interpolating Jacobi ensembles}
We can make use of Corollary \ref{cor1} to determine explicit random
matrices with eigenvalue PDFs which realize (\ref{1.8a}) and (\ref{1.8b}).
Consider first (\ref{1.8a}). Essential to our construction are random
matrices with a doubly degenerate spectrum which have an eigenvalue
PDF of the form (\ref{1.3a}).
The required random matrices are known from \cite{Be97} (see also
\cite{Fo02}). Thus consider a member $S$ of the circular ensemble
CSE${}_{(n^*+n)}$ (for the definition and construction of such matrices
--- in which each element is itself the $2 \times 2$ matrix representation
of a real quaternion --- see e.g.~\cite{Fo02}). Decompose $S$ as
\begin{equation}\label{4.11a}
S = \left [ \begin{array}{cc} r_{n^* \times n^*} & t_{n^* \times n}'
\\ t_{n \times n^*} & r_{n \times n}' \end{array} \right ]
\end{equation}
where $n^* \ge n$ and the subscript on the blocks tells us their
dimension (with the already mentioned qualification that each element
is a $2 \times 2$ matrix). Then we have from \cite{Be97,Fo02} that the
random matrix $t t^\dagger$ has eigenvalue PDF (\ref{1.3a}) with
$a=2(n^*-n)$.
Moreover, if we append to $t$ an extra $n_0$ rows of zeros and denote
this $\tilde{t}$ say, then
$\tilde{t} \, \tilde{t}^\dagger$ has a zero eigenvalue of 
multiplicity $n_0$, and $n$ eigenvalues with PDF (\ref{1.3a}) of
multiplicity 2. Substituting $\tilde{t}\, \tilde{t}^\dagger$ for $A$ in
(\ref{M.1}) we can make use of Corollary \ref{cor1} to deduce that
the PDF for the non-zero eigenvalues of $M$ realize (\ref{1.8a}).

\begin{thm}\label{th1a} 
With $\tilde{t}$ specified above and $\vec{x}$ a normalized 
complex Gaussian column vector 
of $2n+n_0$ rows, the non-zero eigenvalues of the random matrix 
$$ 
M = \Pi \tilde{t} \, \tilde{t}^\dagger \Pi, 
\qquad \Pi = {\bf 1} - \vec{x} \vec{x}^\dagger 
$$ 
have the PDF (\ref{1.3a}) with
\begin{equation}\label{4.13a}
a= 2(n^* - n), \qquad A = 2(n^* - n) -2n_0 + 1.
\end{equation}
\end{thm}

\noindent
Proof. \quad Let the non-zero eigenvalues of $\tilde{t}\, \tilde{t}^\dagger$
be denoted $a_1,a_2,\dots,a_n$. We know they are doubly degenerate and
have distribution (\ref{1.3a}) with $a$ therein given by (\ref{4.13a}),
and furthermore we know that $\tilde{t} \, \tilde{t}^\dagger$ has a zero
eigenvalue of multiplicity $n_0$. According to (\ref{4.5a}) and
(\ref{3.21}) the matrix $M$ then has a zero eigenvalue also of
multiplicity $n_0$, eigenvalues $a_1,\dots,a_n$ with
multiplicity 1 and eigenvalues $\lambda_1,\dots,\lambda_n$ such that
\begin{equation}\label{4.13'}
a_1 > \lambda_1 > a_2 > \cdots > \lambda_n > 0.
\end{equation}
From Corollary \ref{cor1} the conditional PDF of
$\lambda_1,\dots,\lambda_n$ given  $a_1,\dots,a_n$ is proportional to
\begin{equation}\label{sa.3a}
\prod_{i=1}^n a_i^{-2}
\Big ( {\lambda_i \over a_i} \Big )^{n_0-1}
\prod_{1 \le i < j \le n} {(\lambda_i - \lambda_j) \over
(a_i - a_j)^3}
\prod_{i,j=1}^n | a_i - \lambda_j|.
\end{equation}
But the PDF of $a_1,\dots,a_n$ is given by (\ref{1.3a}) (with the $x$'s
replaced by $a$'s and $a$ given by (\ref{4.13a})),
so forming the product with (\ref{sa.3a}) shows
the eigenvalue PDF of $M$ is proportional to
$$
\prod_{i=1}^n a_i^{2(n^*-n)-1} \Big ( {\lambda_i \over a_i}
\Big )^{n_0 - 1} \prod_{1 \le i < j \le n}
(a_i - a_j)(\lambda_i - \lambda_j)
\prod_{i,j=1}^n |a_i - \lambda_j|.
$$
After relabelling we recognize this as the PDF (\ref{1.8a}).
\hfill $\square$

A significant feature of Theorem \ref{th1a} is that by construction
\begin{equation}\label{hm}
{\rm odd}(M) = {\rm JSE}_n \Big |_{a\mapsto a+1 \atop b=0},
\end{equation}
where ${\rm odd}(M)$ refers to the distribution of the odd labelled
eigenvalues of the random matrix $M$.
Choosing any particular value of $A$ allowed by (\ref{4.13a}) and multiplying 
both sides of (\ref{hm}) by $\prod_{l=1}^n x_{2l-1}^{-A}$ we see that
(\ref{hm}) implies
\begin{equation}\label{hm1}
{\rm odd}({\rm JOE}_{2n} |_{a=(c-1)/2 \atop b=0}) =
{\rm JSE}_n |_{a=c+1 \atop b=0}
\end{equation}
where $c=2n_0 -1$.
Identities of this type were classified in \cite{FR01} using
functional properties of the PDFs. In fact it was found that (\ref{hm1})
is one of only two identities relating every second eigenvalue in a matrix
ensemble with orthogonal symmetry and an even number of eigenvalues, to
a matrix ensemble with symplectic symmetry (for the other see
(\ref{loe}) below). A challenge was issued to provide a matrix
derivation of such results; by way of the above construction this
challenge has been answered for the particular identity (\ref{hm1}).

Let us  now seek a realization of (\ref{1.8b}) as an eigenvalue
PDF. Guided by the above construction of random matrices with
eigenvalue PDF (\ref{1.8a}), we first seek random matrices with a
doubly degenerate spectrum which have an eigenvalue PDF equal to
(\ref{1.6aa}) which is the $A \to - \infty$ limit of (\ref{1.8a}).
From \cite{Be97,Fo02} we know that with $S$ a $(n^*+n)\times(n^*+n)$
random unitary matrix decomposed as in (\ref{4.11a}), the random
matrix $t t^\dagger$ has eigenvalue PDF (\ref{1.6aa}) with
\begin{equation}\label{4.11s'}
a = n^* - n,
\end{equation}
although the eigenvalues are all distinct. To obtain a doubly
degenerate spectrum with the same eigenvalue PDF, we simply replace
each complex element $x+iy$ of $t$ by its $2 \times 2$ real matrix
representation
\begin{equation}\label{rms}
\left [ \begin{array}{cc} x & y \\ -y & x \end{array} \right ].
\end{equation}
To this doubly degenerate spectrum with eigenvalue PDF (\ref{1.6aa}) we
want to add a zero eigenvalue of degeneracy $n_0$. As noted below
(\ref{4.11a}), this is achieved by simply appending $n_0$ rows of
zeros; let us denote the real representation of $t$ so modified by
$\hat{t}$. The real symmetric matrix $\hat{t} \hat{t}^T$ then has
a zero eigenvalue of multiplicity $n_0$ and $n$ eigenvalues with
PDF (\ref{1.6aa}) of multiplicity 2. We can now use Corollary
\ref{cor1} to obtain the sought realization of (\ref{1.8b}).

\begin{thm}\label{th2a}
With $\hat{t}$ specified above and $\vec{x}$ a normalized real Gaussian
vector of $2n+n_0$ rows, the non-zero eigenvalues of the random matrix
$$
M = \Pi \hat{t} \, \hat{t}^T \Pi,
\qquad \Pi = {\bf 1} - \vec{x} \vec{x}^T
$$
have PDF (\ref{1.8a}) with
$$
a= n^* - n, \qquad A = n^* - n -n_0 + 1.
$$
\end{thm}  

\noindent 
Proof. \quad Following the reasoning of the proof of Theorem
\ref{th1a}, the matrix $M$ has a zero eigenvalue of multiplicity $n_0$,
a distinct copy of the non-zero eigenvalues $a_1,\dots,a_n$ say of
$\hat{t} \hat{t}^T$, and eigenvalues $\lambda_1,\dots,\lambda_n$
satisfying the interlacing condition (\ref{4.13'}). Corollary \ref{cor1}
with $n \mapsto n+1$, $a_{n+1} = 0$, $s_{n+1} = n_0/2$,
$s_i = 1$ ($i=1,\dots,n$) gives that the conditional PDF of
$\lambda_1,\dots \lambda_n$ given $a_1,\dots,a_n$ is proportional to
$$
\prod_{i=1}^n a_i^{-1} \Big ( {\lambda_i \over a_i} \Big )^{n_0/2-1}
\prod_{1 \le i < j \le n} {(\lambda_i - \lambda_j) \over
(a_i - a_j)} \prod_{i,j=1}^n|a_i - \lambda_j|.
$$
The eigenvalue PDF of $M$ now follows by multiplying this by the PDF of
$a_1,\dots,a_n$ as given by (\ref{1.6aa}) (with the $x$'s replaced by
$a$'s). After relabelling the coordinates the PDF
(\ref{1.8b}) results with the parameters as stated.
\hfill $\square$

Analogous to (\ref{hm}), by construction
\begin{equation}\label{hmy}
{\rm odd}(M) = {\rm JUE}_n \Big |_{b=0}.
\end{equation}
Since with $A=0$ the eigenvalue PDF of $M$ coincides with that of the
matrix ensemble ${\rm JOE}_n |_{a\mapsto (a-1)/2 \atop b=0} \cup
{\rm JOE}_n |_{a\mapsto (a-1)/2 \atop b=0}$ 
(recall sentence below (\ref{1.3b}))
we have a
matrix theoretic understanding of the relation \cite{FR01}
\begin{equation}\label{hmy1}
{\rm odd}\Big ({\rm JOE}_n |_{a\mapsto (a-1)/2 \atop b=0} \cup
{\rm JOE}_n |_{a\mapsto (a-1)/2 \atop b=0} \Big ) =
{\rm JUE}_n  \Big |_{b=0}.
\end{equation}
  
\subsection{A random three term recurrence for interpolating
Jacobi ensembles}\label{s3J}

In this section, inspired by the recent work \cite{DE02}, it will be
shown that $J_{\rm o}^{(n,n-1)}(x,y)$, specified by
(\ref{J.2}), and $J_{\rm e}^{(n,n)}(x,y)$, specified by
(\ref{J.2e}), can be sampled from the zeros of a polynomial which in
turn is specified using a random three term recurrence.
Consider first (\ref{ra}). We begin by noting that (\ref{4.11}),
like (\ref{4.1}) 
is intimately related to the Anderson
density (\ref{An.2}). Thus in the latter put $n\mapsto n+1$, relabel
the $x$'s by $y$'s and the $y$'s by $x$'s, then set $y_1=1$,
$y_{n+1}=0$, relabel $y_{i+1}$ by $y_i$ $(i=1,\dots,n-1)$ and put
$s_1 = \beta+1$, $s_{n+1} = \alpha+1$ and $s_i=k$ $(i=2,\dots,n)$ to
obtain (\ref{4.11}). As a consequence, Anderson's result relating the
random rational function (\ref{Rl}), with coefficients distributed
according to the Dirichlet distribution (\ref{3.22}), to (\ref{An.2})
tells us we can similarly specify a random rational function related
to (\ref{4.11}).

\begin{cor}\label{cor2}
Denote the Dirichlet distribution (\ref{3.22}) by $D_n[s_1,\dots,s_{n-1};
s_n]$.
Let $(w_0,\dots,w_{n-1};w_{n})$ be distributed according to
$D_{n+1}[\beta+1,(k)^{n-1};\alpha+1]$, where the notation
$(k)^{n-1}$ denotes $k$ repeated $n-1$ times. 
We have that the roots of the random
rational function
\begin{equation}\label{R2}
\tilde{R}_{n+1}(x) := {w_0 \over x-1} + {w_n \over x} +
\sum_{i=1}^{n-1} {w_i \over x - y_i}
\end{equation}
are distributed according to the PDF (\ref{4.11}).
\end{cor}

Anderson's result stated below (\ref{3.22}) and Corollary
\ref{cor2} can be used to derive a random
three term recurrence which specifies a polynomial,
the zeros of which sample from the joint PDF
(\ref{J.2}), and also sample from the marginal PDF (\ref{4.10b}).
We will specify the recurrence by first detailing how it leads to
a polynomial with the sought properties in the low degree cases,
before stating its general form.

\smallskip
\noindent 
{\tt Step 1} Consider (\ref{R2}) in the case $n=1$ and write
$w_0 \mapsto w_0^{(1)}$, $w_1  \mapsto w_1^{(1)}$. Let
$(w_0^{(1)}; w_1^{(1)})$ be distributed according to 
$D_2[\beta^{(1)}+1;\alpha^{(1)}+1]$. Let $\lambda_1^{(1)}$ denote the
zero of (\ref{R2}) in this case and form the polynomial
\begin{equation}\label{A1}
A_1(x) := x - \lambda_1^{(1)}.
\end{equation}
It follows from Corollary \ref{cor2} that $\lambda_1^{(1)}$ is distributed
according to
\begin{equation}\label{G.1}
{J_{\rm o}^{(1,0)}(x,y) \over J_{\rm o}^{(\_,0)}(x)}
\Big |_{\alpha=\alpha^{(1)}, \, \beta=\beta^{(1)}} =:
P(\lambda_1^{(1)})
\end{equation}
which is itself the Dirichlet distribution $D_2[\alpha^{(1)}+1;
\beta^{(1)}+1]$.

\smallskip
\noindent 
{\tt Step 2}  Define $A_1(x)$ by (\ref{A1}) and also define
\begin{equation}\label{A0}
A_0(x) := 1.
\end{equation}
Let $(w_0^{(2)}, w_1^{(2)}; w_2^{(2)})$ be distributed according to 
$D_3[\beta^{(2)}+1,k;\alpha^{(2)}+1]$ and construct the random
quadratic polynomial
\begin{equation}\label{A2}
A_2(x) := w_2^{(2)} (x-1) A_1(x) + w_0^{(2)} x A_1(x) + w_1^{(2)} 
x(x-1)A_0(x).
\end{equation}
Dividing both sides by $x(x-1) A_1(x)$, this reads
$$
{A_2(x) \over x(x-1) A_1(x)} = { w_2^{(2)} \over x} +
{w_0^{(2)} \over x - 1} + { w_1^{(2)} \over x - \lambda_1^{(1)}}.
$$
Because $\sum_{\mu=0}^2 w_\mu^{(2)} = 1$ and the $w_\mu^{(2)}$ are
positive, $A_2(x)$ must be monic with real roots and we write
$$
A_2(x) = (x - \lambda_1^{(2)})(x -  \lambda_2^{(2)}).
$$
It follows from Corollary \ref{cor2} that the conditional distribution
of $\{\lambda_1^{(2)},  \lambda_2^{(2)}\}$ given $ \lambda_1^{(1)}$
has the form
\begin{equation}\label{G.2}
{J_{\rm o}^{(2,1)}(x,y) \over J_{\rm o}^{(\_,1)}(y)}
\Big |_{\alpha=\alpha^{(2)}, \, \beta=\beta^{(2)}} =:
P(\lambda_1^{(2)},\lambda_2^{(2)} | \lambda_1^{(1)})
\end{equation} 
and so the joint density of $\lambda_1^{(1)}, \lambda_1^{(2)},\lambda_2^{(2)}$
is
\begin{equation}\label{3.20'}
P(\lambda_1^{(1)}) P(\lambda_1^{(2)},\lambda_2^{(2)} | \lambda_1^{(1)}).
\end{equation}
If we set 
\begin{equation}\label{4.17}
\alpha^{(1)} = \alpha^{(2)} + \alpha_1 +k, \quad
\beta^{(1)} = \beta^{(2)} +\beta_1 +k
\end{equation}
we recognize this as the joint distribution function
$J_{\rm o}^{(2,1)}(x,y)$ with $\alpha=\alpha^{(2)},
\beta = \beta^{(2)}$. According to (\ref{4.10b}), if we now set
$\alpha_1 = \beta_1 = 0$ 
we can compute the
marginal distribution of $\lambda_1^{(2)},\lambda_2^{(2)}$,
\begin{equation}\label{rl}
\int_{\lambda_1^{(2)} > \lambda_1^{(1)} > \lambda_2^{(2)}}
d  \lambda_1^{(1)} \, P(\lambda_1^{(1)}) 
 P(\lambda_1^{(2)},\lambda_2^{(2)} | \lambda_1^{(1)}) 
\Big |_{\alpha_1=\beta_1=0} =
J_{\rm o}^{(2,\_)}(\lambda_1^{(2)*}, \lambda_2^{(2)*})
\Big |_{\alpha=\alpha^{(2)} \atop \beta = \beta^{(2)}},
\end{equation}
where the use of the $*$ on the right hand side
indicates the parameters have been chosen so
that (\ref{4.17}) holds with $\alpha_1 = \beta_1 = 0$. This
distribution is realized
by the roots of $A_2^*(x)$ when
$ \lambda_1^{(1)}$ is not observed.

\medskip
To proceed further requires an extension of the above arguments. 
Let us consider
\begin{equation}\label{3.20d}
{A_1(x) \over A_2^*(x)} = \sum_{l=1}^2 {u_l \over x - \lambda_2^{(l)*}},
\qquad u_1 + u_2 = 1
\end{equation}
and pose the question as to what distribution of $(u_1;u_2)$ is
required so that the distribution of the  zero on the right hand side
has the same distribution as $ \lambda_1^{(1)}$, with
$\lambda_1^{(2)*}, \lambda_2^{(2)*}$ distributed by the right hand side
of (\ref{rl}),
and is thus specified 
by (\ref{G.1})?

According to Anderson's result stated below (\ref{3.22})
 we have that with $(u_1;u_2)$
distributed according to $D_2(k;k)$, the root of (\ref{3.20d})
has conditional distribution
$$
{J_{\rm o}^{(2,1)}(\lambda_1^{(2)*}, \lambda_2^{(2)*};  \lambda_1^{(1)})
\Big |_{\alpha_1 = \beta_1 = 0} 
 \over
J_{\rm o}^{(2,\_)}(\lambda_1^{(2)*}, \lambda_2^{(2)*})
}.
$$
Thus the corresponding marginal distribution of $ \lambda_1^{(1)}$ is
\begin{eqnarray*}
&&
\int_{1 > \lambda_1^{(2)} > \lambda_1^{(1)} >  \lambda_2^{(2)} >0 }
d \lambda_1^{(2)} d \lambda_2^{(2)} \,
J_{\rm o}^{(2,\_)}(\lambda_1^{(2)*}, \lambda_2^{(2)*})
\Big |_{\alpha = \alpha^{(2)} \atop  \beta = \beta^{(2)}  }
{J_{\rm o}^{(2,1)}(\lambda_1^{(2)}, \lambda_2^{(2)};  \lambda_1^{(1)})
\Big |_{\alpha_1 = \beta_1 = 0} 
\over
J_{\rm o}^{(2,\_)}(\lambda_1^{(2)*}, \lambda_2^{(2)*})
}
\nonumber \\
&&
\quad = J_{\rm o}^{(\_,1)}(\lambda_1^{(1)})
\Big |_{\alpha = \alpha^{(2)}+k \atop \beta = \beta^{(2)}+k}
\end{eqnarray*}
where use has been made of (\ref{4.10a}) and (\ref{4.17}). This is indeed the
same distribution as (\ref{G.1}), provided we set $\alpha^{(1)}=
\alpha^{(2)}+k$,
$\beta^{(1)}=
\beta^{(2)}+k$ therein. Let us denote $A_1(x)$ with the parameters so
specialized by $A_1^{\#}(x)$.

As well as making use of (\ref{3.20d}) with $(u_1,u_2)$ distributed
according to $D_2[k;k]$, we require some special properties of the
Dirichlet and beta distributions. First we recall that the
Dirichlet distribution $D_2[\alpha;\beta]$ and the beta distribution
$B[\alpha,\beta]$ are the same thing. We require the fact that if
$(w_0,\dots,w_{n-1};w_n)$ is distributed according to 
$D_{n+1}[\alpha_0,\dots,\alpha_{n-1};\alpha_n]$, then the marginal
distribution of $w_j$ $(j=0,\dots,n-1)$ is given by
$B[\alpha_j,\sum_{i=0, i \ne j}^n \alpha_i]$ and the marginal
distribution of $w_j+w_k$, $(j \ne k, \, j,k\le n)$ is
$B[\alpha_j + \alpha_k, \sum_{i=0, i \ne j,k}^n \alpha_i ]$. We
also require the property of the beta distribution (see
e.g.~\cite[pg.~42]{Ra52})
\begin{equation}\label{bbb}
B[a+b,c] B[a,b] = B[a,b+c]
\end{equation}
where here --- in an abuse of notation --- the left hand side means
the product of random variables from the respective distributions,
and the right hand side tells us the distribution of the product.

\medskip
\noindent
{\tt Step 3} Analogous to the construction of $A_2(x)$, we construct
$A_3(x)$ by the random three term recurrence
$$
A_3(x) := w_2^{(3)}(x-1) A_2^*(x) + w_0^{(3)} x A_2^*(x) +
w_1^{(3)}x(x-1) A_1^{\#}(x)
$$
where $(w_0^{(3)}, w_1^{(3)}; w_2^{(3)})$ is distributed according to
$D_3[\beta^{(3)}+1,2k;\alpha^{(3)}+1]$, or equivalently 
$$
{A_3(x) \over x(x-1) A_2^*(x)} = {w_2^{(3)} \over x} +
{w_0^{(3)} \over x-1} + w_1^{(3)} {A_1^{\#}(x) \over  A_2^*(x)}.
$$
For $A_1^{\#}(x)/ A_2^*(x)$ we substitute (\ref{3.20d}). Now the
theory above (\ref{bbb}) tells us that the marginal distribution of
$w_1^{(3)}$ is $B[2k,\alpha^{(3)}+\beta^{(3)}+2]$, while the
distribution of $u_1$ in (\ref{3.20d}) is $B[k,k]$. Applying
(\ref{bbb}) it follows that we can write
$$
w_1^{(3)} {A_1^{\#}(x) \over	 A_2^*(x)} = \sum_{l=1}^2
{\tilde{u}_l \over x - \lambda_2^{(l)*}}
$$
where $\tilde{u}_1$ has distribution $B[k,\alpha^{(3)}+\beta^{(3)}+k+2]$
and $\tilde{u}_1+\tilde{u}_2$ has distribution
$B[2k;\alpha^{(3)}+\beta^{(3)}+2]$. Consequently we have 
\begin{equation}\label{4.33'}
{A_3(x) \over x(x-1) A_2^*(x)} = {\tilde{w}_3^{(3)} \over x} +
{\tilde{w}_0^{(3)} \over x-1} +
\sum_{l=1}^2 {\tilde{w}_l^{(3)} \over x - \lambda_2^{(l)*}}
\end{equation}
where $(\tilde{w}_0^{(3)},\tilde{w}_1^{(3)},\tilde{w}_2^{(3)};
\tilde{w}_3^{(3)})$ has distribution $D_4[\beta^{(3)}+1,(k)^2;
\alpha^{(3)}+1]$. Arguing now as in the derivation of (\ref{G.2})
that
$$
A_3(x) = (x - \lambda_1^{(3)})(x - \lambda_2^{(3)})
(x - \lambda_3^{(3)})
$$
where the conditional distribution of
$\{\lambda_1^{(3)},\lambda_2^{(3)},\lambda_2^{(3)}\}$ given
$\{\lambda_1^{(2)*}, \lambda_2^{(2)*} \}$ has the form
$$
{J_{\rm o}^{(3,2)}(x,y) \over J_{\rm o}^{(\_,2)}(y)}
\Big |_{\alpha=\alpha^{(3)}, \beta=\beta^{(3)}} =:
P(\lambda_1^{(3)},\lambda_2^{(3)},\lambda_2^{(3)}|
\lambda_1^{(2)*}, \lambda_2^{(2)*}).
$$
The joint density of $\{\lambda_1^{(2)*}, \lambda_2^{(2)*},
\lambda_1^{(3)},\lambda_2^{(3)},\lambda_2^{(3)}\}$ is therefore
\begin{equation}\label{fri}
J_{\rm o}^{(2,\_)}(\lambda_1^{(2)*}, \lambda_2^{(2)*})
P(\lambda_1^{(3)},\lambda_2^{(3)},\lambda_2^{(3)}|
\lambda_1^{(2)*}, \lambda_2^{(2)*})
\end{equation}
which with
$$
\alpha^{(2)} = \alpha^{(3)} + \alpha_1 +k, \quad
\beta^{(2)} = \beta^{(3)} +\beta_1 +k
$$
we recognize as the joint distribution function $J_{\rm o}^{(3,2)}(x,y)$
with $\alpha = \alpha^{(3)}$, $\beta = \beta^{(3)}$.
As with (\ref{rl}), the marginal distribution of
$\lambda_1^{(3)},\lambda_2^{(3)},\lambda_3^{(3)}$ can be computed in the
case $\alpha_1=\beta_1=0$. Thus it follows from (\ref{4.10b}) that
\begin{eqnarray}\label{rll2}
&&
\int_{\lambda_1^{(3)} > \lambda_1^{(2)} > 
\lambda_2^{(3)} > \lambda_2^{(2)} > \lambda_3^{(3)}}
d  \lambda_1^{(2)} d  \lambda_2^{(2)}\,
J_{\rm o}^{(2,\_)} (\lambda_1^{(2)*}, \lambda_2^{(2)*})
 P(\lambda_1^{(3)},\lambda_2^{(3)},\lambda_3^{(3)} | \lambda_1^{(2)*},
\lambda_2^{(2)*}) \Big |_{\alpha_1=\beta_1=0} \nonumber \\ &&\qquad =
J_{\rm o}^{(3,\_)}(\lambda_1^{(3)*}, \lambda_2^{(3)*},  \lambda_3^{(3)*}),
\end{eqnarray}
giving the distribution of the zeros of $A_3^*(x)$ when
$ \lambda_1^{(2)*}, \lambda_2^{(2)*}$ are not observed. 

\medskip
Step 3 is representative of the general step $n$ in generating the
recurrence. Of course we now need inductive hypotheses relating to 
the roots of polynomials generated in earlier steps. In particular,
we suppose that in step $n-2$ a polynomial $A_{n-2}(x)$ has been
generated and the density of its roots is given by
$J^{(\_,n-2)}(y) |_{\alpha_1=\beta_1=0}$ as specified by (\ref{4.10a})
with $\alpha= \alpha^{(n-2)}$, $ \beta=\beta^{(n-2)}$. With the
special choice of parameters $
\alpha^{(n-2)} = \alpha^{(n-1)} +k,  \beta^{(n-2)} = \beta^{(n-1)} +k
$
we denote $A_{n-2}(x)$ by
$A_{n-2}^{\#}(x)$. At step $n-1$ we require that a polynomial $A_{n-1}^*(x)$
has been generated which
has the density of its roots given by $J^{(n-1,\_)}_{\rm o}(x)$ as specified by
(\ref{4.10b}) with $\alpha= \alpha^{(n-1)}$, $ \beta=\beta^{(n-1)}$.

\medskip
\noindent
{\tt Step n} We construct $A_n(x)$ by the random three term recurrence
\begin{equation}\label{stn}
A_n(x) = w_2^{(n)}(x-1) A_{n-1}^*(x) + w_0^{(n)}x A_{n-1}^*(x) +
w_1^{(n)} x(x-1) A_{n-2}^{\#}(x)
\end{equation}
where $(w_0^{(n)}, w_1^{(n)}, w_2^{(n)})$ is distributed according to
$D_3[\beta^{(n)}+1,(n-1)k;\alpha^{(n)}+1]$. Arguing as in the derivation of
(\ref{fri}) we see that with
\begin{equation}\label{abn}
\alpha^{(n-1)}= \alpha^{(n)} +  \alpha_1 + k, \qquad
\beta^{(n-1)}= \beta^{(n)} +  \beta_1 + k
\end{equation}
the joint distribution of the roots of $A_n(x)$ and $A_{n-1}^*(y)$ is given
by $J_{\rm o}^{(n,n-1)}(x,y)$ with $\alpha = \alpha^{(n)}$,
$\beta = \beta^{(n)}$, and that with $\alpha_1=\beta_1=0$ the marginal
distribution of the roots of $A_n(x)$ is given by $J_{\rm o}^{(n,\_)}(x)$
with $\alpha = \alpha^{(n)}$, $\beta = \beta^{(n)}$.

\medskip
From a practical point of view, our objective is to sample from
$J_{\rm o}^{(n,n-1)}(x,y)$ and $J_{\rm o}^{(n,\_)}(x)$ for a fixed value
of $n$ and fixed parameters. To sample from $J_{\rm o}^{(n,\_)}(x)$ with
$\alpha=\alpha_0$, $\beta =\beta_0$ we implement the above steps with
\begin{equation}\label{mr0}
\alpha^{(j)} = (n-j)k+\alpha_0, \qquad 
\beta^{(j)} = (n-j)k+\beta_0.
\end{equation}
We see that in this situation $A_j^{\#}(x) = A_j^*(x)$ and so
$\{A_j^{\#}(x)\}_{j=2,\dots,n}$ is determined by the random recurrence
\begin{equation}\label{mr}
A_j^{\#}(x) = w_2^{(j)}(x-1) A_{j-1}^{\#}(x) + w_0^{(j)}x A_{j-1}^{\#}(x) +
w_1^{(j)} x(x-1) A_{j-2}^{\#}(x)
\end{equation}
where $(w_0^{(j)}, w_1^{(j)}, w_2^{(j)})$ is distributed according to
$D_3[(n-j)k+\beta_0+1,(j-1)k;(n-j)k+\alpha_0+1]$. The initial conditions
for the recurrence are $A_{-1}^{\#}(x)=0$ and $A_0^{\#}(x)=1$.
The zeros of $A_n^{\#}(x)$ then are distributed according to
$J_{\rm o}^{(n,\_)}(x) |_{\alpha=\alpha_0 \atop \beta=\beta_0}$.

If our objective is to sample from $J_{\rm o}^{(n,n-1)}(x,y)$ with
$\alpha=\alpha_0$, $\beta=\beta_0$, we again compute
$\{A_j^{\#}(x) \}_{j=0,\dots,n-1}$
this time replacing $\alpha_0$, $\beta_0$ by $\alpha_0+\alpha_1$,
$\beta_0+\beta_1$ throughout. 
Let us write
$A_j^{\#}(x)$ with these parameters as $\tilde{A}^{\#}_j(x)$. Because we now
have $\alpha^{(n-1)} = k + \alpha_0+\alpha_1$,
$\beta^{(n-1)} = k + \beta_0+\beta_1$ we see that $A_{n-1}^* (x) =
\tilde{A}_{n-1}^{\#}(x)$, so according to (\ref{stn}) the final step is
to compute
\begin{equation}
A_n(x) = w_2^{(n)}(x-1) \tilde{A}_{n-1}^{\#}(x) + 
w_0^{(n)}x \tilde{A}_{n-1}^{\#}(x) +
w_1^{(n)} x(x-1) \tilde{A}_{n-2}^{\#}(x)
\end{equation}
where $(w_0^{(n)}, w_1^{(n)}, w_2^{(n)})$ is distributed according to
$D_3[\beta_0+1,(n-1)k;\alpha_0+1]$. We then have that the zeros of
$(A_n(x), \tilde{A}_{n-1}^{\#}(y))$ 
have the joint distribution $J_{\rm o}^{(n,n-1)}(x,y)$.

Let us now turn our attention to sampling from $J_{\rm e}^{(n,n)}(x,y)$
as specified by (\ref{J.2e}). First we note from Anderson's result stated
below (\ref{3.22}) that the random rational function
\begin{equation}\label{ask}
\hat{R}_{n+1}(x) := {w_{n+1} \over x} + \sum_{i=1}^n {w_i \over
x - y_i},
\end{equation}
where $(w_1,\dots,w_n;w_{n+1})$ is distributed according to
$D_{n+1}[(k)^n;\alpha+1]$, has the PDF for its zeros given by
$J_{\rm e}^{(n,n)}(x,y)/J_{\rm e}^{(\_,n)}(y)$. Let us define
$\{A_j^{\#{\rm e}}(x) \}_{j=0,\dots,n}$ as specified by the
recurrence (\ref{mr}) but with $\alpha_0 \mapsto
\alpha_0 + \alpha_1 + k$, $\beta_0 \mapsto \beta_1$ throughout.
Furthermore, with $(w_1, w_2)$ distributed according to
$B_n[nk,\alpha_0+1]$ define
\begin{equation}\label{vaw}
V_n(x) = w_2 A_n^{\#{\rm e}}(x) + w_1 x A_{n-1}^{\#{\rm e}}(x).
\end{equation}
The significance of $V_n(x)$ is seen by noting from the argument
below (\ref{3.20d}) that
with $y_1,\dots,y_n$ denoting the zeros of $A_j^{\#{\rm e}}(x)$ we have 
$$
{A_{n-1}^{\#{\rm e}}(x) \over A_n^{\#{\rm e}}(x)} =
\sum_{l=1}^n {u_l \over x - y_l}
$$
where $(u_1,\dots,u_n)$ is distributed according to $D_n[(k)^{n-1};k]$,
and then proceeding as in the derivation of (\ref{bbb}) to deduce
from this the expansion
\begin{equation}\label{val}
{V_n(x) \over x A_n^{\#{\rm e}}(x)} =
{\tilde{w}_{n+1} \over x} + \sum_{l=1}^n {\tilde{w}_l \over
x - y_l^{(j)}}
\end{equation}
where $(\tilde{w}_1,\dots, \tilde{w}_n; \tilde{w}_{n+1})$ is distributed
according to $D_{n+1}[(k)^n;\alpha_0+1]$. The right hand side of (\ref{val})
is just the rational function (\ref{ask}), and so the PDF for its zeros,
given $\{y_1,\dots,y_n\}$, is
\begin{equation}\label{aa0}
{J_{\rm e}^{(n,n)}(x,y) \over J_{\rm e}^{(\_,n)}(y)}
\Big |_{\alpha = \alpha_0}.
\end{equation}
But the marginal distribution of $\{y_1,\dots,y_n\}$ is 
$J_{\rm o}^{(n,\_)}(y)$ with $\alpha = \alpha_0+\alpha_1+k$, $\beta = \beta_1$.
Multiplying this by (\ref{aa0}) shows that the joint distribution of the
zeros of $\{V_n(x),A_n^{\#{\rm e}}(y) \}$ is given by
$J_{\rm e}^{(n,n)}(x,y)|_{\alpha = \alpha_0}$.

We note from (\ref{4.10ae'}) that if we set $\alpha_1=0$,
$\beta_1=\beta_0-k$ in the construction of $\{A_j^{\# {\rm e}}(x) \}_{j=0,
\dots,n}$ and then compute $V_n(x)$ according to (\ref{vaw}),
the marginal  distribution of the zeros of $V_n(x)$ are given by
$J_{\rm e}^{(n,\_)}(x) |_{\alpha = \alpha_0 \atop \beta_1 = \beta_0-k}$.
But according to (\ref{4.10ae'}) and (\ref{4.10b}) the latter is
identical to $J_{\rm o}^{(n,\_)}(x) |_{\alpha = \alpha_0 \atop \beta = \beta_0}
$ and so the marginal  distribution of the zeros of $V_n(x)$ in this case
is the same as that for the zeros  of $A_n^{\# }(x)$. Thus with
$(w_1, w_2)$ as in (\ref{vaw}) we have
\begin{equation}\label{mrr}
A_n^{\# }(x) = \Big ( w_2 A_n^{\# }(x) 
+ w_1 x
A_{n-1}^{\# }(x) \Big ) \Big |_{\alpha_0 \mapsto
\alpha_0 +k \atop \beta_0 \mapsto \beta_0 -k}
\end{equation}

The random recurrences (\ref{mr}) and  (\ref{mrr}) assume
definite forms if we write $\alpha_0 = ak$, $\beta_0 = bk$ and take the
limit $k \to \infty$. Thus the random variables
$(w_0^{(j)}, w_1^{(j)}, w_2^{(j)})$ in
(\ref{mr}) crystallize to the definite value
$((n-j+b)/d, (j-1)/d, (n-j+a)/d)$ where $d=2n-j-1+a+b$ and so (\ref{mr})
reads
\begin{equation}\label{fe}
(2n-j-1+a+b) A_j^{\#}(x) = (n-j+a)(x-1) A_{j-1}^{\#}(x) +
(n-j+b) x A_{j-1}^{\#}(x) + (j-1)x(x-1)  A_{j-2}^{\#}(x) 
\end{equation}
with initial conditions $A_{-1}^{\#}(x)=0$, $A_0^{\#}(x) = 1$.
Similarly (\ref{mrr}) reads
\begin{equation}\label{fe2}
(a+n) A_{n}^{\# }(x) = \Big ( a A_{n}^{\# }(x) + n x A_{n-1}^{\# }(x)
\Big ) \Big |_{a \mapsto
a +1 \atop \beta_0 \mapsto b -1} .
\end{equation}
Using standard Jacobi polynomial recurrences we can show that the
solution of (\ref{fe}) is
given by
\begin{equation}\label{fe3}
A_{j}^{\# }(x) = \tilde{P}_j^{(n+a-1-j, n+b-1-j)}(x),
\end{equation}
where the use of $\tilde{}$ indicates the monic version of the
corresponding polynomial. Furthermore, (\ref{fe3}) satisfies
(\ref{fe2}). The fact from (\ref{fe3}) that $A_n^{\# }(x) =
 \tilde{P}_n^{(a-1,b-1)}(x)$ can be anticipated.
Thus in general
the PDF for the marginal distribution of
the zeros of $A_n^{\# }(x)$ is given by
$J_{\rm o}^{(n,\_)}(x)$, and if we put 
$\alpha_0 = ak$, $\beta_0 = bk$ and take the
limit $k \to \infty$ it is a known result \cite{Sz75} that this
PDF crystallizes at the zeros of the Jacobi polynomial
$P_n^{(a-1,b-1)}(x)$.

In Appendix D we make use of our ability to sample from $J_0^{(n,\_)}$
to give a Monte Carlo evaluation of a multidimensional integral
formula \cite{Fo93aa} for the bulk two-point correlation function
of matrix ensembles with symplectic symmetry.

\section{Random matrix realizations of the Laguerre interpolating
ensembles}
\setcounter{equation}{0}
Consider  the Dirichlet distribution (\ref{3.22}). Let $s_n=L$ and
scale $w_1,w_2,\dots,w_{n-1}$ so that $w_i \mapsto w_i/L$,
($i=1,\dots,n-1$). Then in the limit $L \to \infty$ (\ref{3.22})
reduces to the product of independent gamma distributions
\begin{equation}\label{4.9l}
{\sigma^{-(n-1)} \over \Gamma(s_1) \cdots \Gamma(s_{n-1})}
\prod_{i=1}^{n-1}(w_i/\sigma)^{s_i-1} e^{-w_i/\sigma}, \quad w_i>0
\end{equation}
with $\sigma = 1$.
If we also scale $a_i \mapsto a_i/L$ $(i=1,\dots,n-1)$, $\lambda
\mapsto \lambda/L$ and set $a_n=1$ then the limiting form of the
random rational function (\ref{Rl}) reads
\begin{equation}\label{4.6l}
R^{\rm L}(\lambda) := 1 + \sum_{i=1}^{n-1} {w_i \over a_i - \lambda}
\end{equation}
where the $w_i$ are distributed according to (\ref{4.9l}) (the superscipt
``L'' denotes Laguerre). The distribution of the roots of (\ref{4.6l}) is
given by the appropriate limiting form of the Anderson density
(\ref{An.2}), in accordance with Anderson's result stated below
(\ref{3.22}). Let us take note of the explicit form.

\begin{cor}\label{cor3}
Consider the random rational function (\ref{4.6l}) with the coefficients
$w_1,\dots,w_{n-1}$ distributed according to (\ref{4.9l}). 
This has exactly $n-1$ roots, which
since the
$w_i$ are positive,  are real.
For given $\{a_i\}$ these roots have the PDF
\begin{equation}\label{ss2}
{1 \over  \Gamma(s_1) \cdots \Gamma(s_{n-1})}
e^{-\sum_{j=1}^{n-1}(\lambda_j - a_j)}
\prod_{1 \le i < j \le n-1} {(\lambda_i - \lambda_j) \over
(a_i - a_j)^{s_i+s_j-1}}
\prod_{i,j=1 \atop i\ne j}^n |\lambda_i - a_j|^{s_j - 1}
\end{equation}
where
\begin{equation}
\lambda_1 > a_1 > \lambda_2 > a_2 > \cdots > \lambda_n  > a_n .
\end{equation}
\end{cor}

\noindent
We remark that this PDF is implicit in the work of Evans
\cite{Ev94}, who was studying Laguerre analogues of the Selberg
integral using the method of Anderson. Also, special cases of
Corollary \ref{cor3} are known from \cite{AM98}. 

A matrix structure for which the eigenvalue condition reduces to the
calculation of the roots of (\ref{4.6l}) (with $n$ replaced by
$n+1$ for convenience) is easy to specify. One approach would be to
consider the appropriate limiting form of (\ref{M.1}). Alternatively
we can write down the required matrix and check directly that it has the
sought property. Thus let $A$ be a real symmetric (complex Hermitian)
matrix with eigenvalues $a_1>a_2>\dots>a_n$ of multiplicities
$m_1,\dots,m_n$ respectively. Let $X$ be a vector of independent real
standard Gaussians (complex Gaussians) having the same number of rows as
$A$, and consider the matrix
\begin{equation}\label{M.1l}
M = A + b \vec{x} \vec{x}^\dagger, \qquad b>0.
\end{equation}
A simple calculation along the lines of Lemma \ref{lem8} and the
derivation of (\ref{M.1}) shows that $M$ has eigenvalues $a_i$ with
multiplicities $m_i-1$, and the remaining $n$ eigenvalues are given by the
zeros of the rational function (\ref{4.6l}) (with $n \mapsto n+1$).
The $w_i$ in the latter are distributed according to (\ref{4.9l}), 
with $\sigma = 2b/\beta$ and
and 
$s_i = \beta m_i/2$ where $\beta=1$ (real case), $\beta=2$ (complex case).
Thus (\ref{ss2}) with $a_i \mapsto \beta a_i/2b$, $\lambda_i \mapsto \beta
\lambda_i/2b$, gives the eigenvalue PDF of $M$.

The eigenvalue PDF of (\ref{M.1l}) in the
complex case with all eigenvalues $a_i$ distinct can be derived in a
different way which has the advantage of applying to a more general matrix
structure. This is given in Appendix E.

\subsection{Construction of interpolating Laguerre matrix ensembles}
We can use knowledge of the eigenvalue PDF of (\ref{M.1l}) to construct
matrix ensembles which have as their eigenvalue PDF (\ref{1.1}) and
(\ref{1.5}). For the parameter dependent PDF (\ref{1.1}) we begin by
recalling (see e.g.~\cite{Fo02}) that (\ref{1.3}) is realized
as the eigenvalue distribution
of the matrix ensemble of
$4n \times 4n$ antisymmetric Gaussian random
matrices, in which the elements are
pure imaginary numbers with each $2 \times 2$ block having a real
quaternion structure, and one changes variables $\lambda_i^2
\mapsto \lambda_i$. Such matrices are equivalent to block
matrices of the form
\begin{equation}\label{1.3'}
\left [ \begin{array}{cc} 0_{2n \times 2n} & X_{2n \times 2n} \\
 X_{2n \times 2n}^\dagger &  0_{2n \times 2n} \end{array} \right ]
\end{equation}
where $X$ is an antisymmetric Gaussian complex matrix, real and imaginary
parts having variance $1/2$, so $X^\dagger X$ has eigenvalue PDF
(\ref{1.3}) with each eigenvalue doubly degenerate.

\begin{thm}\label{th1}
Let $X$ be a $2n \times 2n$ antisymmetric complex Gaussian matrix,
and let $\vec{x}$ be a $2n \times 1$ complex Gaussian vector, where
the real and imaginary parts of the complex Gaussians have variance
$1/2$. The random matrix
\begin{equation}\label{5.5'}
M = X^\dagger X + b \vec{x} \vec{x}^{\, \dagger}
\end{equation}
has eigenvalue PDF (\ref{1.1}) with $A=1-2/b$. 
\end{thm}

\noindent
Proof. \quad The eigenvalues of $X^\dagger X$ have multiplicity 2, and
thus according to the result noted below (\ref{M.1l}) the
eigenvalues of $M$ consist of the eigenvalues of $X^\dagger X$ with
multiplicity 1 $(y_1,\dots,y_n$ say), as well as $n$ additional
eigenvalues $x_1,\dots,x_n$ which must satisfy the interlacing condition
\begin{equation}\label{5.5a}
x_1 > y_1  > x_2 > y_2 > \cdots > x_n > y_n.
\end{equation}
Setting $m_i=2$, $\beta=2$ in the result noted below (\ref{M.1l}), it follows
from (\ref{ss2}) that the conditional distribution of the $x$'s
given the $y$'s is proportional to
$$
e^{-\sum_{i=1}^n(x_i-y_i)/b}
\prod_{1 \le i < j \le n} {(x_i - x_j) \over (y_i - y_j)^3}
\prod_{i,j=1}^n |x_i - y_j|.
$$
Multiplying this by (\ref{1.3}) (with the $x$'s relabelled
$y$'s) gives that the eigenvalue PDF of $M$
is proportional to
$$
e^{-\sum_{i=1}^n(y_i + (x_i-y_i)/b)}
\prod_{1 \le i < j \le n} (x_i - x_j) (y_i - y_j)
\prod_{i,j=1}^n |x_i - y_j|,
$$
and relabelling the eigenvalues gives the desired result.
\hfill $\square$

By construction
$$
{\rm even}(M) = {\rm LSE}_n |_{a=0}
$$
where ${\rm even}(M)$ refers to the distribution of the even labelled
eigenvalues of the random matrix $M$. 
The result holds independent of the parameter
$b$ in (\ref{5.5'}). Because Theorem \ref{th1} also tells us that when
$b=2$ the random matrix $M$ has the same PDF as LOE${}_{2n}|_{a=0}$ matrices,
we thus have a matrix theoretic understanding of the relation
\begin{equation}\label{loe}
{\rm even}({\rm LOE}_{2n}|_{a=0}) = {\rm LSE}_n|_{a=0}.
\end{equation}

We remark that with $\tilde{X}$ denoting the $(2n+1) \times 2n$ random
matrix which is obtained from $X$ in (\ref{5.5'}) by adjoining an
extra row $\sqrt{b} \vec{x}$, we have
\begin{equation}\label{jam}
\tilde{X}^\dagger \tilde{X} = X^\dagger X + b \vec{x} \vec{x}^\dagger.
\end{equation}
Thus we can interpret Theorem \ref{th1} as applying to the 
square of the singular
values of $\tilde{X}$. This latter viewpoint indicates a special
property of the case $b=1$. Then $\tilde{X}$ is identical to the
$(2n+1)\times (2n+1)$ version of $X$ with the last column removed.
But it is a standard result that the singular values of a matrix
interleave with those of a matrix obtained by removing a single
row or column (starting with the largest singular value of the larger
matrix). 
Now we know that a $(2n+1)\times(2n+1)$ antisymmetric complex
Gaussian matrix is such that $\tilde{X}^\dagger \tilde{X}$ has one
zero eigenvalue, and $n$ doubly degenerate eigenvalues, the latter
having PDF LSE${}_n |_{a=2}$. The interlacing property now implies
\begin{equation}\label{lam}
{\rm odd}(M |_{b=1}) = {\rm LSE}_{n} |_{a=2}.
\end{equation}
As done in \cite{FR02}, this can be checked directly from the PDF
(\ref{1.1}).

Consider next the PDF (\ref{1.5}). We know that as $A \to  - \infty$
this reduces to (\ref{1.6}), which is the matrix ensemble
 LUE${}_n|_{a=0}$. It is well known (see e.g.~\cite{Fo02}) 
that the eigenvalue PDF for the LUE${}_n$ is realized
 by matrices of the form $X^\dagger X$ where $X$ is an $n \times n$ matrix
with i.i.d.~complex Gaussian entries. Let us replace each complex element
$x+iy$ of $X$ by its $2 \times 2$ real matrix representation
(\ref{rms})
and denote the corresponding $2n \times 2n$ matrix by $\bar{X}$. The
eigenvalues of $\bar{X}$ are the eigenvalues of $X$ except that in
$\bar{X}$ each has multiplicity 2.

\begin{thm}\label{th2b}
Let $\bar{X}$ be the $2n \times 2n$ real matrix constructed from the
$n \times n$ complex Gaussian matrix as specified above, and let
$\vec{x}$ denote a $2n \times 1$ real Gaussian vector, where in the
Gaussians each independent part has variance $1/2$. The Gaussian
random matrix
$$
M = \bar{X}^T \bar{X} + b \vec{x} \vec{x}^T
$$
has eigenvalue PDF (\ref{1.5}) with $A=1-2/b$.
\end{thm}

\noindent
Proof. \quad Arguing as in the proof of Theorem \ref{th1} we see that
the eigenvalues of $M$ consist of the eigenvalues of $\bar{X}^T \bar{X}$
with multiplicity 1, $y_1,\dots,y_n$ say, as well as $n$ additional
eigenvalues $x_1,\dots,x_n$ which must satisfy the interlacing
condition (\ref{5.5a}). Making use of the result noted below (\ref{M.1l})
with $m_i=2$, $\beta=1$, it follows from (\ref{ss2}) 
that the conditional distribution of the $x$'s
given the $y$'s is proportional to
$$
e^{-\sum_{i=1}^n(x_i-y_i)/b}
\prod_{1 \le i < j \le n} {(x_i - x_j) \over (y_i - y_j)}.
$$
Relabelling the $x$'s in (\ref{1.3}) by $y's$, and forming the product
with the conditional  distribution
shows that the eigenvalue PDF of $M$
is proportional to
$$
e^{-\sum_{i=1}^n(y_i + (x_i-y_i)/b)}
\prod_{1 \le i < j \le n} (x_i - x_j) (y_i - y_j)
\prod_{i,j=1}^n |x_i - y_j|,
$$
and relabelling this coincides with (\ref{1.5}) with $A=1-2/b$ therein.
\hfill $\square$

Analogous remarks made after Theorem \ref{th1} also apply to
Theorem \ref{th2}. Thus by construction
$$
{\rm even}(M) = {\rm LUE}_n |_{a=0},
$$
and because when $b=2$ the random matrix $M$ has the same
eigenvalue PDF as LOE${}_n |_{a=0}\cup$LOE${}_n |_{a=0} $ matrices, we thus have
a matrix theoretic understanding of the relation \cite{FR01}
$$
{\rm even}({\rm LOE}_n |_{a=0}\cup {\rm LOE}_n |_{a=0}) = 
{\rm LUE}_n |_{a=0}.
$$
Also, the equation (\ref{jam}) holds with $\bar{X}$ replacing
$X$. Here the matrix $\tilde{X}$ in the case $b=1$ is equivalent to
the $(2n+2) \times 2n$ version of $\bar{X}$ with the last row removed.
Now before removing the row, such Gaussian matrices multiplied by
their transpose have a doubly degenerate zero eigenvalue and
$n$ doubly degenerate eigenvalues with PDF ${\rm LUE}_n |_{a=1}$.
Arguing as in the derivation of (\ref{lam}) we therefore conclude
\begin{equation}\label{lam1}
{\rm odd}(M|_{b=1}) = {\rm LUE}_n |_{a=1},
\end{equation}
which like (\ref{lam}) can  be verified by direct integration of
the PDF (\ref{1.5}) with $A=-1$ \cite{FR02}.

\subsection{Laguerre limit of the three term recurrences}
The joint PDF (\ref{J.2}) has a well defined Laguerre limit, specified
by changing variables $x_i \mapsto x_i/L$, $y_i \mapsto y_i/L$, setting
$\beta = L/b, \beta_1 = L/b_1$ and taking the limit $L \to \infty$. This
gives
\begin{eqnarray}
L_{\rm o}^{(n,n-1)}(x,y) & := & {1 \over \Gamma(1+\alpha)
(\Gamma(k))^{n-1}} {1 \over \widetilde{W}_{n-1}(\alpha+\alpha_1+k,k;
bb_1/(b+b_1))}
\prod_{i=1}^n x_i^\alpha e^{-x_i/b}\prod_{1 \le i < j \le n} |x_j - x_i|
\nonumber \\&& \times
\prod_{i=1}^{n-1} y_i^{\alpha_1} e^{-y_i/b_1} \prod_{1 \le i < j \le n}
|y_j - y_i| \prod_{i=1}^n \prod_{j=1}^{n-1}|x_j - y_i|^{k-1}
\end{eqnarray}
where
\begin{eqnarray*}
\lefteqn{
\widetilde{W}_n(a,k;b) =
\int_0^\infty dx_1 \cdots \int_0^\infty dx_n \,
\prod_{l=1}^n x_l^a e^{-x_l/b} \prod_{1 \le j < k \le n}
|x_k - x_j|^k } \\ && =
b^{n(a+1+(n-1)k)} \prod_{j=1}^n {\Gamma(1+kj) \Gamma(1+a+k(j-1)) \over
\Gamma(1+k)}
\end{eqnarray*}
and the $x$'s and $y$'s are interlaced according to
\begin{equation}\label{iiu}
x_1 > y_1 > x_2 > y_2 > \cdots > y_{n-1} > x_n > 0.
\end{equation}
Also of interest is the Laguerre limit of the marginal distribution
(\ref{4.10b}),
\begin{equation}\label{5.14}
\int_R dy_1 \cdots dy_{n-1} \, L_{\rm o}^{(n,n-1)}(x,y) 
\Big |_{\alpha_1=0
\atop 1/b_1=0}
= {1 \over \widetilde{W}_n(\alpha,k;b)}
\prod_{i=1}^n x_i^\alpha e^{-x_i/b} \prod_{1 \le j < k \le n} |x_k - x_j|^k
=: L_{\rm o}^{(n,\_)}(x)
\end{equation}
where $R$ refers to the region (\ref{iiu}).

Sampling from $L_{\rm o}^{(n,n-1)}(x,y)$ and
$L_{\rm o}^{(n,\_)}(x)$ can be undertaken by taking the Laguerre
limit of the three term recurrences in Section \ref{s3J}. Consider first
the recurrence (\ref{mr}) determining the polynomial with zeros
realizing the PDF $J_{\rm o}^{(n,\_)}(x)$. The Laguerre limit is
obtained by scaling $x \mapsto x/L$, $w_1^{(j)} \mapsto v_1^{(j)}/L$,
$w_2^{(j)} \mapsto v_2^{(j)}/L$, $w_0^{(j)} = 1$, where the 
$v_1^{(j)}, v_2^{(j)}$ are distributed according to the gamma
distributions $\Gamma[b;(j-1)k]$, $\Gamma[b;(n-j)k+\alpha_0+1]$
respectively (here the notation $\Gamma[\sigma;s]$ refers to the
density function proportional to $x^{s-1} e^{-s/\sigma}$).
With $v_1^{(j)}, v_2^{(j)}$ so specified, and introducing the
further scaling $A_j^{\#}(x) = L^{-j} B_j^{\#}(x)$ we see that the
Laguerre limit of (\ref{mr}) reads
\begin{equation}\label{Bj}
B_j^{\#}(x) = (x - v_2^{(j)}) B_{j-1}^{\#}(x) - x v_1^{(j)}
B_{j-2}^{\#}(x).
\end{equation}
This recurrence is to be solved subject to the initial conditions
$B_{-1}^{\#}(x) = 0$ and $B_0^{\#}(x) = 1$.
The zeros of $B_n^{\#}(x)$ are then distributed according to
$L_{\rm o}^{(n,\_)}(x) |_{ \alpha = \alpha_0}$.

To sample from $L_{\rm o}^{(n,n-1)}(x,y)$ we take an appropriate
Laguerre limit of the procedure to sample from
$J_{\rm o}^{(n,n-1)}(x,y)$ detailed below  (\ref{mr}). Thus we use
(\ref{Bj}) to first compute $\{ \tilde{B}_j^{\#}(x) \}_{j=0,\dots,
n-1}$ where $\tilde{B}_j^{\#}(x)$ refers to ${B}_j^{\#}(x)$ with
parameters $\alpha_0 \mapsto \alpha_0 + \alpha_1$,
$1/b \mapsto 1/b + 1/b_1$. We then form the random polynomial
\begin{equation}
B_n(x) = (x - v_2^{(n)}) B_{n-1}^{\#}(x) - x v_1^{(n)}
B_{n-2}^{\#}(x) 
\end{equation}
with $(v_1^{(n)}, v_2^{(n)})$ distributed according to
$(\Gamma[1/b;(n-1)k], \Gamma[1/b;\alpha_0+1])$. The zeros of
$(B_n(x), B_{n-1}^{\#}(y))$ then have the joint PDF
$L_{\rm o}^{(n,n-1)}(x,y)$.

A recurrence to sample from $L_{\rm o}^{(n,\_)}(x) |_{ \alpha = a -k(n-1)-1
\atop b=2}$ has been given by Dumitriu and Edelman \cite{DE02}, as a
corollary of their construction of a random tridiagonal matrix with
this eigenvalue PDF. 
Denote by $\chi_p^2$ the gamma distribution $\Gamma[2;p/2]$,
and let $\chi_p$ denote the square root of a random variable with
distribution $\chi_p^2$. It was shown in \cite{DE02} that
the symmetric $n \times n$ random tridiagonal matrix
\begin{equation}\label{tri}
\left [ \begin{array}{ccccc} a_n & b_{n-1} & & & \\
b_{n-1} & a_{n-1} & b_{n-2} & & \\
 & b_{n-2} & a_{n-2} & b_{n-3} & \\
& \ddots & \ddots & \ddots & \\
&&b_{2} & a_2 & b_{1} \\
& & & b_{1} & a_1 \end{array} \right ],
\end{equation}
with the distribution of the elements specified by
$$
a_n \sim \chi_{2a}^2, \quad
a_i \sim \chi_{ki}^2 + \chi_{2a-k(n-i)}^2, \quad
b_i \sim \chi_{ki} \chi_{2a - k(n-i-1)} \: \: (i=n-1,\dots,1)
$$ 
has eigenvalue PDF given by
$L_{\rm o}^{(n,\_)}(x) |_{ \alpha = a -k(n-1)-1
\atop b=2}$. The random recurrence  now follows from the fact that
in general the characteristic polynomial of the bottom right
$k \times k$ submatrix of (\ref{tri}) satisfies the three term
recurrence
\begin{equation}\label{Bj1}
P_j(x) = (x-a_j) P_{j-1}(x) - b_{j-1}^2 P_{j-2}(x),
\end{equation}
subject to the initial conditions $P_{-1}(x) = 0$, $P_0(x)=1$.
Note that (\ref{Bj1}) differs from (\ref{Bj}).

The distribution $J_{\rm e}^{(n,n)}(x,y)$ as given by (\ref{J.2e}) also
has a well defined Laguerre limit, obtained by writing
$x_i \mapsto (1 - x_{n+1-i}/L)$, $y_i \mapsto (1 - y_{n+1-i}/L)$,
setting $\alpha = L/b$, $\alpha_1 = L/b_1$, $\beta_1 = \alpha_1$
and taking $L \to \infty$. This gives
\begin{eqnarray}
\lefteqn{L_{\rm e}^{(n,n)}(x,y) = {1 \over \Gamma(1+\alpha)
(\Gamma(k))^n} {1 \over \widetilde{W}_n(\alpha_1,k,(b+b_1)/bb_1)}}
\nonumber \\
&& \times \prod_{i=1}^n e^{-x_i/b} y_i^{\alpha_1} e^{-y_i/b_1}
\prod_{1 \le i < j \le n} |x_j - x_i | |y_j - y_i|
\prod_{i,j=1}^n | x_j - y_i |^{k-1}
\end{eqnarray}
where the $x$'s and $y$'s are interlaced according to
\begin{equation}\label{iux}
x_1 > y_1 > x_2 > y_2 > \cdots > x_n > y_n > 0.
\end{equation}
We make note too of the Laguerre limit of (\ref{4.10ae'}),
\begin{equation}\label{iax1}
 \int_R dy_1 \cdots dy_n \, L_{\rm e}^{(n,n)}(x,y)
\Big |_{1/b_1 = 0} = {1 \over \widetilde{W}_n(\alpha_1 + k, k,
b)} \prod_{i=1}^n x_i^{\alpha_1+k} e^{-x_i/b}
\prod_{1 \le i < j \le n} |x_i - x_j|^{2k} =:
L_{\rm e}^{(n,\_)}(x)
\end{equation}
where $R$ denotes the region (\ref{iux}).

To sample from $L_{\rm e}^{(n,n)}(x,y)$, we note that the Laguerre limit of
$\{A_j^{\#\rm e}(x) \}_{j=0,\dots,n}$, used in (\ref{vaw}) to sample
from $J_{\rm e}^{(n,n)}(x,y)$, is given by 
$\{B_j^{\# \rm e}(x)\}_{j=0,\dots,n}$ where the $B_j^{\#\rm e}(x)$ are
specified by the recurrence (\ref{Bj}) but with $1/b \mapsto 1/b + 1/b_1$,
$\alpha_0 \mapsto \alpha_1$ in the specification of the distribution
of $(v_1^{(j)}, v_2^{(j)})$. Then taking the Laguerre limit of
(\ref{vaw}) tells us if we define
\begin{equation}\label{iax2}
U_n(x) = u_2 B_n^{\#\rm e}(x) + u_1 B_{n-1}^{\#\rm e}(x), \qquad
u_1 + u_2 = 1
\end{equation}
where $u_1$ is distributed according to $\Gamma[b;nk]$, then the joint
distribution of the zeros $\{U_n(x), B_n^{\# \rm e}(y) \}$ is given by
$L_{\rm e}^{(n,n)}(x,y)$. We remark that according to (\ref{iax1})
the marginal distribution of $U_n(x)$ in the case $1/b_1 = 0$ is given
by $L_{\rm e}^{(n,\_)}(x,y)$. But $B_n^{\#\rm e}(x)$ with $1/b_1=0$,
$\alpha_0 \mapsto \alpha_1$ is the same as $B_n^{\#}(x)$ with
$\alpha_0 \mapsto \alpha_1$ and thus has the PDF for its zeros given
by (\ref{5.14}) with $\alpha = \alpha_1$, and this in turn is
identical to $L_{\rm e}^{(n,\_)}(x) |_{\alpha_1  \mapsto \alpha_1 -k}$.
Thus we have the Laguerre limit of (\ref{mrr}),
$$
B_n^{\#}(x) = \Big ( u_2 B_n^{\#}(x) + u_1 x B_{n-1}^{\#}(x)
\Big ) \Big |_{\alpha_1 \mapsto \alpha_1 - k}.
$$

\section{Gaussian interpolating ensembles}
\setcounter{equation}{0}
The random rational functions (\ref{Rl}) and (\ref{4.6l}) have a counterpart
\begin{equation}\label{RG}
R^{\rm G}(\lambda) := \lambda + \sum_{j=1}^n {w_j \over a_j - \lambda},
\end{equation}
where $w_j$ is distributed according to $\Gamma[1;s_j]$, which is
closely related to the Gaussian ensembles. Thus implicit in the work of
Evans \cite{Ev94} on Gaussian analogues of the Selberg integral according
to the method of Anderson is the following result for the PDF of the
zeros of $R^G(\lambda)$.

\begin{prop}\label{punsun}
Consider the random rational function (\ref{RG}). This has $n+1$ zeros
$\lambda_1,\dots,\lambda_{n+1}$ restricted by the interlacing condition
\begin{equation}\label{RG1}
\lambda_1 > a_1 > \lambda_2 > a_2 > \cdots > a_n > \lambda_{n+1}
\end{equation}
and the further requirement that
\begin{equation}\label{RG2}
\sum_{l=1}^{n+1} \lambda_l = \sum_{l=1}^n a_l.
\end{equation}
Subject to (\ref{RG1}) and (\ref{RG2}) the PDF of the zeros of (\ref{RG})
is given by
\begin{equation}\label{RG3}
{1 \over \Gamma(s_1) \cdots \Gamma(s_n)}
{\prod_{1 \le j < k \le n+1} (\lambda_j - \lambda_k) \over
\prod_{1 \le j < k \le n} (a_j - a_k)^{s_j+s_k-1}}
\prod_{j=1}^{n+1} \prod_{p=1}^n | \lambda_j - a_p |^{s_p-1}
\exp \Big ( - {1 \over 2} \Big ( \sum_{j=1}^{n+1} \lambda_j^2 -
\sum_{j=1}^n a_j^2 \Big ) \Big ).
\end{equation}
\end{prop}

Also of interest is the random rational function
\begin{equation}\label{RGt}
\tilde{R}^{\rm G}(\lambda) := \lambda - w_0
+ \sum_{j=1}^n {w_j \over a_j - \lambda},
\end{equation}
where the $w_j$ $(j=1,\dots,n)$ are distributed as in (\ref{RG}) while
$w_0$ is distributed according to N[0,1]. The PDF for the zeros of
(\ref{RGt}) is readily deduced from Proposition \ref{punsun}.

\begin{cor}
The zeros of the random rational function (\ref{RGt}) have PDF
(\ref{RG3}) multiplied by $1/\sqrt{2 \pi}$, except that the condition
(\ref{RG2}) is no longer required.
\end{cor}

\noindent
Proof. \quad The rational function $\tilde{R}^{\rm G}(\lambda)$ results
from $\tilde{R}(\lambda)$ by making the replacements $\lambda \mapsto
\lambda - w_0$, $a_j \mapsto a_j - w_0$. We then have that
$$
\exp \Big ( - {1 \over 2} \Big ( \sum_{j=1}^{n+1} \lambda_j^2 -
\sum_{j=1}^n a_j^2 \Big ) \Big ) \delta \Big (
\sum_{j=1}^{n+1} \lambda_j -
\sum_{j=1}^n a_j \Big ) \mapsto
e^{w_0^2/2}
\exp \Big ( - {1 \over 2} \Big ( \sum_{j=1}^{n+1} \lambda_j^2 -
\sum_{j=1}^n a_j^2 \Big ) \Big ) \delta \Big (
\sum_{j=1}^{n+1} \lambda_j -
\sum_{j=1}^n a_j - w_0 \Big ).
$$
Multiplying this by ${1 \over \sqrt{2 \pi}} e^{-w_0^2/2}$ (the
distribution of $w_0$) and integrating over $w_0$ eliminates the
delta function (and thus the restriction (\ref{RG2})) but leaves
all other terms unchanged. \hfill $\square$ 

We remark that the random rational function (\ref{RGt}) can be
derived as a limit of the random rational function (\ref{R2}) with
$n \mapsto n+1$ and $(w_0,\dots, w_n;w_{n+1})$ distributed according
to $D_{n+2}[\alpha/2,s_1,\dots,s_n;\alpha/2]$. Thus if we write
$\alpha = L^2$ and take $L \to \infty$ then the marginal
distribution of $w_0$ and $w_{n+1}$ have the asymptotic form
${1 \over 2} + {1 \over 2L} {\rm N}[0,1]$ while the $w_i$
$(i=1,\dots,n)$ have to leading order the marginal distribution ${1 \over L^2}
\Gamma[1;s_i]$. It then follows from (\ref{R2}) with
$x \mapsto {1 \over 2}(1 - {\lambda \over L})$ and
$y_i \mapsto {1 \over 2}(1 - {a_i \over L})$ that we have
\begin{equation}\label{Lr}
{L \over 2} \tilde{R}_{n+2}(x) \: \mathop{\sim}\limits_{L \to \infty}
\tilde{R}_n^{\rm G}(\lambda).
\end{equation}

The random rational functions (\ref{RG}) and (\ref{RGt}) occur in two
closely related eigenvalue problems (see e.g.~\cite{BGS01}). Thus let
$A$ be a real symmetric (complex Hermitian) matrix with
eigenvalues $a_1 > a_2 > \cdots > a_n$ of multiplicities
$m_1,\dots, m_n$. From $A$ form a random matrix $M$ of one extra
column and one extra row by bordering $A$ by a constant $\sqrt{b}$
times a vector of independent real standard Gaussians (complex Gaussians)
as the final column, and the Hermitian conjugate of this as the final
row (therefore in both the real ($\beta = 1$)
and complex case ($\beta = 2$) we require the final
entry of the vector to be real; let it have distribution 
N[0,$\sqrt{2/\beta}$]).
Thus if the final column of $A$ is number $n^*$, then
\begin{equation}\label{pers}
[M]_{i,j} = A, \: \: [M]_{i,n^*+1} = [M^*]_{n^*+1,i} = \sqrt{b}
[\vec{x}]_i, \: \: (1 \le i,j \le n^*) \quad 
[M]_{n^*+1,n^*+1} \sim N[0,\sqrt{2b/\beta}],
\end{equation}
where here the symbol $\sim$ denotes `has distribution'. A straight forward
calculation shows that $M$ has eigenvalues $a_i$ with multiplicities
$m_i-1$, and $n+1$ further eigenvalues given by the zeros of the
rational function (\ref{RGt}) with $w_0$ ($w_j$) distributed according
to N$[0,\sqrt{2b/\beta}]$ ($\Gamma[2b/\beta,\beta m_j/2]$). It follows
by scaling (\ref{RGt}) that if we choose $c=2b$ then the eigenvalue
PDF of $M$ is given by (\ref{RG3}) with $\lambda_j \mapsto
\sqrt{\beta/2b} \lambda_j$, $a_j \mapsto \sqrt{\beta/2b} a_j$.

If in the prescription (\ref{pers}) we choose $[M]_{n^*+1,n^*+1}=0$ we
find that $M$ has eigenvalues $a_i$ with multiplicities
$m_i-1$, and $n+1$ further eigenvalues given by the zeros of the
random rational function (\ref{RG}) with $w_i$ as specified in the
above paragraph. Note that the condition (\ref{RG2}) then has the
interpretation as the statement that Tr$(A) = {\rm Tr}(M)$.

\subsection{Construction of Gaussian interpolating matrix ensembles}
Following a strategy analogous to that used in the construction of
random matrices realizing the Jacobi and Laguerre interpolating 
ensembles, we can use the eigenvalue problem relating to (\ref{RGt})
to construct random matrices with eigenvalue PDFs realizing certain
Gaussian interpolating ensembles. In particular we can construct random
matrices with eigenvalue PDF of the form
\begin{equation}\label{g1g}
{1 \over C} \prod_{i=1}^{n+1} e^{-c_1 x_i^2/2}
\prod_{1 \le i < j \le n+1} (x_i - x_j)
\prod_{i=1}^n e^{-c_2 y_i^2/2} \prod_{1 \le i < j \le n}
(y_i - y_j),
\end{equation}
where
\begin{equation}\label{g1g'}
x_1 > y_1 > \cdots > y_n > x_{n+1},
\end{equation}
which with $c_2=0$ reduces to (\ref{bary}), and the eigenvalue PDF
\begin{equation}\label{g2g}
{1 \over C} \prod_{i=1}^{n+1} e^{-c_1 x_{2i-1}^2/2}
\prod_{i=1}^n e^{-c_2 x_{2i}^2/2}
\prod_{1 \le i < j \le 2n+1} (x_i - x_j)
\end{equation}
where
\begin{equation}\label{g2g'}
x_1 > x_2 > \cdots > x_{2n+1}.
\end{equation}

To obtain (\ref{g1g}) we choose the matrix $A$ in (\ref{pers}) to be
an $n \times n$ member of the GUE (see e.g.~\cite{Fo02} for the
precise definition of such matrices), and we extend $A$ so specified to
a $2n \times 2n$ real matrix by replacing each complex element by its
$2 \times 2$ real matrix representation (\ref{rms}). Following the
strategies of the proofs of Theorems \ref{th2a} and \ref{th2b} it
follows that $M$ as specified by (\ref{pers}) with $n^* = 2n$,
$\beta = 1$ has the $n$ eigenvalues of $A$ with multiplicity 1,
$y_1,\dots,y_n$ say, and a further $n+1$ eigenvalues $x_1,\dots,x_{n+1}$
say interlaced according to (\ref{g1g'}). Furthermore it follows that
the joint eigenvalue PDF is given by (\ref{g1g}) with
\begin{equation}
c_1 = {1 \over 2b}, \qquad c_2 = - {1 \over 2b} + 2.
\end{equation}

The above construction gives
\begin{equation}
{\rm even}(M) = {\rm GUE}_n.
\end{equation}
Now with $b=1/2$ we have $c_1 = c_2 = 1$ and we recognize (\ref{g1g}) as
the eigenvalue PDF for ${\rm GOE}_{n+1}
 \cup {\rm GOE}_n$ (see e.g.~\cite{FR01}).
Thus we have a matrix theoretic understanding of the identity
\cite{FR01}
\begin{equation}
{\rm even}({\rm GOE}_{n+1} \cup {\rm GOE}_n) = {\rm GUE}_n.
\end{equation}
We note too that with $b=1/4$ the matrix $M$ coincides with the upper
left $(2n+1) \times (2n+1)$ block of the real matrix representation of a
$(n+1) \times (n+1)$ GUE matrix. By an argument analogous to the
derivation of (\ref{lam}) we must therefore have 
\begin{equation}
{\rm odd}(M |_{b = 1/4}) = {\rm GUE}_{n+1}.
\end{equation}
As $b=1/4$ corresponds to $c_2=0$, this identity is relevant to
(\ref{bary}).

For the realization of (\ref{g2g}) we choose the matrix $A$ in
(\ref{pers}) to be a $n \times n$ member of the GSE (by definition
--- see e.g.~\cite{FR02} --- the elements of such matrices are real
quaternions, so as a complex matrix $A$ is $2n \times 2n$). The
eigenvalues are doubly degenerate, with the independent eigenvalues
$y_1,\dots,y_n$ say having distribution
$$
{1 \over C} \prod_{l=1}^n e^{-y_l^2} \prod_{1 \le j < k \le n}
(y_j - y_k)^4.
$$
Here we follow the strategy of the proofs of Theorems \ref{th1a} and
\ref{th1} to conclude that (\ref{pers}) with this choice of $A$ and
$n^* = 2n$, $\beta = 2$ has the $n$ eigenvalues of $A$ and
with multiplicity 1, $x_2,x_4,\dots,x_{2n}$ say, and a further $n+1$
eigenvalues $x_1,x_3,\dots,x_{2n+1}$ say, interlaced according to
(\ref{g2g'}) and with eigenvalue PDF (\ref{g2g}) with
$$
c_1 = {1 \over b}, \qquad c_2 = -{1 \over b} + 2.
$$
Since by construction
$$
{\rm even}(M) = {\rm GSE}_n,
$$
and with $b=1$ and thus $c_1=c_2=1$ the PDF (\ref{g2g}) reduces to the
PDF for GOE${}_{2n+1}$, we thus have a matrix theoretic understanding of
the relation \cite{FR02}
$$
{\rm even}({\rm GOE}_{2n+1}) = {\rm GSE}_n.
$$
Furthermore, with $b=1/2$ the matrix $M$ coincides with the upper left
$(2n+1) \times (2n+1)$ block of the complex representation of a
$(n+1) \times (n+1)$ GSE matrix, and so we must have
$$
{\rm odd}(M|_{b=1/2}) = {\rm GSE}_{n+1}.
$$

\subsection{Gaussian limit of the three term recurrences}
The PDFs (\ref{g1g}) and (\ref{g2g}) are special cases of a
limiting form of the joint PDF (\ref{J.2}). Thus in (\ref{J.2})
let us change variables $x_i \mapsto ( {1 \over 2} - {x_i \over
2L} )$, $y_i \mapsto ( {1 \over 2} - {y_i \over 2L})$, set
$\alpha = \beta = aL^2$, $\alpha_1 = \beta_1 = a_1 L^2$ and take
$L \to \infty$. We then obtain the joint PDF
\begin{eqnarray}
\lefteqn{
G_{\rm o}^{(n,n-1)}(x,y) := \Big ( {a \over \pi} \Big )^{1/2}
{(2a)^{(n-1)k} \over (\Gamma(k))^{n-1} }
{1 \over M_{n-1}(k;2(a+a_1))} } \nonumber \\&&
\times \prod_{i=1}^n e^{-a x_i^2}
\prod_{1 \le i < j \le n} |x_j - x_i|
\prod_{i=1}^{n-1} e^{-a_1 y_i^2} \prod_{1 \le i < j \le n-1} |y_j - y_i|
\prod_{i=1}^n \prod_{j=1}^{n-1} |x_j - y_i|^{k-1}
\end{eqnarray}
where
\begin{eqnarray*}
M_n(k;c)  & := & \int_{-\infty}^\infty dx_1 \cdots \int_{-\infty}^\infty dx_n
\, e^{-(c/2) \sum_{l=1}^n x_l^2}
\prod_{1 \le i < j \le n} |x_j - x_i|^{2k} \nonumber \\
& = & c^{-n/2 - kn(n-1)/2} (2 \pi)^{n/2}
\prod_{j=0}^{n-1} {\Gamma(1+(j+1)k) \over \Gamma(1+k)}
\end{eqnarray*}
and the $x$'s and $y$'s are interlaced according to
\begin{equation}\label{6.16'}
\infty > x_1 > y_1 > x_2  > y_2 > \cdots > y_{n-1} > x_n > - \infty.
\end{equation} 
We make note of the special marginal distribution
\begin{equation}\label{6.16a}
\int_R dy_1 \cdots dy_{n-1} \, G_{\rm o}^{(n,n-1)}(x,y)
\Big |_{a_1 = 0} = {1 \over M_n(k;2a)}
\prod_{i=1}^n e^{-2ax_i^2} \prod_{1 \le i < j \le n}
|x_i - x_j|^{2k} =: G_{\rm o}^{(n,\_)}(x),
\end{equation}
which is equivalent to the integration formulas (\ref{4.10b})
and (\ref{5.14}).

To sample from $G_{\rm o}^{(n,\_)}(x)$ we can take the Gaussian limit
of the three term recurrence (\ref{mr}). First we note that with
$(w_0^{(j)},w_1^{(j)},w_2^{(j)} )$ distributed as specified below
(\ref{mr}), setting $\alpha_0 = \beta_0 = aL^2$ and taking $L \to
\infty$, the marginal distributions of $w_0^{(j)}$ and $w_2^{(j)}$ have
the asymptotic form ${1 \over 2} + {1 \over 2L} {\rm N}[0,{1 \over \sqrt{2a}}]$
while $w_1^{(j)}$ has the leading order marginal  distribution
${1 \over L^2} \Gamma[{1 \over 2a};(j-1)k]$ (c.f.~the statements
above (\ref{Lr})). Thus by also writing $x \mapsto {1 \over 2}(1 -
{x \over L} )$, $A_j^{\#}(x) \mapsto (-2L)^{-j} C_j^{\#}(x)$, we
see that in the Gaussian limit (\ref{mr}) reduces to
\begin{equation}\label{mrg}
C_j^{\#}(x) = (x - r) C_{j-1}^{\#}(x) - s^{(j-1)} C_{j-2}^{\#}(x)
\end{equation}
where $r$ has distribution ${\rm N}[0,{1 \over \sqrt{2a}}]$ while
$s^{(j-1)}$ has distribution $ \Gamma[{1 \over 2a};(j-1)k]$. With the
initial conditions $C_{-1}^{\#}(x)=0$, $C_0^{\#}(x)=1$, we have
that the zeros of $C_n^{\#}(x)$ have PDF $G_{\rm o}^{(n,\_)}(x)$.
The recurrence (\ref{mrg}) has the structure (\ref{Bj1}) and thus
can be viewed as specifying the characteristic polynomial for a
corresponding random tridiagonal matrix (\ref{tri}). In fact this
is precisely the random tridiagonal matrix found by
Dumitriu and Edelman \cite{DE02} and shown to have eigenvalue PDF given
by $G_{\rm o}^{(n,\_)}(x)$. 

To sample from $G_{\rm o}^{(n,n-1)}(x,y)$, the Gaussian limit of the
procedure to sample from $J_{\rm o}^{(n,n-1)}(x,y)$ detailed below
(\ref{mr}). Thus we use (\ref{mrg}) to generate $\{\tilde{C}_j^{\#}(x)
\}_{j=0,\dots,n-1}$ where $\tilde{C}_j^{\#}(x)$ refers to
${C}_j^{\#}(x)$ with parameter $a \mapsto a+a_1$. We then form
the random polynomial
$$
C_n(x) = (x-r) \tilde{C}_{n-1}^{\#}(x) - s^{(n-1)} 
\tilde{C}_{n-2}^{\#}(x)
$$
with $(r,s^{(n-1)})$ distributed according to
$({\rm N}[0, {1 \over \sqrt{2a}}], \Gamma[{1 \over 2a};
(n-1)k])$.  The PDF $G_{\rm o}^{(n,n-1)}(x,y)$ is then realized by
the zeros of $(C_n(x), \tilde{C}_{n-1}^{\#}(y))$.
Equivalently we can realize the PDF in terms of the eigenvalues
of a random $n \times n$ tridiagonal matrix and its lower right
$(n-1) \times (n-1)$ submatrix.

\begin{thm}
Consider the symmetric tridiagonal matrix (\ref{tri}). Let the
elements be random with distributions
\begin{eqnarray*}
&&a_i \sim {\rm N} [ 0, {1 \over \sqrt{2(a+a_1)}}], \quad
b_{i-1}^2  \sim \Gamma [{1 \over 2(a+a_1)}; (i-1)k] \quad (i=1,\dots,n-1)\\
&&a_n  \sim {\rm N} [ 0, {1 \over \sqrt{2a}}], \quad
b_{n-1}^2  \sim \Gamma [{1 \over 2a};(n-1)k].
\end{eqnarray*}
The joint distribution of the eigenvalues $x_1,\dots,x_n$ 
of this matrix, and the eigenvalues $y_1,\dots,y_{n-1}$  of the
$(n-1) \times (n-1)$ bottom right submatrix, is given by
$G_{\rm o}^{(n,n-1)}(x,y)$.
\end{thm}

\section*{Appendix A}
\subsection*{Geometrical RSK}
\renewcommand{\theequation}{A.\arabic{equation}}
\setcounter{equation}{0}
The Robinson-Schensted-Knuth correspondence between non-negative
integer matrices and pairs of semi-standard tableaux of the same
shape has a geometrical representation relating the matrix to
paths defining the interface of a sequence of growth models
\cite{Jo02} (for closely related geometrical representations
see \cite{Fu97} and \cite{Fom95}). Here will will show that
coordinates specifying the paths can be used to deduce the
equation (\ref{6.5}) and also the joint probability (\ref{2.19}).

\begin{figure}[th]
\epsfxsize=10cm
\centerline{\epsfbox{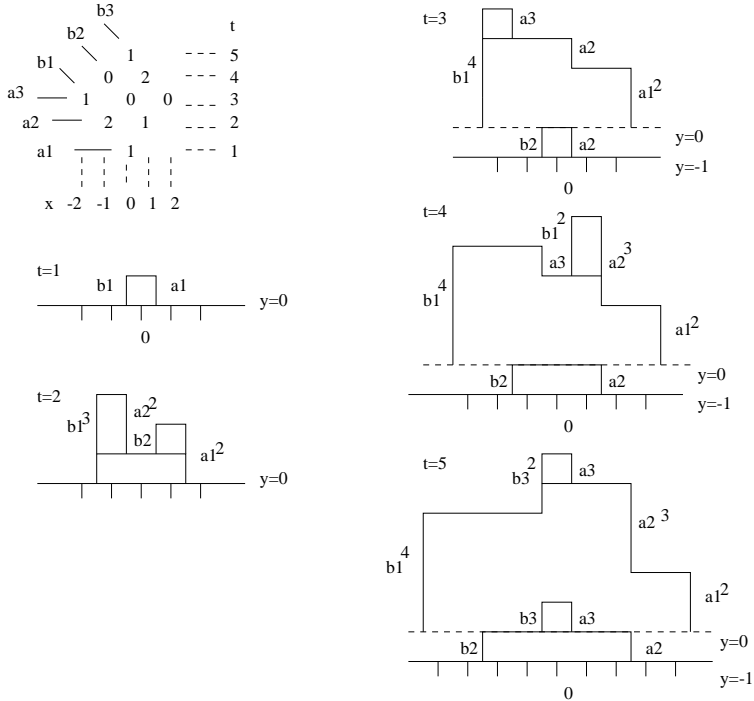}}
\caption{\label{f1} The RSK mapping from a weighted integer matrix to
a set of weighted non-intersecting paths. The mapping is invertible and so
is a bijection.}
\end{figure}

First we revise the construction of \cite{Jo02}, wherein  each distinct
$n \times n$ non-negative square matrix $X = [x_{i,j}]_{i,j=1,\dots,
n}$ (rows counted from the bottom), with entries
$x_{i,j}$  weighted 
$(1-a_i b_j)(a_i b_j)^{x_{i,j}}$, is put into a one-to-one
correspondence with a set of at most $n$ non-intersecting
weighted lattice paths, 
starting 
at $(x,y) = (-(2n-1/2), l-1)$ and finishing at
$(x,y) = ((2n-1/2), l-1)$ 
($l=1,2,\dots$). The $l$th member of the set --- the path 
starting and finishing along $y=-(l-1)$ ---
will be referred to as the
level-$l$ path. Each level-$l$ path can be regarded as a pair of paths
because the weights and the allowed steps are different depending
on $x<0$ or $x>0$. Thus the first (second) member of the pair
starts at $x=-(2n-1/2)$ $(x=(2n-1/2))$ and goes 
either right (left) in steps of two units,
or  up in
integer amounts at $x=-(2n+3/2-2j)$
($x=(2n+3/2-2j)$
with each unit regarded as a step weighted by
$b_j$ ($a_j$), until it reaches $x=-1/2$ ($x=1/2$) where both
paths must have the same final $y$-coordinate (see Figure \ref{f1}).

Let $\mu_l$ denote the maximum height of the level-$l$ path,
which is the  displacement of this path at $x = \pm 1/2$.
Because the paths cannot intersect we must have
$\mu_1 \ge \mu_2 \ge \cdots \ge \mu_n$ so
$\mu = (\mu_1,\mu_2,\dots, \mu_n)$ forms a partition. It is
a standard result (see e.g.~\cite{Sa01}) that the total weight
of all non-intersecting paths of the specified type
for $x<0$, initially
equally spaced at $y=0,\dots,-(n-1)$ along
$x=-(2n-1/2)$ and finishing at $y=\mu_1, \mu_2 - 1,
\dots, \mu_n - (n-1)$ along $x=-1/2$ is given by the Schur
polynomial $s_\mu(b_1,\dots,b_n)$. Similarly the total weight
of all non-intersecting paths of the specified type for $x>0$,
initially equally spaced at $y=0,\dots,-(n-1)$ along
$x = (2n-1/2)$ and finishing at $y=\mu_1, \mu_2 - 1,
\dots, \mu_n - (n-1)$ along $x=1/2$ is given by the Schur polynomial
$s_\mu(a_1,\dots,a_n)$. Furthermore it is another
standard result (see e.g.~\cite{Sa01}) that each set of non-intersecting
paths from $x=-(2n-1/2)$ to $x=-1/2$ is equivalent to a
semi-standard tableau of shape $\mu$, content $n$, as each set 
of paths from $x = (2n-1/2)$ to $x=1/2$. Thus for the
 non-intersecting lattice paths of Figure \ref{f1} we obtain
the pair of tableaux 

\begin{figure}[th]
\epsfxsize=10cm
\centerline{\epsfbox{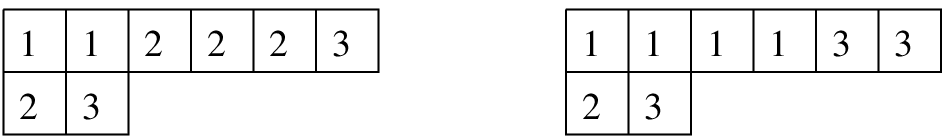}}
\end{figure}

\noindent
the first corresponding  to the paths $x>0$ and the second the paths $x<0$.
In
the former (latter) each occurence of $a_j$ ($b_j$) in the level-$l$
path is recorded as a box labelled $j$ in the $l$th row of the
tableau.

 Thus if we
accept the mapping, the probability (\ref{6.3}) is
immediate (an unequal number of $a$ and $b$ weights can be
achieved by simply setting some of them equal to zero
and using the stability property of the Schur polynomial,
$s_\mu(a_1,\dots,a_{n-1},0) = s_\mu(a_1,\dots,a_{n-1})$).

To derive (\ref{2.19}) and (\ref{6.5}) we must investigate
the details of the mapping, which takes the form of a cascade
of polynuclear growth models (see Figure \ref{f1}). We rotate the non-negative
matrix $X = [x_{i,j}]_{i,j=1,\dots,n}$
45${}^\circ$
anti-clockwise and label the horizontal rows of the rotated matrix by
$t=1,2,\dots,2n-1$ and the vertical columns by $x=0,\pm 1, \dots,
\pm (n-1)$ where $x=0$ corresponds to the diagonal $i=j$ of $X$
(recall that the rows are being counted from the bottom).
The entries $x_{i,j}$ in the matrix
for successive $t$ values $(t=i+j-1)$ 
are heights of weighted `nucleation
events' --- columns of unit width and height $x_{i,j}$ centred
about the corresponding $x$-coordinate which are placed on top of the
profile formed by earlier nucleation events and their growth
and weighted by $(a_i b_j)^{x_{i,j}}$ (in addition the matrix has
a normalization weighting of $\prod_{i,j=1}^n (1 - a_i b_j)$
independent of the entries).
Thus at $t=1$ there is a nucleation event at $x=0$ which consists of
a column of width 1, height $x_{11}$ and weight
$(a_1 b_1)^{x_{11}}$ marked on the line at $y=0$
in the $xy$-plane.
In general, as $t \mapsto t+1$ the profile of all
nucleation events so far recorded is to `grow' one unit in the
$-x$ direction and one unit in the $+x$ direction. Thus in going
from $t=1$ to $t=2$ the nucleation event centred at $x=0$ of
height $x_{11}$ now has width 3 units. On top of this profile,
centred at $x=-1$ and $x=1$ nucleation events of unit width
and height $x_{21}, x_{12}$ and weight $(a_2 b_1)^{x_{21}}$,
$(a_1 b_2)^{x_{12}}$ respectively are then drawn.
In now
going from $t=2$ to $t=3$ this new profile is to grow one unit
to the left and one unit to the right. In so doing we see that
an overlap of width one unit and height $\min (x_{21}, x_{12})$
will occur. This overlap is ignored in the first diagram
(profile on $y=0$), and recorded instead as a profile on the line
immediately below (here $y=-1$). The process is repeated with
these rules until the nucleation event of height $x_{nn}$,
weight $(a_n b_n)^{x_{nn}}$ at
$t=2n-1$ has been recorded above $x=0$ on the first diagram.
In this way we obtain the sought mapping from a weighted
non-negative integer matrix to weighted non-intersecting
lattice paths. The mapping is easily seen to be
invertible, and so is a bijection. With each path
considered as a pair of paths depending on whether $x<0$ or $x>0$,
and then the set of paths for $x<0$ and the set of paths for
$x>0$ recorded
as a pair of semi-standard tableaux, this gives the same
correspondence between non-negative integer matrices and
pairs of semi-standard tableaux of the same shape as the
Robinson-Schensted-Knuth algorithm (this last point follows
because, as noted in \cite{Jo02},
 the above algorithm can be viewed as  a graphical
presentation of the matrix-ball construction of Fulton
\cite{Fu97}, which has been shown to give the RSK
correspondence).

\begin{figure}[th]
\epsfxsize=15cm
\centerline{\epsfbox{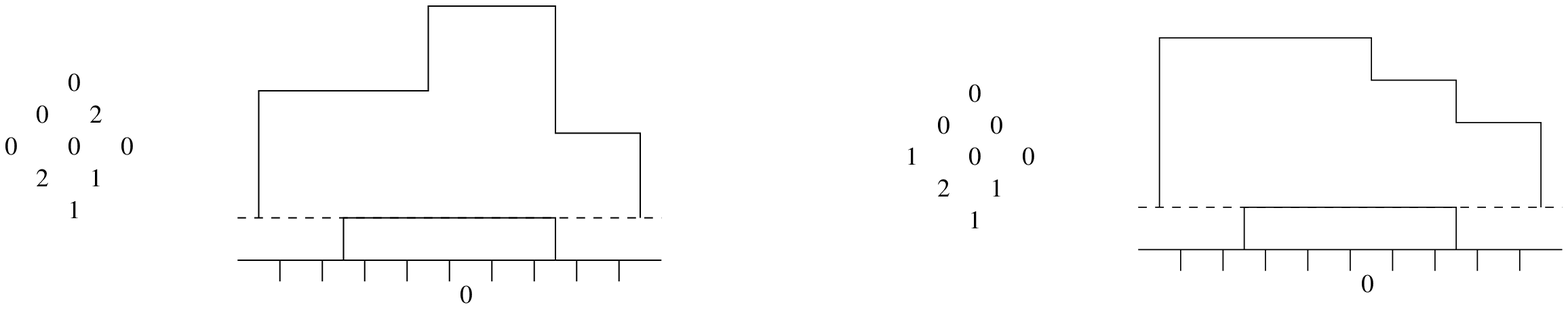}}
\caption{\label{f4} The path diagram for the matrix of Figure \ref{f1}
with the final row (column) set equal to zero. This can be deduced
from the $t=4$ diagram of Figure \ref{f1} with $a_3$ ($b_3$) set
equal to zero. The maximum heights therefore coincide with the
maximum height at $x=1/2$ $(x=-1/2)$ in this diagram.
}
\end{figure}

Of crucial interest to us is the sequence of maximum displacements
$\lambda_l(n_1,n_2)$ of the level-$l$ path obtained by applying
the growth process to the truncation $X_{n_1,n_2}$ say
of the matrix X to the
first $n_1$ rows and $n_2$ columns, extended to a square
matrix of dimension
$\max(n_1,n_2) \times \max(n_1,n_2)$ by appending rows of zeros 
to the top (or columns of zeros to the right, as appropriate).
Consider first the level-1 path.
It follows from the rules of the growth process that $\lambda_1(n_1,
n_2)$ results by adding $x_{n_1,n_2}$ to the maximum of the
height at $x=-1$ and the height at
$x=1$ in the previous time step ($h_1(n_1,n_2-1)$ and
$h_1(n_1-1,n_2)$ respectively say; see the change in the
height at the origin in going from $t=4$ to $t=5$ in
Figure \ref{f1}). Thus
$$
\lambda_1(n_1,n_2) = \max \Big ( h_1(n_1,n_2-1), h_1(n_1-1,n_2) \Big ) +
x_{n_1,n_2}.
$$
But again from the rules of the growth process
\begin{equation}\label{hl}
h_1(n_1,n_2-1) = \lambda_1(n_1,n_2-1), \qquad h_1(n_1-1,n_2) =
\lambda_1(n_1-1,n_2)
\end{equation}
since the nucleation events in column $n_2$, and rows
$1,2,\dots, n_1-1$  from the bottom cannot contribute to
$h_1(n_1,n_2-1)$, 
and similarly the nucleation events in row $n_1$ from the
bottom and columns $1,2,\dots,n_2-1$ cannot contribute to
$h_1(n_1-1,n_2)$
(see Figure \ref{f4}). Therefore we obtain the recurrence
\begin{equation}\label{r1}
\lambda_1(n_1,n_2) = \max \Big (\lambda_1(n_1,n_2-1),
\lambda_1(n_1-1,n_2) \Big ) + x_{n_1,n_2}
\end{equation}
which with the boundary condition
$$
\lambda_1(0,j) = \lambda_1(i,0) = 0
$$
uniquely specifies $\{\lambda_1(i,j)\}_{i,j=1,\dots,n}$.
One observes that $L(n_1,n_2)$ as specified by (\ref{6.2})
satisfies the very same recurrence (\ref{r1}), and indeed 
one of the primary relations in the RSK correspondence is
\begin{equation}\label{r2}
\lambda_1(n_1,n_2) = L(n_1,n_2).
\end{equation}

For the maximum displacements $\lambda_l(n_1,n_2)$ of the
level-$l$ path, $l > 1$, we see from the derivation of
(\ref{r1}) that
$$
\lambda_l(n_1,n_2) = \max \Big (\lambda_l(n_1,n_2-1),
\lambda_l(n_1-1,n_2) \Big ) + x_{n_1,n_2}^{(l-1)}
$$
where $x_{n_1,n_2}^{(l-1)}$ is the height of an overlap event
(if any) which occurs in the growth of the nucleation events
corresponding to $x_{n-1,n_2-1}$ or $x_{n_1-1,n_2}$. 
We remark that $[ x_{i,j}^{(l-1)}]_{i,j=1,\dots,n}$ defines the $l$th
member of the sequence of matrices in Fulton's matrix ball
construction \cite{Fu97}.
The rules
of the growth process give
\begin{eqnarray*}
x_{n_1,n_2}^{(l-1)} & = & \min \Big ( h_{l-1}(n_1,n_2-1),
h_{l-1}(n_1-1,n_2) \Big ) - h_{l-1}(n_1-1,n_2-1) \\
& = & \min \Big ( \lambda_{l-1}(n_1,n_2-1),
\lambda_{l-1}(n_1-1,n_2) \Big ) - \lambda_{l-1}(n_1-1,n_2-1)
\end{eqnarray*}
(see the nucleation events created in level 2 in going from
$t=2$ to $t=3$, and going from $t=4$ to $t=5$ in
Figure \ref{f1}) and so for $l>1$
\begin{eqnarray}\label{r3}
\lambda_l(n_1,n_2) & = &
\max \Big (\lambda_{l}(n_1,n_2-1),
\lambda_{l}(n_1-1,n_2) \Big )  \nonumber \\ &&
+
\min  \Big ( \lambda_{l-1}(n_1,n_2-1),
\lambda_{l-1}(n_1-1,n_2) \Big ) - \lambda_{l-1}(n_1-1,n_2-1),
\end{eqnarray}
which with the boundary condition
\begin{equation}\label{r4}
\lambda_l(0,j) = \lambda_l(i,0) = 0
\end{equation}
and knowledge of $\{\lambda_1(i,j)\}_{i,j=1,\dots,n}$ from
(\ref{r1})
uniquely specifies $\{\lambda_l(i,j)\}_{i,j=1,\dots,n}$. For recurrences
closely related to (\ref{r3}), see \cite{Ki00,NY02}.

We have defined $\lambda_l(n_1,n_2)$ as the maximum height of the
level-$l$ path resulting from applying the growth process to
the truncation $X_{n_1,n_2}$ of the original matrix $X$. 
But there is another equally important interpretation of
$\lambda_l(n_1,n_2)$. Thus consider $n_1,n_2$ as fixed and apply the
growth process to $X_{n_1,n_2}$. Then we can see, arguing as in the
justification of the equalities (\ref{hl}), 
that $\lambda_l(n_1,j)$ is equal to the displacement
of the level-$l$ path at $x=-2n^*-3/2+2j$, while $\lambda_l(i,n^*)$
is equal to the displacement of the level-$l$ path at
$x=2n^*+3/2-2i$, where $n^* = \max (n_1, n_2)$.
It follows immediately from
this interpretation  that (see Figure \ref{f1.5}) 
\begin{eqnarray}\label{2.44}
&& \lambda_l(i,j) = 0 \qquad {\rm for} \quad
l > \max (n_1, n_2) \nonumber \\
&&\lambda_l(i,j) \ge \lambda_l(i,j-1) \ge \lambda_{l+1}(i,j) \nonumber \\
&&\lambda_l(i,j) \ge \lambda_l(i-1,j) \ge \lambda_{l+1}(i,j),
\end{eqnarray}
and furthermore
\begin{eqnarray}
\sum_{l=1}^n\Big (\lambda_l(n,j) - \lambda_l(n,j-1) \Big ) & = &
\sum_{i=1}^n x_{ij} \nonumber \\
\sum_{l=1}^n\Big (\lambda_l(i,n) - \lambda_l(i-1,n) \Big ) & = &
\sum_{j=1}^n x_{ij}.
\end{eqnarray}

\begin{figure}[ht]
\begin{center}

\setlength{\unitlength}{0.00055555in}
\begingroup\makeatletter\ifx\SetFigFont\undefined%
\gdef\SetFigFont#1#2#3#4#5{%
  \reset@font\fontsize{#1}{#2pt}%
  \fontfamily{#3}\fontseries{#4}\fontshape{#5}%
  \selectfont}%
\fi\endgroup%
{
\begin{picture}(2655,1971)(0,-10)
\put(1200,1362){\blacken\ellipse{76}{76}}
\put(1200,1362){\ellipse{76}{76}}
\put(750,912){\blacken\ellipse{76}{76}}
\put(750,912){\ellipse{76}{76}}
\path(300,1812)(750,1812)(750,1362)
	(1200,1362)(1200,912)(1200,462)(1725,462)
\put(750,1812){\blacken\ellipse{76}{76}}
\put(750,1812){\ellipse{76}{76}}
\path(300,912)(750,912)(750,12)(1725,12)
\put(0,1020){\makebox(0,0)[lb]{\smash{{{\SetFigFont{12}{14.4}{\rmdefault}{\mddefault}{\updefault}{\small $\lambda_{l+1}^*(i,j)$}}}}}}
\put(900,1812){\makebox(0,0)[lb]{\smash{{{\SetFigFont{12}{14.4}{\rmdefault}{\mddefault}{\updefault}{\small $\lambda_l(i,j)$}}}}}}
\put(1425,1287){\makebox(0,0)[lb]{\smash{{{\SetFigFont{12}{14.4}{\rmdefault}{\mddefault}{\updefault}{\small $\lambda_l(i-1,j)$}}}}}}
\end{picture}
}
\end{center}
\caption{\label{f1.5} Graphical demonstration of the final inequality in
(\ref{2.44}), where here all marked displacements are with respect
to the line $y=-(l-1)$, and $\lambda_{l+1}^*(i,j) :=
\lambda_{l+1}(i,j) + 1$.} 

\end{figure}

We are now in a position to derive the joint probability (\ref{2.19}).
Let $X_{n_1,n_2+1}$ be an integer matrix mapping to 
a pair of semi-standard tableaux of shape
$\mu$ under the RSK correspondence. 
Now each part $\mu_l$ of $\mu$ is equal to the maximum
displacement $\mu_l$ of the level-$l$ path, and so from the
above discussion $\mu_l = \lambda_l(n_1,n_2+1)$. Similarly, with
$\kappa$ denoting the shape of the pair of tableaux resulting from
applying the RSK correspondence to the truncation $X_{n_1,n_2}$ of
$X_{n_1,n_2+1}$ obtained by deleting the rightmost column, the
above discussion shows $\kappa_l = \lambda_l(n_1,n_2)$. The relations
(\ref{2.44}) immediately give (\ref{2.11}) in the case $n_2 \ge n_1$,
and (\ref{2.11a}) for $n_2 < n_1$. Furthermore, from the geometrical
RSK mapping
illustrated in Figure \ref{f1}, the joint probability is seen to be equal
to $\prod_{i=1}^{n_1} \prod_{j=1}^{n_2+1}(1 - a_i b_j)$ (the normalization
weighting), times the total weight of all non-intersecting paths from
$(2n^* + 1/2, l-1)$ to $(1/2, l-1)$ $(l=1,\dots,n^*)$ with
$n^* = \max (n_1, n_2+1)$, weighting $a_j$ for each unit up step
at $x = 2n -3/2 + 2j$, $(j=1,\dots,n_1)$, times the total weight of
all non-intersecting paths from $(-2n^* - 1/2, l-1)$ to $(-3/2, l-1)$,
weighting $b_j$ for each unit up step at $x = 2n^* - 3/2 + 2j$
$(j=1,\dots,n_1)$, times the step weight $b_{n_2+1}$ raised to the power
of the difference in the maximum height of the level-$l$ path for
$x>0$ ($\mu_l$) and the level-$l$ path for $x<0$ ($\kappa_l$)
summed over $l$. Writing the total weights of the paths in terms of
Schur polynomials we see that (\ref{2.19}) results for $n_2 \ge n_1$, and
(\ref{2.19}) modified by (\ref{2.11a}) results for $n_2 < n_1$.

To derive (\ref{6.5}) we first note that in the special case
$X =[x_{i,j}]_{i,j=1,\dots,n}$ is symmetric about $i=j$, we must have
$\lambda_l(i,j) = \lambda_l(j,i)$. Hence it follows from (\ref{r1})
and (\ref{r3}) that then
\begin{eqnarray*}
\lambda_1(i,i) & = & \lambda_1(i,i-1) + x_{i,i} \\
\lambda_l(i,i) & = & \lambda_l(i,i-1) + \lambda_{l-1}(i,i-1) -
\lambda_{l-1}(i-1,i-1), \quad l > 1.
\end{eqnarray*}
Forming appropriate linear combinations of these equations shows
$$
\sum_{l=1}^i (-1)^{l-1} \lambda_l(i,i) -
\sum_{l=1}^{i-1} (-1)^{l-1}  \lambda_l(i-1,i-1) = x_{i,i},
$$
and summing this equation over $i$ from 1 to $n$ gives
(\ref{6.5}).

\subsection*{Continuous RSK}
In the above description of the RSK correspondence a weighted
non-negative integer matrix $X=[x_{ij}]_{i,j=1,\dots,n}$ has been
mapped bijectively to a set of non-intersecting weighted lattice paths.
Because for given maximum displacements $\mu_1,\dots,\mu_n$ the total
weight of the non-intersecting paths given by a product of Schur
polynomials of the same index $\mu$, this allows the probability that
the integer matrix matrix maps to such  lattice paths to be specified by
(\ref{6.3}).

Also of interest is the case when the matrix $X$ consists of non-negative
real valued random variables $x_{ij}$ distributed according to the
exponential distribution
\begin{equation}\label{dis}
{\rm Pr}(x_{ij} \in [y,y+dy]) = (\alpha_i + \beta_j)
e^{-(\alpha_i + \beta_j)y} dy, \quad y \ge 0.
\end{equation}
Using this distribution to define a probability measure on $X$, we see
(as noted in \cite{Jo02}) that the RSK correspondence gives a bijective
mapping to a set of non-intersecting paths with a certain probability
measure. The description of the lattice paths differs in some details
to the discrete case. First, their steps are continuous in the
$y$-direction and discrete in the $x$-direction. All paths start along
$y=0$, with the level-$l$ path  starting at $(x,y)=(-(2n+3/2-2l),0)$ and
finishing at $(x,y)=(2n+3/2-2l,0)$. Vertical steps can occur at
$x=-(2n+3/2-2j)$ ($x=2n+3/2-2j$), $j=l,\dots,n$, with the constraints
that the non-zero height of the level-$l$ path is greater than that of
the level-$(l+1)$ path, and the height of each level-$l$ path is
weakly increasing going from $x=-(2n-1/2)$ to $x=-1/2$ and weakly
decreasing going from $x=1/2$ to $x= 2n-1/2$.

We are interested in the probability density that the level-$l$
paths will have maximum displacement $y_l$, $l=1,\dots,n$. By the
choice of the exponential distribution (\ref{dis}), the RSK
correspondence shows that at $x=-(2n+3/2-2j)$ ($x=2n+3/2-2j$) the
vertical increment of each level-$l$ path with $l\le j$ is a random
variable proportional to $e^{-\beta_j y}$ ($e^{-\alpha_j y}$) --- the
normalization $\prod_{i,j=1}^n(\alpha_i+\beta_j)$ is taken as an
overall factor --- which is conditioned so that the paths for
levels $1,2,\dots,j$ do not intersect, and furthermore at level-$l$
the sum of the vertical increments for both $x>0$ and $x<0$ is
equal to $y_l$.

The total weight of a single path with vertical increments of length
$v_j$ at $x=-(2n+3/2-2j)$ ($j=l,\dots,n$), weighted by
$e^{-\beta_j v_j}$ and constrained so that $\sum_{j=l}^n v_j
= y_k$ is given by 
$$
\int_0^\infty d \delta_l \, e^{-\beta_l v_l} \cdots
\int_0^\infty d \delta_n \, e^{-\beta_n v_n} \,
\delta\Big (y_k - \sum_{j=l}^n v_j\Big ) =
\sum_{j=l}^n {e^{-\beta_j y_k} \over \prod_{\mu = l \atop \mu \ne j}^n
(\beta_j - \beta_\mu)} =: u_l(\{\beta_j \}_{j=l,\dots,n};y_k).
$$
The extension of the Karlin-McGregor theorem used in \cite{Jo02} gives
that the corresponding total weight of the set of continuous
non-intersecting paths for $x<0$ is
$\det[ u_l(\{\beta_j \}_{j=l,\dots,n};y_k)]_{k,l=1,\dots,n}$.
Similarly for $x>0$ the total weight of the set of continuous paths is
$\det[ u_l(\{\alpha_j \}_{j=l,\dots,n};y_k)]_{k,l=1,\dots,n}$. Thus
the sought probability density is
\begin{equation}\label{crsk}
\prod_{i,j=1}^n(\alpha_i + \beta_j) \,
\det[ u_l(\{\alpha_j \}_{j=l,\dots,n};y_k)]_{k,l=1,\dots,n}
\det[ u_l(\{\beta_j \}_{j=l,\dots,n};y_k)]_{k,l=1,\dots,n}.
\end{equation}
In the special case
$$
\alpha_i = a + (i-1), \qquad \beta_j = \bar{a} + (j-1)c
$$
we have the simplifications
$$
u_l(\{\alpha_j \}_{j=l,\dots,n};y_k) =
{e^{-(a + (l-1)) y_k} \over  (n-l)!}
(1 - e^{- y_k})^{n-l}, \quad
u_l(\{\beta_j \}_{j=l,\dots,n};y_k) =
{e^{-(\bar{a} + (l-1)c) y_k} \over c^{n-l} (n-l)!}
(1 - e^{-c y_k})^{n-l}
$$
and, after making use of the Vandermonde determinant identity,
we obtain the corresponding simplification of (\ref{crsk})
\begin{equation}\label{crsk1}
\prod_{j=1}^n {\Gamma(a+ \bar{a} + (j-1)c +n) \over
\Gamma(a+ \bar{a} + (j-1)c )c^{j-1} \Gamma^2(j)}
e^{-(a+\bar{a} ) \sum_{j=1}^n y_j}
\prod_{1 \le i < j \le n} (e^{-cy_j} - e^{-cy_i})(
e^{-y_j} - e^{-y_i}).
\end{equation}

\section*{Appendix B}
\subsection*{The summation identity (\ref{sps4})}
\renewcommand{\theequation}{B.\arabic{equation}}
\setcounter{equation}{0}
Substituting (\ref{pro}) in (\ref{sps4}) shows that
we have the normalization identity
\begin{equation}\label{B.1}
{1 \over \varepsilon_{t^n,t}(P_\mu)}
\sum_\kappa \varepsilon_{t^{n-1},t}(P_\kappa)
\psi_{\mu/\kappa}(q,t) t^{|\kappa|} = 1
\end{equation}
where the summation over $\kappa$ is over the region (\ref{pro2}).
Making use of (\ref{rr2}) and (\ref{3.7}) we find that with
\begin{equation}\label{B.1a}
t = q^k, \quad y_{n-i} := q^{\kappa_i} t^{n-1-i}, \quad
x_{n+1-i} := q^{\mu_i} t^{n-i},
\end{equation}
(\ref{B.1}) can be rewritten to read
\begin{eqnarray}\label{B.2}
&& \sum_y \prod_{j=1}^{n-1} y_j
\prod_{1 \le i < j \le n-1} (y_i - y_j)
\prod_{1 \le i \le j \le n-1} (qy_j/x_i;q)_{k-1}
(qy_i/x_{j+1};q)_{k-1} \nonumber \\
&& \qquad =
{((q;q)_{k-1})^n \over (q;q)_{kn-1}}
\prod_{1 \le i < j \le n}(x_i-x_j)(qx_i/x_j;q)_{k-1}
(qx_j/x_i)_{k-1}
\end{eqnarray}
where the summation over $y$ is over the regions
\begin{equation}\label{B.2a} 
y_i = x_i, qx_i, q^2x_i,q^3x_i, \dots, x_{i+1} \quad
(i=1,\dots,n-1).
\end{equation}
This can be viewed as a multidimensional $q$-integral. It is the
special case $s_1=\cdots=s_n=k$ of Evans \cite{Ev92}
$q$-integral
\begin{eqnarray}\label{B.3}
&& \sum_y \prod_{j=1}^{n-1} y_j
\prod_{1 \le i < j \le n-1} (y_i - y_j)
\prod_{1 \le i \le j \le n-1} (qy_j/x_i;q)_{s_i-1}
(qy_i/x_{j+1};q)_{s_{j+1}-1} \nonumber \\
&& \qquad =
{(\prod_{l=1}^n(q;q)_{s_l-1})^n \over (q;q)_{\sum_{l=1}^n s_l -1}}
\prod_{1 \le i < j \le n}(x_i-x_j)(qx_i/x_j;q)_{s_j-1}
(qx_j/x_i)_{s_i-1},
\end{eqnarray}
and it is also the special case $\nu = \emptyset$ of Okounkov's
$q$-integral formula for the Macdonald polynomials \cite{Ok98}
\begin{eqnarray}\label{B.4}
&&\Big ( \prod_{1 \le i < j \le n} (x_i - x_j)
(qx_i/x_j;q)_{k-1} (qx_j/x_i)_{k-1} \Big )^{-1}
\sum_y P_\nu(y_1,\dots,y_{n-1};q,t) \prod_{j=1}^{n-1} y_j
\prod_{1 \le i < j \le n-1}(y_i - y_j) \nonumber \\
&& \quad \times
\prod_{1\le i \le j \le n-1}(qy_j/x_i;q)_{k-1}
(qy_i/x_{j+1};q)_{k-1} =
{((q;q)_{k-1})^n \over (q;q)_{kn-1}}
{\varepsilon_{t^{n-1},t}(P_\nu) \over
\varepsilon_{t^{n},t}(P_\nu)}
P_\nu(x_1,\dots,x_n;q,t).
\end{eqnarray}

This latter observation allows us to write (\ref{B.4}) in a structured
form from which a simple derivation in the general $\nu$ case
follows. Recalling (\ref{B.1a}) and (\ref{B.2a}), and introducing
the notation $u_\mu^{(n)}$ ($u_\kappa^{(n-1)}$) from
\cite[pg.~331]{Ma95} to denote the evaluation map on polynomials in
$n$-variables ($(n-1)$-variables) which sets $x_i=q^{\mu_i} t^{n-i}$
($y_i=q^{\kappa_i} t^{n-1-i}$), we see that (\ref{B.4}) can be
rewritten
\begin{equation}\label{B.5}
\sum_\kappa {u_\kappa^{(n-1)}(P_\nu) \over
u_0^{(n-1)}(P_\nu)}
{u_0^{(n-1)}(P_\kappa) \over u_0^{(n)}(P_\mu)}
\psi_{\mu/\kappa}(q,t) t^{|\kappa|} =
{u_\mu^{(n)}(P_\nu) \over u_0^{(n)}(P_\nu)}.
\end{equation}
According to \cite[(6.6)~pg.~332]{Ma95}, the evaluation map satisfies
the symmetry relation
\begin{equation}\label{B.s}
{u_\kappa^{(n-1)}(P_\nu) \over
u_0^{(n-1)}(P_\nu)} =
{u_\nu^{(n-1)}(P_\kappa) \over
u_0^{(n-1)}(P_\kappa)}
\end{equation}
and so we have 
\begin{equation}\label{B.s1}
{\rm LHS(\ref{B.5})} =
{1 \over  u_0^{(n)}(P_\mu)}
u_\nu^{(n-1)} \Big (
\sum_{\kappa} P_\kappa(y_1,\dots,y_{n-1})
\psi_{\mu/\kappa}(q,t) t^{|\kappa|} \Big ).
\end{equation}
The fundamental recurrence (\ref{rr1}) allows the summation over $\kappa$
in this expression to be performed, leaving us with
\begin{equation}\label{B.s2}
{\rm LHS(\ref{B.5})} =
{1 \over  u_0^{(n)}(P_\mu)}
u_\nu^{(n-1)} \Big ( P_\mu(ty_1,\dots,ty_{n-1},1) \Big ).
\end{equation}
But according to the definitions of $u_\nu^{(n-1)}$ and $u_\nu^{(n)}$
we have
$$
u_\nu^{(n-1)} \Big ( P_\mu(ty_1,\dots,ty_{n-1},1) \Big ) =
u_\nu^{(n)}( P_\mu).
$$
Substituting this in (\ref{B.s2}) then using the symmetry relation
(\ref{B.s}) in the form
$$
{u_\nu^{(n)}(P_\mu) \over
u_0^{(n)}(P_\mu)} = {u_\mu^{(n)}(P_\nu) \over
u_0^{(n)}(P_\nu)}
$$
gives the RHS of (\ref{B.5}).

\section*{Appendix C}
\subsection*{Matrix Bessel functions}
\renewcommand{\theequation}{C.\arabic{equation}}
\setcounter{equation}{0}
Guhr and Kohler \cite{GK02} defined the matrix Bessel function
$$
\Phi_N^{(\beta)}(x,k) = \int d\mu(U) \, \exp \Big ( i {\rm Tr} \,
U^\dagger x U k \Big )
$$
where $U \in U(N;\beta)$ with $U(N;1) = O(N)$ (orthogonal group),
$U(N;2) = U(N)$ (unitary group) and $U(N;4) = Sp(N)$
(unitary symplectic group). Furthermore $x={\rm diag}(x_1,\dots,x_N)$,
$k={\rm diag}(k_1,\dots,k_N)$ for $\beta=1,2$ while
$x={\rm diag}(x_1,x_1,\dots,x_N,x_N)$, $k={\rm diag}(k_1,k_1\dots,k_N,k_N)$
for $\beta = 4$ (and thus each eigenvalue is doubly degenerate in this case).
Let $U_N$ denote the $N$th column of $U$ (for $\beta = 4$ each element is
itself a $2 \times 2$ matrix representing a real quaternion) and construct
the random corank 1 projection of $x$ by
\begin{equation}\label{xp}
\Pi x \Pi, \qquad \Pi := {\bf 1} - U_N U_N^\dagger
\end{equation}
where ${\bf 1}$ denotes the identity matrix.
In  \cite{GK02} the non-zero eigenvalues of this projection are termed
the radial Gelfand-Tzeltin coordinates.

Let $A$ denote the $N \times N-1$ matrix with 1's down its diagonal
and 0's elsewhere. Then we can write
$$
\Phi_N^{(\beta)}(x,k) = \int d\mu(U) \, \exp \Big ( i {\rm Tr} \,
U^\dagger x U k A A^T\Big ) \exp \Big ( i {\rm Tr} \,
U^\dagger x U k ({\bf 1} - A A^T)\Big ).
$$
Noting that $kA=A\tilde{k}$ with $\tilde{k} = {\rm diag}(k_1,\dots,k_{N-1})$
while $k({\bf 1} - A A^T) = ({\bf 1} - A A^T) k_N$, and defining $x'$ as the
$(N-1) \times (N-1)$ matrix
\begin{equation}\label{xd}
x' := A^T U^\dagger x U A
\end{equation}
we see that we have
$$
\Phi_N^{(\beta)}(x,k) = \int d\mu(U) \, \exp \Big ( i {\rm Tr} \, i
x' \tilde{k} \Big )
\exp \Big ( i k_N ({\rm Tr} \, x - {\rm Tr} \, x') \Big ).
$$
A simple calculation (see \cite{GK02}) shows that the eigenvalues of
(\ref{xd}) coincide with the non-zero eigenvalues of (\ref{xp}).
Consequently the second exponential in the integrand only depends on
$U_N$, so if we decompose $d \mu(U)$ as
$$
d \mu(U) = d\mu(\tilde{U}) dU_N 
$$
where $\tilde{U} \in U(N-1;\beta)$ we have the factorization \cite{GK02}
\begin{eqnarray*}
\Phi_N^{(\beta)}(x,k) & = &
\int d U_N \, \exp \Big ( i k_N ({\rm Tr} \, x - {\rm Tr} \, x') \Big )
\int  d\mu(\tilde{U}) \,  \exp \Big ( i {\rm Tr} \, i
x' \tilde{k} \Big ) \nonumber \\
 & = & 
\int d U_N \, \exp \Big ( i k_N ({\rm Tr} \, x - {\rm Tr} \, x') \Big )
\Phi_{N-1}^{(\beta)}(x,k).
\end{eqnarray*}
Finally, the change of variables from $U_N$ to the eigenvalues of (\ref{xp}),
with the $x$'s given,
is carried out according to Corollary \ref{cor1}. 

\section*{Appendix D}
\subsection*{Sampling from $J_0^{(n,\_)}$ in a Monte Carlo calculation}
\renewcommand{\theequation}{D.\arabic{equation}}
\setcounter{equation}{0}
The eigenvalue PDF for Dyson's circular ensembles of random unitary
matrices with orthogonal $(\beta = 1)$, unitary $(\beta = 2)$ and
symplectic $(\beta = 4)$ symmetry is given by
\begin{equation}
{1 \over C_N} \prod_{1 \le j < k \le N} |e^{i \theta_k} -
e^{i \theta_j} |^\beta.
\end{equation}
The corresponding scaled two point correlation function is defined as
\begin{eqnarray}
\lefteqn{
\rho_2(x) := \lim_{N \to \infty}
{N(N-1) \over C_N}
|1 - e^{2 \pi i x/N}|^\beta} \nonumber \\
&&
\times
 \int_0^{2 \pi} d \theta_3 \cdots
\int_0^{2 \pi} d \theta_N \, \prod_{l=3}^N |1 - e^{i \theta_l}|^\beta
|e^{2 \pi i x/N} - e^{i \theta_l} |^\beta
\prod_{3 \le j < k \le N}
| e^{i \theta_k} -  e^{i \theta_j}|^\beta
\end{eqnarray}
(note that with this scaling the mean eigenvalue spacing is unity).
For general even $\beta > 0$ it was shown in \cite{Fo93aa} that
\begin{equation}\label{e3a}
\rho_2(x) = (\beta/2)^\beta {((\beta/2)!)^3 \over \beta !
(3 \beta / 2)!} (2 \pi x)^\beta
\Big \langle \cos 2 \pi x(\sum_{j=1}^\beta x_j - \beta / 2)
\Big \rangle
\end{equation}
where the average is over the PDF $J_0^{(\beta,\_)}(x)$ (\ref{4.10b})
with $k=2/\beta$, $\alpha = \beta = -1+2/\beta$.
But we know the polynomial $A_\beta^{\#}(x)$ computed from the random
recurrence (\ref{mr}) with $k=2/\beta$, $\alpha_0 = \beta_0 = -1+2/\beta$
has zeros with this distribution, and so 
$\sum_{j=1}^\beta x_j =
- [x^{\beta - 1}] A_\beta^{\#}(x) =: \gamma_\beta$ where the symbol
$[x^p]$ denotes the coefficient of $x^p$. According to the general
theory of Monte Carlo integration, if we sample $\gamma_\beta$
a total of $M$ times we then have
\begin{equation}\label{e3b}
\rho_2(x) = (\beta/2)^\beta {((\beta/2)!)^3 \over \beta !
(3 \beta / 2)!} (2 \pi x)^\beta
\Big ( {1 \over M} \sum_{j=1}^M 
\cos (2 \pi x (\gamma_\beta^{(j)} - \beta/2) +
O\Big ( {1 \over \sqrt{M}} \Big ) \Big ).
\end{equation}

It is illustrative to compute (\ref{e3b}) in the case $\beta = 4$,
since then we have an alternative formula to (\ref{e3a})
for the exact evaluation \cite{Me91},
\begin{equation}\label{e3c}
\rho_2(x) = 1 - \Big ( {\sin 2 \pi x \over 2 \pi x} \Big )^2 +
{1 \over 2 \pi} {d \over dx}  \Big ( {\sin 2 \pi x \over x}
\Big ) \int_0^{2 \pi x}  {\sin t \over t} \, dt,
\end{equation}
from which $\rho_2(x)$ can be computed to essentially arbitrary precision,
thereby allowing the accuracy of (\ref{e3b}) to be assertained.
The results are listed in Table \ref{t1}.

\begin{table}
\begin{center}
\begin{tabular}{l|l|l}
$x$ & Exact & Monte Carlo \\ \hline
0.2 & 0.01687 & 0.01690 \\
0.4 & 0.2059 & 0.207 \\
0.6 & 0.6641 & 0.674 \\
0.8 & 1.1173 & 1.15 \\
1.0 & 1.2257 & 1.30 \\
1.2 & 1.0208 & 1.15 \\
1.4 & 0.8308 & 1.05 
\end{tabular}
\end{center}
\caption{\label{t1} The value of $\rho_2(x)$ in the
case $\beta = 4$ as computed from (\ref{e3b}) using
$M = 22,500$ distinct samples of $\gamma_\beta$, tabulated against
the exact value computed from (\ref{e3c}). The deterioration of the
accuracy of the former as $x$ increases is due in part to
amplification of the error caused by the factor of $x^\beta$ in
(\ref{e3b}). }
\end{table}

\section*{Appendix E}
\subsection*{Eigenvalue PDF for a generalization of the matrix structure
(\ref{M.1l})}
\renewcommand{\theequation}{E.\arabic{equation}}
\setcounter{equation}{0}
Here we will compute the eigenvalue PDF for the random matrix
\begin{equation}\label{ms0}
M = A + XBX^\dagger + \sqrt{t} Y
\end{equation}
where $A$ ($B$) is a fixed $n\times n$ ($m \times m$) Hermitian matrix
with distinct eigenvalue $a_1,\dots,a_n$ ($b_1,\dots, b_m$), $X$ is a
$n \times m$ complex Gaussian matrix with real and imaginary parts having
variance $1/2$ and $Y$ is a random Hermitian matrix from the $n \times
n$ GUE (see e.g.~\cite{Fo02} for the definition of the latter). Our
expression will involve the operator
$$
{\cal L}^{-1}[f(s)](x) = \lim_{\tau \to 0} \int_{{\rm Re}(s)=0}
e^{s x + \tau s^2/2} f(s) \, {ds \over 2 \pi i}, \quad x \in \mathbb R,
$$
which is the inverse of the two sided Laplace transform
$$
{\cal L}[f(x)](s) = \int_{-\infty}^\infty e^{-sx} f(x) \, dx.
$$
We will show that in the special case that $t=0$ and $B$ has rank 1 the
term involving ${\cal L}^{-1}$ can be evaluated explicitly, reclaiming
(\ref{ss2}) in the case $s_i=1$ ($i=1,\dots,n$). 

\begin{thm}\label{th2}
The eigenvalue PDF of the random matrix (\ref{ms0}) is given by
\begin{equation}\label{D.1a}
{1 \over n!} \det \Big [ {\cal L}^{-1} [e^{t s^2/2} \det(1 + Bs)^{-1}]
(x_i-a_j) \Big ]_{i,j=1,\dots,n}
{\det [x_i^{j-1}]_{i,j=1,\dots,n} \over \det [a_i^{j-1}]_{i,j=1,\dots,n}}.
\end{equation}
\end{thm}

\noindent
Proof. \quad Clearly, applying a unitary change of basis to $A$ leaves
the eigenvalue distribution of $M$ unchanged; we may thus consider the
random matrix
$$
M' := UAU^\dagger + X B X^\dagger + \sqrt{t} Y
$$
for $U$ Haar-distributed from $U(n)$. Denote by ${\bf E}$ the operation
of averaging over random variables, and consider the moment generating
function (multivariate Laplace transform)
$$
f_{M'} : \: C \to {\bf E}(\exp(-{\rm Tr}(M'C)))
$$
for Hermitian matrices $C$. We must therefore compute
$$
f_{M'}(C) = {\bf E}(\exp(-{\rm Tr}(UAU^\dagger C)))
{\bf E}(\exp(-{\rm Tr}(XBX^\dagger C)))
{\bf E}(\exp(-{\rm Tr}(\sqrt{t}YC)))
$$
Now, if $C$ has eigenvalues $c_1,\dots,c_n$ then
$$
{\bf E}(\exp(-{\rm Tr}(XBX^\dagger C))) = \prod_{i=1}^m \prod_{j=1}^n
{\bf E}(\exp(-b_ic_j|X_{ij}|^2))= \prod_{i=1}^m \prod_{j=1}^n
{1 \over 1 + b_i c_j} = \prod_{i=1}^n \det (1 + Bc_i)^{-1}
$$
and
$$
{\bf E}(\exp(-{\rm Tr}((\sqrt{t}YC))) =
{\bf E}(\exp(-\sum_{i=1}^n
((\sqrt{t}Y_{ii} c_i)) = \prod_{i=1}^n \exp(-tc_i^2/2).
$$
Finally, by the Harish-Chandra/Itzykson-Zuber formula
(see e.g.~\cite{Me91})
\begin{equation}\label{D.2}
{\bf E}(\exp(-{\rm Tr}(UAU^\dagger C))) = (-1)^{n(n-1)/2}
\prod_{i=1}^n(i-1)!
{\det [ \exp(-a_j c_i) ]_{i,j=1,\dots,n} \over
\det [ a_i^{j-1} ]_{i,j=1,\dots,n} \det [c_i^{j-1}]_{i,j=1,\dots,n}}
\end{equation}
so we have
\begin{equation}\label{D.3}
f_{M'}(C) = (-1)^{n(n-1)/2}
\prod_{i=1}^n(i-1)!
{\det [ \exp(-a_j c_i) \exp(-tc_i^2/2)
 \det (1 + Bc_i)^{-1} ]_{i,j=1,\dots,n} \over
\det [ a_i^{j-1} ]_{i,j=1,\dots,n} \det [c_i^{j-1}]_{i,j=1,\dots,n}}. 
\end{equation}
Note that this 
is analytic for $C$ in a neighbourhood of 0.

On the other hand, suppose we know that a random matrix $\tilde{M}$ has
distribution invariant under unitary change of basis, and has
eigenvalue PDF
\begin{equation}\label{D.4}
{1 \over Z} \det [x_i^{j-1}]_{i,j=1,\dots,n}
 \det [g_j(x_i)]_{i,j=1,\dots,n} 
\end{equation}
for some functions $g_j$ with Laplace transform defined in a neighbourhood
of $0$. Then
\begin{eqnarray}\label{D.5}
\lefteqn{{\bf E}(\exp(-{\rm Tr}(C\tilde{M})))} \nonumber \\
&& = {\bf E}({\bf E}_{U \in U(n)}(\exp(-{\rm Tr}(CU\tilde{M}U^\dagger))))
\nonumber \\
&& =  {(-1)^{n(n-1)/2}
\prod_{i=1}^n(i-1)! \over Z  \det [c_i^{j-1}]_{i,j=1,\dots,n}}
\int_{-\infty}^\infty dx_1 \cdots \int_{-\infty}^\infty dx_n \,
\det [ \exp(-x_j c_i) ]_{i,j=1,\dots,n}
\det [g_j(x_i)]_{i,j=1,\dots,n} \nonumber \\
&& =  {(-1)^{n(n-1)/2}
\prod_{i=1}^{n+1}(i-1)! \over Z  \det [c_i^{j-1}]_{i,j=1,\dots,n}}
\int_{-\infty}^\infty dx_1 \cdots \int_{-\infty}^\infty dx_n \,
\det \Big [ \int_{-\infty}^\infty e^{-x c_i} g_j(x) \, dx 
\Big ]_{i,j=1,\dots,n}
\nonumber \\
&& = {(-1)^{n(n-1)/2}
\prod_{i=1}^{n+1}(i-1)! \over Z  \det [c_i^{j-1}]_{i,j=1,\dots,n}}
\int_{-\infty}^\infty dx_1 \cdots \int_{-\infty}^\infty dx_n \,
\det [{\cal L}[g_j(x)](c_i)]_{i,j=1,\dots,n}
\end{eqnarray}
where to obtain the second equality, use has been made of the
Harish-Chandra/Itzykson-Zuber formula (\ref{D.2}). Comparing
(\ref{D.5}) with (\ref{D.2}) allows us to deduce the value of $Z$
and $g(x)$, which when substituted in (\ref{D.4}) gives
(\ref{D.1a}).
\hfill $\square$

In the special case $t=0$, $B = {\rm diag}(b,0,\dots,0)$, the random matrix
in (\ref{D.1a}) coincides with (\ref{M.1l}). In this case
$$
{\cal L}^{-1}[e^{t s^2/2} \det(1 + Bs)^{-1}]
(x_i-a_j) = {\cal L}^{-1}[(1+bs)^{-1}](x_i-a_j)
= \chi_{x_i - a_j > 0} {1 \over b} e^{-(x_i - a_j)/b}, \quad b>0
$$
and so
$$
\det\Big [{\cal L}^{-1}[e^{t s^2/2} \det(1 + Bs)^{-1}]
(x_i-a_j) \Big ]_{i,j=1,\dots,n} = b^{-n}
e^{-\sum_{i=1}^n (x_i - a_j)/b}
\det [ \chi_{x_i \ge a_j} ]_{i,j=1,\dots,n}.
$$
For $a_1 > a_2 > \cdots > a_n$ the latter determinant is non-zero only
when
$$
x_1 \ge a_1 \ge x_2 \ge \cdots x_n \ge a_n
$$
in which case it is 1. Thus we see that (\ref{D.1a}) reproduces the
eigenvalue PDF for the matrices (\ref{M.1l}) in the case that
$\vec{x}$ is complex and the eigenvalues of $A$ distinct. 

Another special case of interest is $B=0$. Noting that
$$
{\cal L}^{-1}[e^{t s^2/2}](x_i-a_j) = (2 \pi t)^{-1/2}
e^{-(x_i - a_j)^2/2t}
$$
we see from Theorem \ref{th2} that $A + \sqrt{t} Y$ for $Y$ an element
of the GUE has eigenvalue density
$$
{1 \over n! (2 \pi t)^{n/2}}
\det [ e^{-(x_i - a_j)^2/2t} ]_{i,j=1,\dots,n}
{\det [ x_i^{j-1} ]_{i,j=1,\dots,n} \over
\det [ a_i^{j-1} ]_{i,j=1,\dots,n} }.
$$
This is a well known consequence of the Harish-Chandra/Itzykson-Zuber
formula (see e.g.~\cite{Me91}).

%\bibliographystyle{plain}
%\bibliography{book}

\end{document}